\PassOptionsToPackage{dvipsnames}{xcolor}
\documentclass[sigconf,nonacm,balance=false]{aamas}
\setcitestyle{nosort,open={(},close={)},aysep={,}}%
\usepackage{mathtools,nicefrac,xspace}
\usepackage{tikz,array,makecell}
\usetikzlibrary{arrows.meta,positioning}
\usepackage{xurl}
\usepackage{algorithm}
\usepackage{algpseudocode}
\usepackage{longtable}

\setcopyright{none}
\AtBeginDocument{%
  \fancypagestyle{firstpagestyle}{\fancyhf{}%
    \fancyfoot[C]{\footnotesize\thepage}}%
  \fancypagestyle{standardpagestyle}{\fancyhf{}%
    \fancyhead[L]{\footnotesize\shorttitle}%
    \fancyhead[R]{\footnotesize\shortauthors}%
    \fancyfoot[C]{\footnotesize\thepage}}%
  \pagestyle{standardpagestyle}}

\makeatletter\def\@titlefont{\Huge\bfseries}\makeatother
\title[Local Work, Global Progress]{Local Work, Global Progress:\texorpdfstring{\\}{} A Convergence Theory of AI Agent Collaboration}

\author{Ivan Ilin}
\affiliation{%
  \institution{KAUST}}
\email{ivan.ilin@kaust.edu.sa}

\author{Peter Richt\'arik}
\affiliation{%
  \institution{KAUST}}
\email{peter.richtarik@kaust.edu.sa}

\renewcommand{\shortauthors}{Ilin and Richt\'arik}
\hypersetup{pdftitle={Local Work, Global Progress: A Convergence Theory of AI Agent Collaboration},
  pdfauthor={Ivan Ilin, Peter Richtarik},pdfsubject={},
  pdfkeywords={multi-agent collaboration, inexact block coordinate descent, correction learning}}

\definecolor{repolink}{RGB}{128,40,56}
\newcommand{\repourl}{\href{https://github.com/vectozavr/local-work-global-progress}{\color{repolink}\texttt{https://}\allowbreak\texttt{github.com/}\allowbreak\texttt{vectozavr/}\allowbreak\texttt{local-work-global-progress}}}

\newcommand{\eqdef}{\coloneqq}
\makeatletter\newcommand{\keepwithlist}{\@beginparpenalty=\@M}
\newcommand{\holdheading}{\@secpenalty=\@M}\newcommand{\releaseheading}{\@secpenalty=-300\relax}\makeatother

\newcommand{\cX}{\mathcal X}

\DeclareMathOperator*{\argmax}{arg\,max}
\DeclareMathOperator*{\argmin}{arg\,min}
\theoremstyle{acmdefinition}
\newtheorem{assumption}{Assumption}
\theoremstyle{acmplain}
\newtheorem{theorem}{Theorem}
\newtheorem{lemma}[theorem]{Lemma}
\newtheorem{proposition}[theorem]{Proposition}
\newtheorem{corollary}[theorem]{Corollary}


\begin{abstract}
Teams of large language model (LLM) agents are increasingly applied to complex tasks, but their collaboration protocols are mostly designed empirically, without convergence guarantees. We formalize collaborative problem solving as block optimization: each agent improves one part of a shared solution, guided by its local objective plus a correction for how its changes affect the other parts. A coordinator checks every proposal against the global objective and merges proposals that do not interfere with each other. We prove linear convergence up to an error floor that vanishes when the corrections are exact. Interfering proposals can also be merged once their combination has been checked; the rate then depends on how well the proposals actually combine, not on worst-case interactions. With GPT-OSS-120B agents on robot routing, a three-agent team finds collision-free plans in every run, better than those of a single agent that controls all robots. On a larger routing task, a six-agent team completes its search about twice as fast as the single agent (in median). Merging the best checked combination of proposals, rather than only proposals that do not interfere, cuts the team's search time by about 30\%. With constants measured on the runs, our convergence bound holds in every round and follows the observed progress. We also test the theory on query planning in SQLite and on a synthetic task.
\end{abstract}
\begin{document}
\maketitle
\section{Introduction}
Large language model (LLM) agents are increasingly asked to produce artifacts that are too large for a single request, such as software with many modules or plans for many robots. Human organizations handle such work by division of labor, and dividing work among agents promises two benefits. First, each request becomes smaller and easier, so a task that overwhelms one agent can become solvable. Second, the parts can be searched in parallel, so a solution arrives sooner. We do not aim to reduce monetary cost: like a construction crew compared with a single builder, a team spends more in total but completes work that one agent would finish slowly or not at all.

Not every task benefits from a team, and not every team needs coordination. A simple task, like changing a light bulb, is best done by one worker. If the parts do not interact, for example when one agent waters a garden while another goes shopping, agents can simply work in parallel. The interesting and common case lies in between: large tasks whose parts are coupled, such as the modules of a program, the components of an aircraft, or the routes of robots that share a floor. Each part is manageable alone, but a change to one part can change what is best for the others.

Coupling creates a coordination problem: improving each part in isolation need not improve the whole. A module can pass its own tests while violating another module's interface; a robot can reach its goal along a route that collides with a robot controlled by another agent. A global check can reject a harmful proposal, but rejection does not tell the agent which alternative to try next: it needs information that changes how it ranks its own candidates.

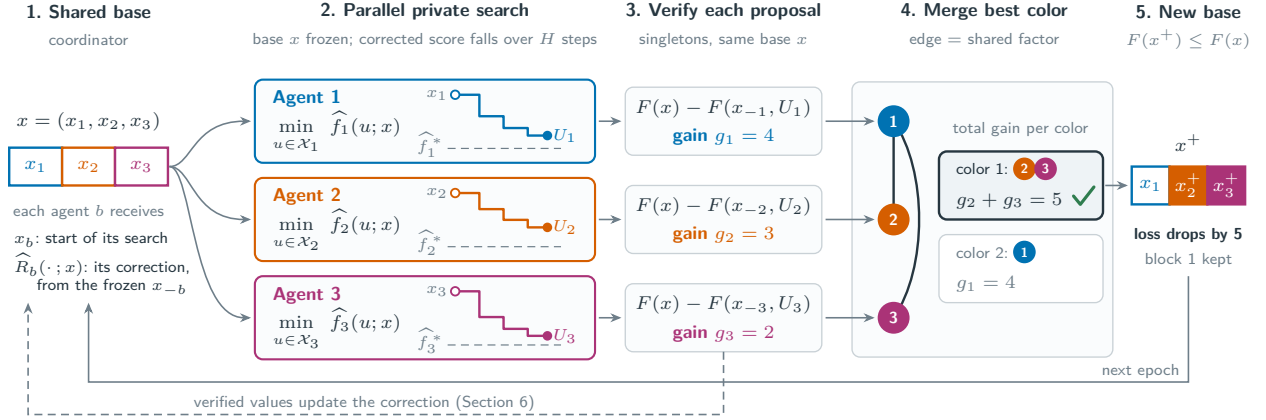
\begin{figure*}[t]
\centering
\providecommand{\cX}{\mathcal X}
\expandafter\ifx\csname ifprotocolsearchplot\endcsname\relax
  \expandafter\newif\csname ifprotocolsearchplot\endcsname
  \csname protocolsearchplottrue\endcsname
\fi
\begingroup
\definecolor{agentA}{HTML}{0072B2}%
\definecolor{agentB}{HTML}{D55E00}%
\definecolor{agentC}{HTML}{AA3377}%
\definecolor{ink}{HTML}{263640}%
\definecolor{rulegray}{HTML}{C9D1D6}%
\definecolor{mutegray}{HTML}{6E7E89}%
\definecolor{paper}{HTML}{FAFCFD}%
\definecolor{pick}{HTML}{EEF3F6}%
\definecolor{okgreen}{HTML}{2F7D4F}%
\begin{tikzpicture}[x=1cm,y=1cm,font=\sffamily\footnotesize,
  >={Stealth[length=4.2pt,width=3.4pt]},
  flow/.style={->,draw=mutegray,line width=.6pt},
  lane/.style={draw=#1,line width=.8pt,rounded corners=3pt,fill=paper},
  card/.style={draw=rulegray,line width=.5pt,rounded corners=3pt,fill=white},
  title/.style={font=\sffamily\footnotesize\bfseries,text=ink},
  sub/.style={font=\sffamily\scriptsize,text=mutegray},
  blocknode/.style={circle,fill=#1,draw=none,minimum size=12pt,inner sep=0pt,
    text=white,font=\sffamily\scriptsize\bfseries},
  mininode/.style={circle,fill=#1,draw=none,minimum size=8.5pt,inner sep=0pt,
    text=white,font=\sffamily\tiny\bfseries}]

\node[title] at (1.25,5.45) {1. Shared base};
\node[sub]   at (1.25,5.08) {coordinator};
\node[title] at (5.7,5.45) {2. Parallel private search};
\node[sub]   at (5.7,5.08) {base $x$ frozen; corrected score falls over $H$ steps};
\node[title] at (9.65,5.45) {3. Verify each proposal};
\node[sub]   at (9.65,5.08) {singletons, same base $x$};
\node[title] at (13.1,5.45) {4. Merge best color};
\node[sub]   at (13.1,5.08) {edge $=$ shared factor};
\node[title] at (15.8,5.45) {5. New base};
\node[sub]   at (15.8,5.08) {$F(x^+)\leq F(x)$};

\node[text=ink] at (1.25,4.0) {$x=(x_1,x_2,x_3)$};
\foreach \i/\c in {1/agentA,2/agentB,3/agentC}{
  \draw[draw=\c,line width=.9pt,fill=white] ({0.2+0.7*(\i-1)},3.15) rectangle ++(0.7,0.5);
  \node[text=\c] at ({0.55+0.7*(\i-1)},3.4) {$x_{\i}$};
}
\node[sub] at (1.25,2.80) {each agent $b$ receives};
\node[anchor=west,text=ink,font=\sffamily\scriptsize] at (0.15,2.44) {$x_b$: start of its search};
\node[anchor=west,text=ink,font=\sffamily\scriptsize] at (0.15,2.10) {$\widehat R_b(\cdot\,;x)$: its correction,};
\node[anchor=west,text=ink,font=\sffamily\scriptsize] at (0.45,1.80) {from the frozen $x_{-b}$};
\foreach \y in {4.0,2.7,1.4}{\draw[flow] (2.32,3.4) .. controls (2.62,3.4) and (2.62,\y) .. (3.4,\y);}

\foreach \y/\c/\b/\fa/\fb/\fc in {4.0/agentA/1/.40/.62/.76, 2.7/agentB/2/.30/.50/.64, 1.4/agentC/3/.48/.70/.83}{
  \draw[lane=\c] (3.45,\y-0.55) rectangle (7.95,\y+0.55);
  \node[anchor=west,text=\c,font=\sffamily\footnotesize\bfseries] at (3.57,\y+0.3) {Agent \b};
  \node[anchor=west,text=ink] at (3.57,\y-0.15) {$\min\limits_{u\in\cX_{\b}}\; \widehat f_{\b}(u;x)$};
  \ifprotocolsearchplot
    \pgfmathsetmacro{\ytop}{\y+0.35}\pgfmathsetmacro{\ybot}{\y-0.36}\pgfmathsetmacro{\yspan}{\ytop-\ybot}
    \pgfmathsetmacro{\va}{\ytop-\fa*\yspan}\pgfmathsetmacro{\vb}{\ytop-\fb*\yspan}
    \pgfmathsetmacro{\vc}{\ytop-\fc*\yspan}
    \draw[densely dashed,draw=mutegray,line width=.45pt] (6,\ybot) -- (7.5,\ybot);
    \node[anchor=base east,text=mutegray,font=\sffamily\scriptsize,inner sep=1.5pt] at (6,\ybot-0.03) {$\widehat f^{\,*}_{\b}$};
    \draw[draw=\c,line width=.9pt,line join=round]
      (6.1,\ytop) -- (6.42,\ytop) -- (6.42,\va) -- (6.74,\va) -- (6.74,\vb) -- (7.06,\vb)
      -- (7.06,\vc) -- (7.32,\vc);
    \draw[draw=\c,fill=white,line width=.7pt] (6.1,\ytop) circle (1.6pt);
    \fill[\c] (7.32,\vc) circle (1.8pt);
    \node[anchor=east,text=mutegray,font=\sffamily\scriptsize,inner sep=2.5pt] at (6.1,\ytop) {$x_{\b}$};
    \node[anchor=west,text=\c,font=\sffamily\scriptsize,inner sep=2pt] at (7.32,\vc) {$U_{\b}$};
  \else
    \node[text=mutegray] (s\b) at (6.2,\y-0.15) {$x_{\b}$};
    \node[text=\c] (e\b) at (7.5,\y-0.15) {$U_{\b}$};
    \draw[->,draw=\c,line width=.8pt] (s\b) -- node[above,font=\sffamily\scriptsize,text=mutegray]{$H$ steps} (e\b);
  \fi
  \draw[flow] (8,\y) -- (8.3,\y);
}

\foreach \y/\c/\b/\g in {4.0/agentA/1/4, 2.7/agentB/2/3, 1.4/agentC/3/2}{
  \draw[card,fill=paper] (8.35,\y-0.45) rectangle (10.95,\y+0.45);
  \node[text=ink] at (9.65,\y+0.17) {$F(x)-F(x_{-\b},U_{\b})$};
  \node[text=\c,font=\sffamily\footnotesize\bfseries] at (9.65,\y-0.2) {gain $g_{\b}=\g$};
}

\draw[card,fill=paper] (11.35,0.88) rectangle (14.8,4.52);
\foreach \y in {4.0,2.7,1.4}{\draw[flow] (11.0,\y) -- (11.65,\y);}
\draw[draw=ink,line width=.9pt] (11.9,4.0) -- (11.9,2.7);
\draw[draw=ink,line width=.9pt] (11.9,4.0) .. controls (12.35,3.35) and (12.35,2.05) .. (11.9,1.4);
\foreach \y/\c/\b in {4.0/agentA/1, 2.7/agentB/2, 1.4/agentC/3}{
  \node[blocknode=\c] at (11.9,\y) {\b};
}
\node[sub] at (13.58,3.9) {total gain per color};
\draw[card,draw=ink,line width=.9pt,fill=pick] (12.5,2.7) rectangle (14.67,3.6);
\node[anchor=west,text=ink,font=\sffamily\scriptsize] at (12.61,3.37) {color 1:};
\node[mininode=agentB] at (13.63,3.37) {2};
\node[mininode=agentC] at (13.91,3.37) {3};
\node[anchor=west,text=ink] at (12.61,2.95) {$g_2+g_3=5$};
\draw[okgreen,line width=1.2pt,line cap=round,line join=round]
  (14.28,3.01) -- (14.37,2.9) -- (14.55,3.13);
\draw[card] (12.5,1.6) rectangle (14.67,2.5);
\node[anchor=west,text=mutegray,font=\sffamily\scriptsize] at (12.61,2.27) {color 2:};
\node[mininode=agentA] at (13.63,2.27) {1};
\node[anchor=west,text=mutegray] at (12.61,1.85) {$g_1=4$};

\draw[flow] (14.7,3.15) -- (15,3.15);
\node[text=ink] at (15.8,3.72) {$x^+$};
\draw[draw=agentA,line width=.9pt,fill=white] (15.05,2.9) rectangle ++(0.5,0.5);
\node[text=agentA] at (15.3,3.15) {$x_1$};
\draw[draw=agentB,line width=.9pt,fill=agentB] (15.55,2.9) rectangle ++(0.5,0.5);
\node[text=white] at (15.8,3.15) {$x_2^+$};
\draw[draw=agentC,line width=.9pt,fill=agentC] (16.05,2.9) rectangle ++(0.5,0.5);
\node[text=white] at (16.3,3.15) {$x_3^+$};
\node[text=ink,font=\sffamily\scriptsize\bfseries] at (15.8,2.5) {loss drops by 5};
\node[text=mutegray,font=\sffamily\scriptsize] at (15.8,2.17) {block 1 kept};

\draw[flow] (15.8,1.95) -- (15.8,0.55) -- (1.25,0.55) -- (1.25,1.6);
\node[sub,anchor=south east,inner sep=0.6pt] at (15.71,0.57) {next epoch};
\draw[flow,densely dashed] (9.65,0.95) -- (9.65,0.12) -- (0.45,0.12) -- (0.45,1.6);
\node[sub,anchor=south,inner sep=0.6pt] at (4.9,0.14)
  {verified values update the correction (Section~\ref{sec:main-learning})};
\end{tikzpicture}
\endgroup
\Description{A five-stage pipeline. Stage one: a coordinator holds the shared solution x with three blocks; each agent b receives its own block, where its search starts, and its correction, computed from the other blocks, which stay frozen. Stage two: three agents, drawn in blue, orange and magenta, each minimize their local objective plus correction over their own block; a small staircase shows the corrected score of the best candidate dropping at each of three private steps, from zero at the current block to the proposal, which stays above a dashed line marking the minimum. Stage three: a verifier evaluates each proposal alone against the same base, giving gains four, three and two. Stage four: in the dependency graph block one is connected to blocks two and three, which are not connected to each other. Color one contains blocks two and three with total gain five and is marked as selected; color two contains block one with gain four. Stage five: the new base replaces blocks two and three and keeps block one, and the loss drops by five. A solid arrow returns to the start for the next epoch; a dashed arrow shows that verified values update the correction.}
\caption{One epoch of Algorithm~\ref{alg:protocol} on a constructed example with illustrative gains. Staircases show each agent's best corrected score so far, which need not reach the minimum $\widehat f_b^{\,*}$ of the corrected objective. Each proposal is verified alone, and the coordinator accepts the color with the largest total gain: blocks 2 and 3 ($3+2=5$) rather than the single largest gain ($4$).}
\label{fig:protocol}
\end{figure*}

We study this problem through finite block optimization. An agent is a local search procedure, for example an LLM, that owns one block of a shared solution and searches it while the other blocks stay frozen. A \emph{correction} added to its local objective tells the agent how its changes affect the other blocks. Agents search their corrected objectives in parallel. A verifier then scores each proposal against the current shared solution, and a coloring of the dependency graph selects proposals that do not interfere (Figure~\ref{fig:protocol}). The correction steers the proposals, while verification and coloring ensure that the shared solution never gets worse.

We test the protocol with GPT-OSS-120B agents on robot routing, where robots of different agents can collide. A team of three agents finds collision-free plans in every run, better than those of a single agent that controls all robots at once. The same team without corrections rarely avoids collisions. On a larger task, a team of six agents also searches about twice as fast as the single agent in median. The coordinator can even merge interfering proposals once the verifier confirms that together they lower the loss more than the non-interfering ones alone. This rule, which our theory covers, cuts the team's search time by about 30\% and lets it reach low losses sooner than single agents searching in parallel. With constants measured on the runs, the convergence bound holds in every round, so theory and practice agree. Two smaller studies, on query planning in SQLite and on a synthetic objective, explore the theory and its limits further.

\holdheading
\paragraph{Contributions.}
{\keepwithlist\begin{enumerate}
\item We formalize collaborative problem solving as block optimization with corrected local objectives, and introduce a protocol of parallel private search, exact verification and compatible merging (Section~\ref{sec:main-setup}).
\item We prove linear convergence with an explicit error floor (Theorem~\ref{thm:rate}). The rate separates task geometry, local search quality and merge compatibility.
\item We show that the coordinator can accept the best verified combination of the agents' proposals, keeping the compatible work of interfering blocks, with the same guarantees and a rate set by how well the proposals actually combine (Section~\ref{sec:main-combination}).
\item We show how agents can learn reusable corrections from global evaluations, with a joint convergence rate for the solution and the correction (Appendix~\ref{sec:learn}), and account for calibration, local work and synchronization costs (Sections~\ref{sec:main-learning} and~\ref{sec:main-cost}).
\item We evaluate the protocol with teams of three and six LLM agents on robot routing and relate the observed convergence to the theorem; a database study shows when acquiring correction information pays off, and a synthetic study confirms the error floor (Section~\ref{sec:experiments}).
\end{enumerate}}\releaseheading
The appendices contain further theory and examples (Appendix~\ref{app:development}), extended related work (Appendix~\ref{app:related}), and complete experimental protocols and outcomes.

The implementation and code for reproducing our experiments are available at \repourl.

\subsection{Notation}
All key notation used in this paper is summarized in tabular form in the \hyperref[supp:notation]{Notation} section before the appendices; see Table~\ref{tab:notation}.

\section{Related work}
\label{sec:main-related}

\paragraph{Block optimization and distributed coordination.}
Block coordinate descent and its convergence theory are classical \citep{bertsekas1989,tseng2001,nesterov2012}, as are its parallel and inexact variants \citep{richtarik2016,tappenden2016}. When parallel coordinate updates interfere through shared terms, prior analyses bound the interference \citep{bradley2011,richtarik2016} or group weakly coupled coordinates \citep{scherrer2012}. We instead verify every proposal and then either merge only blocks of one color, whose gains add exactly, or check the merged combination as a whole. Distributed constraint optimization assigns coupled variables to agents: MGM coordinates improving moves, DSA randomizes local updates, and DPOP propagates conditional utility tables \citep{maheswaran2004,zhang2005,petcu2005}. Our addition is an explicit interface for correcting candidate values and a bound that converts the correction error into lost global progress.

\paragraph{Persistent corrections and identification.}
ProxSkip and SCAFFOLD correct local work through control variates \citep{mishchenko2022,karimireddy2020}. Our corrections play the analogous role for discrete search, adjusting candidate scores of disjoint blocks rather than gradients; whether they also admit ProxSkip-type acceleration remains open. Our update for learning corrections (Section~\ref{sec:main-learning}) is a randomized Kaczmarz, or sketch-and-project, step \citep{strohmer2009,gower2015}, and additive Bayesian optimization and linear-bandit identification also exploit known structure and informative measurements \citep{kandasamy2015,soare2014}.

\paragraph{LLMs as evaluator-guided optimizers.}
LLMs can serve as search operators when paired with an automatic evaluator. OPRO proposes new solutions from a history of scored ones, and FunSearch and AlphaEvolve evolve programs under automated evaluation \citep{yang2024opro,romeraparedes2024,novikov2025}. Our agents are such evaluator-guided searchers, each restricted to one block. The theory asks what feedback the evaluator must provide so that parallel block searches add up to global progress.

\paragraph{LLM collaboration and multi-robot planning.}
MetaGPT and ChatDev coordinate software-development artifacts, and AutoGen composes conversing agents \citep{hong2024,qian2024,wu2024}. GPTSwarm optimizes agent graphs, and AgentPrune removes redundant communication \citep{zhuge2024,agentprune2025}. Reflexion retains feedback-derived text across trials, and EvoMAC adapts collaboration using verification feedback \citep{shinn2023,hu2025}. Decomposed prompting splits complex tasks into simpler subtasks \citep{khot2023}. In robotics, one LLM agent per robot can negotiate a joint plan through dialogue and re-plan from motion-planner feedback such as collisions \citep{mandi2024roco}, and LLMs can decompose instructions and allocate the subtasks among robots \citep{kannan2024smart}. Centralized, decentralized and hybrid LLM planners have also been compared as teams grow \citep{chen2024scalable}. LLMs solve multi-agent path finding on empty maps but still fail on standard maps with obstacles \citep{chen2024mapfllm}, which is the difficulty our routing study targets. These works show empirically that decomposition and feedback help; our analysis states which properties of the agents and of the feedback yield convergence. Appendix~\ref{app:related} gives a broader comparison.

\section{Model and protocol}
\label{sec:main-setup}

This section turns the coordination problem from the introduction into an optimization model. We first define the shared task and the corrected objective that each agent searches (Section~\ref{sec:main-task}), then the dependency graph that determines which changes can be combined safely (Section~\ref{sec:main-coloring}), and finally the protocol that combines them (Section~\ref{sec:main-protocol}, Algorithm~\ref{alg:protocol}).

\subsection{Task and corrected objectives}
\label{sec:main-task}
The shared solution is $x=(x_1,\ldots,x_m)$: agent $b$ owns block $x_b$, which takes values in a nonempty finite set $\cX_b$. The team solves
\begin{equation}\label{eq:problem}
\min_{x\in\cX}F(x),\qquad \cX=\cX_1\times\cdots\times\cX_m,
\end{equation}
for a loss $F:\cX\to\mathbb R$. An exact verifier returns $F(x)$ for any solution, although its formula need not be known to the agents. Each agent can also cheaply evaluate a local objective $L_b:\cX_b\to\mathbb R$ of its candidate blocks $u\in\cX_b$, which may ignore how $u$ interacts with the other blocks.

The optimal loss is $F^*$, and the \emph{conditional change} $f_b(u;x)$ is how much the global loss changes if agent $b$ alone replaces its block by $u$:
\begin{equation}\label{eq:global}
F^*\eqdef\min_{x\in\cX}F(x),\qquad f_b(u;x)\eqdef F(x_{-b},u)-F(x).
\end{equation}
These quantities serve the analysis; agents need not know them. The \emph{exact correction} $R_b$ is the part of the conditional change that the local objective misses, and an agent searches the \emph{corrected objective} $\widehat f_b$ built from an available approximation $\widehat R_b$:
\begin{align}
R_b(u;x)&\eqdef f_b(u;x)-[L_b(u)-L_b(x_b)],\label{eq:correction}\\
\widehat f_b(u;x)&\eqdef L_b(u)-L_b(x_b)+\widehat R_b(u;x).\label{eq:surrogate}
\end{align}
The four functions $f_b,R_b,\widehat R_b,\widehat f_b:\cX_b\times\cX\to\mathbb R$ take a candidate block and the current solution. If $\widehat R_b=R_b$, then $\widehat f_b=f_b$. We require \emph{anchoring}, $\widehat R_b(x_b;x)=0$, so that keeping the current block always scores zero. A useful correction $\widehat R_b(u;x)$ must vary with the candidate $u$, because adding the same constant to every candidate cannot change their ranking; constructing and communicating it costs work, which Section~\ref{sec:main-cost} accounts for.

Neither $L_b$ nor $\widehat R_b$ has to be given to the agent as a formula. An LLM agent receives its task and the information behind the correction in its prompt, for example which collisions the verifier has found, and proposes candidates that it judges better. The functions $L_b$ and $\widehat R_b$ describe the preferences that this information conveys; they let us state what the search must achieve (Assumption~\ref{ass:local}) without modeling how the agent reasons.

\paragraph{Example.}
In the routing task of Section~\ref{sec:main-routing}, $x_b$ lists the moves of the three robots that agent $b$ controls, and $L_b$ adds their delivery costs and the collisions among them. The exact correction $R_b$ is the change in the penalty for collisions with the other agents' robots. The implemented correction $\widehat R_b$ computes this change only for the robot pairs whose collisions the verifier has already reported, using the other robots' current routes.

\subsection{Dependency graph and coloring}
\label{sec:main-coloring}
We write the loss as a sum of $M$ real-valued terms, called \emph{factors}:
\begin{equation}\label{eq:factor-definition}
F(x)=\sum_{a=1}^M\ell_a(x_{N(a)}),
\end{equation}
where factor $\ell_a$ depends only on the blocks in a set $N(a)\subseteq\{1,\ldots,m\}$, and $x_{N(a)}\eqdef(x_b)_{b\in N(a)}$ denotes these blocks. Every loss can be written in this form, trivially as the single factor $\ell_1=F$ with $N(1)=\{1,\ldots,m\}$. Only the sets $N(a)$ must be known; the factors themselves, like the formula of $F$, may stay hidden in the verifier.

For example, the loss $F(x)=(x_1-x_2)^2+(x_1-x_3)^2$ of three blocks has two factors, $\ell_1(x_1,x_2)=(x_1-x_2)^2$ and $\ell_2(x_1,x_3)=(x_1-x_3)^2$, so $N(1)=\{1,2\}$ and $N(2)=\{1,3\}$.

The \emph{dependency graph} has one vertex per block and joins blocks $b\neq b'$ whenever some factor depends on both, that is, $b,b'\in N(a)$ for some $a$. It records how the objective couples the blocks, not which agents communicate. A \emph{proper coloring} with $\kappa$ colors is a map $\chi:\{1,\ldots,m\}\to\{1,\ldots,\kappa\}$ that gives adjacent blocks different colors; it splits the blocks into the \emph{color classes} $\mathcal C_j\eqdef\{b:\chi(b)=j\}$. In the example above, the graph has the edges $\{1,2\}$ and $\{1,3\}$, and $\chi(1)=2$, $\chi(2)=\chi(3)=1$ is a proper coloring with $\kappa=2$, as in step~4 of Figure~\ref{fig:protocol}.

The blocks of one factor are pairwise adjacent, so they all have different colors. If several blocks of one color change at once, each factor therefore sees at most one of the changes, and the individual changes add exactly: for every set $S\subseteq\mathcal C_j$ and candidates $U_b$, $b\in S$,
\begin{equation}\label{eq:additive}
F(x_{-S},U_S)-F(x)=\sum_{b\in S}f_b(U_b;x),
\end{equation}
where $(x_{-S},U_S)$ is $x$ with each block $x_b$, $b\in S$, replaced by $U_b$. This makes the merge of Section~\ref{sec:main-protocol} safe.

Fewer colors are better, because this merge keeps at least a fraction $1/\kappa$ of the total verified gain (see~\eqref{eq:merge} in Section~\ref{sec:main-protocol}); a greedy coloring needs at most one more color than the maximum degree of the graph. At one extreme, a separable task with exact local objectives, $F(x)=\sum_bL_b(x_b)$ up to a constant, has a graph without edges: $\kappa=1$, the exact correction is zero, every agent can optimize alone and all improvements can be merged at once. At the other extreme, the single factor $\ell_1=F$ gives a complete graph and $\kappa=m$, so only one block can safely be replaced per epoch: changes of two blocks that share a factor may each lower the loss and yet raise it together (Appendix~\ref{sec:obstacle}). Corrections and coloring matter exactly when blocks share factors. This safety has a price: the coloring separates any two blocks that could interact for some candidates, even when the proposals of an epoch do not. Section~\ref{sec:main-combination} lets the coordinator accept proposals of different colors together whenever the verifier confirms that they lower the loss more than the best color alone.

\subsection{Search, verification and merging}
\label{sec:main-protocol}
Algorithm~\ref{alg:protocol} states the protocol, and Figure~\ref{fig:protocol} illustrates one \emph{epoch}: a round of private search, verification and merging. Within an epoch, all agents search from the same frozen solution $x$, called the \emph{base}. Before the search, the correction is updated with the latest verified values, in one of the ways described in Section~\ref{sec:main-learning}. Each agent returns a proposal $U_b$ after $H$ private steps: the best candidate it has found, or its current block if no candidate improves on it ($\mathrm{Search}_H$ in Algorithm~\ref{alg:protocol}). The verifier then records its gain over the base,
\begin{equation}\label{eq:gain}
g_b\eqdef\max\{0,-f_b(U_b;x)\}.
\end{equation}
The coordinator then accepts the strictly improving proposals of the color with the largest total gain, which gives the new base $x^+$. By~\eqref{eq:additive}, these gains add exactly:

\begin{algorithm}[t]
\caption{Corrected local search with verified merging}
\label{alg:protocol}
\small
\begin{algorithmic}[1]
\Require $x^0\in\cX$; local objectives $L_b$; proper coloring $\chi$ with $\kappa$ colors; private horizon $H$; number of epochs $K$
\For{$k=0,1,\ldots,K-1$}
  \State Update the correction with verified values (Section~\ref{sec:main-learning})
  \For{$b=1,\ldots,m$ \textbf{in parallel}}
    \State $\widehat f_b(u;x^k)\gets L_b(u)-L_b(x^k_b)+\widehat R_b(u;x^k)$
    \State $U_b\gets\mathrm{Search}_H\bigl(\widehat f_b(\cdot\,;x^k),\,x^k_b\bigr)$ \Comment{$H$ private steps}
    \State $g_b\gets\max\{0,\,F(x^k)-F(x^k_{-b},U_b)\}$ \Comment{verify}
  \EndFor
  \State $j^k\gets\argmax_{1\leq j\leq\kappa}\sum_{b\in\mathcal C_j}g_b$ \Comment{best color}
  \State $S^k\gets\{b\in\mathcal C_{j^k}:g_b>0\}$
  \State $x^{k+1}\gets x^k$
  \State $x^{k+1}_b\gets U_b$ for all $b\in S^k$ \Comment{merge (or Alg.~\ref{alg:best-combination})}
\EndFor
\end{algorithmic}
\end{algorithm}

\begin{equation}\label{eq:merge}
F(x)-F(x^+)=\max_{1\leq j\leq\kappa}\sum_{b\in\mathcal C_j}g_b
\geq\frac1\kappa\sum_bg_b.
\end{equation}
The coordinator thus keeps at least the average gain of a color, and the shared objective never increases. Applying all individually helpful proposals at once could increase it, because adjacent blocks share factors.

The best color discards the verified improvements of all other colors, even when they would combine well. Section~\ref{sec:main-combination} shows that the coordinator may instead accept the best verified combination of the proposals (Algorithm~\ref{alg:best-combination}), with the same guarantees and a rate set by how well the proposals actually combine.

\section{Convergence and the error floor}
\label{sec:main-convergence}
The guarantee rests on three ingredients: accurate corrections $\widehat R_b$ in~\eqref{eq:surrogate}, effective local search and favorable task geometry. Throughout, $\mathcal H$ denotes everything fixed before a search: the base, correction and past randomness. Expectations are over the agents' randomness, and all constants must hold at every base the run reaches.

\paragraph{Correction accuracy where it matters.}
An approximate correction $\widehat R_b$ can be badly wrong for poor candidates that no reasonable search returns. What matters is its accuracy at two candidates: a true block minimizer $u_b^*$ of $f_b(\cdot\,;x)$, which the agent aims to find, and the proposal $U_b$ that it actually returns.

\begin{assumption}[Anchored correction accuracy]\label{ass:correction}
There are constants $\delta_b^*,\delta_b^U\geq0$ such that at every epoch the correction is anchored, $\widehat R_b(x_b;x)=0$, and its errors at some true block minimizer $u_b^*$ and at the returned proposal $U_b$ satisfy
\begin{equation}\label{eq:error}
\begin{gathered}
|\widehat R_b(u_b^*;x)-R_b(u_b^*;x)|\leq\delta_b^*,\\
\mathbb E\bigl[|\widehat R_b(U_b;x)-R_b(U_b;x)|\,\big|\,\mathcal H\bigr]\leq\delta_b^U.
\end{gathered}
\end{equation}
\end{assumption}
A uniform bound $\sup_{u\in\cX_b}|\widehat R_b(u;x)-R_b(u;x)|\leq\delta_b$ implies~\eqref{eq:error} with $\delta_b^*=\delta_b^U=\delta_b$, but it can be far larger. In the routing experiments of Section~\ref{sec:routing-assumptions}, for example, a uniform bound exceeds $10^4$, whereas the errors in~\eqref{eq:error} measured during the runs are zero at the true block minimizers and about 40 on average at the returned proposals. To verify the assumption, one need not find $u_b^*$, which may be hard or impossible to compute; a bound on the error there suffices, and it can follow from the structure of the objective.

Next, local search must make progress on the corrected objective. Let $\widehat f_b^*\eqdef\min_{u\in\cX_b}\widehat f_b(u;x)$. Since the unchanged block scores zero, an agent can gain at most $0-\widehat f_b^*$, and we require its $H$ private steps to capture, in expectation, a fixed fraction of this amount. The staircases in step~2 of Figure~\ref{fig:protocol} depict such progress.

\begin{assumption}[Conditional local contraction]\label{ass:local}
The solver starts from the unchanged block, and for some $0<\rho_b<1$ its $H$ steps satisfy
\begin{equation}\label{eq:local}
\mathbb E[\widehat f_b(U_b;x)-\widehat f_b^*\mid\mathcal H]
\leq(1-\rho_b)^H(0-\widehat f_b^*).
\end{equation}
\end{assumption}
With $q_b(H)\eqdef1-(1-\rho_b)^H$ and $q(H)\eqdef\min_b q_b(H)$, $q_b(H)$ is the guaranteed fraction of the available corrected improvement that agent $b$ captures in $H$ steps; in the routing task of Section~\ref{sec:routing-assumptions}, LLM agents capture on average about $\nicefrac{2}{3}$ of this improvement in $H=3$ steps.

Finally, improving single blocks must be enough to make global progress. Let the \emph{best block decrease} $D_b(x)$ be the largest decrease that agent $b$ alone can achieve, and $\Delta(x)$ the remaining gap to the optimum:
\begin{equation}\label{eq:gap}
D_b(x)\eqdef-\min_{u\in\cX_b}f_b(u;x)\geq0,\qquad \Delta(x)\eqdef F(x)-F^*.
\end{equation}
Whenever the solution is suboptimal, the agents together must be able to improve it in proportion to the remaining gap.

\begin{assumption}[Block error bound]\label{ass:eb}
There is a constant $\mu\in(0,1]$ such that every reached base satisfies
\begin{equation}\label{eq:eb}
\sum_bD_b(x)\geq\mu\Delta(x).
\end{equation}
\end{assumption}
This excludes \emph{single-block traps}, where only a coordinated change of several blocks helps, for example two modules whose shared interface can be changed only by editing both at once. On a finite domain, nothing more is needed: some $\mu>0$ exists if and only if every solution that no single agent can improve is optimal. Rearranged,~\eqref{eq:eb} bounds the unknown gap, $\Delta(x)\leq\mu^{-1}\sum_bD_b(x)$, so upper bounds on the block improvements certify how close a solution is to the optimum, although $F^*$ is unknown (Proposition~\ref{prop:geometry}). In the routing experiments, the ratio $\sum_bD_b(x)/\Delta(x)$ is at least 0.43 at every recorded base from which the team starts a round (Section~\ref{sec:routing-assumptions}), and on the synthetic task all three assumptions are certified (Section~\ref{sec:synthetic-assumptions}).

\begin{lemma}[Correction error becomes lost progress]\label{lem:local}
Under Assumptions~\ref{ass:correction} and~\ref{ass:local},
\begin{equation}\label{eq:localgain}
\mathbb E[g_b\mid\mathcal H]\geq q_b(H)D_b(x)-q_b(H)\delta_b^*-\delta_b^U.
\end{equation}
\end{lemma}
\begin{proof}
Condition on $\mathcal H$, and write the correction error as $\varepsilon_b(u)\eqdef\widehat R_b(u;x)-R_b(u;x)$. By~\eqref{eq:correction} and~\eqref{eq:surrogate}, $\widehat f_b=f_b+\varepsilon_b$. At the true minimizer, $\widehat f_b^*\leq\widehat f_b(u_b^*;x)=-D_b(x)+\varepsilon_b(u_b^*)\leq-D_b(x)+\delta_b^*$, and local contraction with anchoring gives $\mathbb E\widehat f_b(U_b;x)\leq q_b(H)\widehat f_b^*$. Hence
\[
\begin{aligned}
\mathbb E f_b(U_b;x)&=\mathbb E[\widehat f_b(U_b;x)-\varepsilon_b(U_b)]\leq q_b(H)\widehat f_b^*+\delta_b^U\\
&\leq-q_b(H)D_b(x)+q_b(H)\delta_b^*+\delta_b^U,
\end{aligned}
\]
and $g_b\geq-f_b(U_b;x)$ proves the claim. Verification also ensures $g_b\geq0$ when this lower bound is negative.
\end{proof}

\begin{theorem}[Convergence with a correction-error floor]\label{thm:rate}
Under the factor structure, proper coloring and Assumptions~\ref{ass:correction}--\ref{ass:eb}, let
\begin{equation}\label{eq:constants}
\alpha\eqdef\frac{\mu q(H)}{\kappa},\qquad
B\eqdef\frac1\kappa\sum_b\bigl(q_b(H)\delta_b^*+\delta_b^U\bigr).
\end{equation}
From a deterministic $x^0$, the iterates of Algorithm~\ref{alg:protocol} satisfy, for every $K\geq0$,
\begin{equation}\label{eq:rate}
\boxed{\mathbb E\Delta(x^K)\leq(1-\alpha)^K\Delta(x^0)
+\frac B\alpha\bigl[1-(1-\alpha)^K\bigr]}\,.
\end{equation}
Exact corrections give geometric decay to zero expected gap.
\end{theorem}
\begin{proof}
By~\eqref{eq:merge}, Lemma~\ref{lem:local} and the block error bound,
\[
\mathbb E[F(x)-F(x^+)\mid\mathcal H]
\geq\frac1\kappa\sum_bq_b(H)D_b(x)-B
\geq\alpha\Delta(x)-B.
\]
Total expectation gives $\mathbb E\Delta(x^{k+1})\leq(1-\alpha)\mathbb E\Delta(x^k)+B$. Iterating and summing the geometric series proves~\eqref{eq:rate}.
\end{proof}

The rate has three distinct factors: useful block directions ($\mu$), finding them ($q$), and retaining compatible gains ($1/\kappa$); Section~\ref{sec:main-combination} sharpens the third. More local work improves only the second. The floor $B/\alpha$ comes from correction error: more private search cannot bring it below $\sum_b(\delta_b^*+\delta_b^U)/\mu$, but a more accurate correction lowers it. The floor belongs to an upper bound, not a level at which the search must stall. If the correction error does not change which candidate scores best, the agent returns the same proposal as with an exact correction and loses nothing; in the synthetic task of Section~\ref{sec:main-synthetic}, for example, small correction biases leave every run at the optimum. Figure~\ref{fig:routing-theory} evaluates the recurrence behind~\eqref{eq:rate} with per-round values measured on LLM runs, and Figure~\ref{fig:main-synthetic-known}(b) compares observed mean convergence with~\eqref{eq:rate} using certified constants.

\section{Merging interfering proposals}
\label{sec:main-combination}
Algorithm~\ref{alg:protocol} accepts only the improving proposals of one color. Blocks of one color share no factors, so this merge is safe for every candidate and keeps at least a fraction $1/\kappa$ of the verified gain~\eqref{eq:merge}. The price is that the verified improvements of all other colors are thrown away. The coloring is built for the worst case: it separates every pair of blocks that could interact for some candidates, whether or not the proposals of a given epoch actually do. In practice, proposals of adjacent blocks often do not interfere at all, and even proposals that do interfere can lower $F$ more together than either does alone. Throwing all of this work away is wasteful. This section shows how the coordinator can keep it and what the theory then guarantees.

\paragraph{Accepting the best combination.} A merge accepts the proposals of some set $S\subseteq\{1,\ldots,m\}$ of blocks and keeps the other blocks unchanged, giving $(x_{-S},U_S)$ as in~\eqref{eq:additive}; $S=\emptyset$ leaves $x$ unchanged. Algorithm~\ref{alg:protocol} uses $S=S^k$, the improving blocks of the best color (line 9). Instead, the coordinator can evaluate $\Phi(S)\eqdef F(x_{-S},U_S)$ for every $S$ and accept the set with the lowest loss, even if it contains proposals that interfere (Algorithm~\ref{alg:best-combination}). Because $S^k$ is one of these candidates, this \emph{best-combination} merge never gains less than the best color: the equality in~\eqref{eq:merge} becomes an inequality, and the gain is still at least $\kappa^{-1}\sum_bg_b$.

With many blocks, checking all $2^m$ sets is impractical, so the coordinator can use a greedy search instead. Starting from $S^k$, it repeatedly adds the block whose proposal lowers $\Phi$ most, as long as one does; it never raises the loss of the best-color plan, so the same bound holds. Theorem~\ref{thm:rate} and the convergence results for learned corrections in Appendix~\ref{sec:learn} use the merge only through the inequality $F(x)-F(x^+)\geq\kappa^{-1}\sum_bg_b$ in~\eqref{eq:merge}, so they hold for both versions of Algorithm~\ref{alg:best-combination}.

\begin{algorithm}[!ht]
\caption{Best-combination merge}
\label{alg:best-combination}
\small
\begin{algorithmic}[1]
\Require base $x^k$, proposals $U_1,\ldots,U_m$, $S^k$ (line 9 of Algorithm~\ref{alg:protocol})
\If{all $2^m$ sets can be checked}
  \State $S^k\gets\argmin_S\Phi(S)$ \Comment{all combinations}
\Else
  \While{$\min_{b\notin S^k}\Phi(S^k\cup\{b\})<\Phi(S^k)$} \Comment{greedy}
    \State $S^k\gets S^k\cup\bigl\{\argmin_{b\notin S^k}\Phi(S^k\cup\{b\})\bigr\}$
  \EndWhile
\EndIf
\State $x^{k+1}\gets(x^k_{-S^k},U_{S^k})$ \Comment{replaces lines 10--11 of Alg.~\ref{alg:protocol}}
\end{algorithmic}
\end{algorithm}

\paragraph{Realized number of colors.} Theorem~\ref{thm:rate} needs the merge to keep at least a fraction $1/\kappa$ of the total verified gain $\sum_bg_b$. How much Algorithm~\ref{alg:best-combination} keeps in epoch $k$ is measured by its \emph{realized number of colors}
\begin{equation}\label{eq:realized-colors}
\kappa^*_k\eqdef\frac{\sum_bg_b}{F(x^k)-F(x^{k+1})},
\end{equation}
and $\kappa^*_k\eqdef1$ if no block improves. In words, the merge gains $\sum_bg_b/\kappa^*_k$, whereas the best color is only guaranteed $\sum_bg_b/\kappa$; since the merge never gains less than the best color, $\kappa^*_k\leq\kappa$. When all $2^m$ sets are checked, the merge accepts the combination with the largest gain, so no other way of combining this epoch's proposals gives a smaller $\kappa^*_k$. The realized number $\kappa^*_k$ is not the number of accepted groups: if three proposals that gain 100 each gain only 110 together, accepting all three can still be best, and then $\kappa^*_k=300/110\approx2.7$; if proposals help each other, $\kappa^*_k<1$. In the routing experiments of Section~\ref{sec:main-routing}, where any two agents' robots can collide and Algorithm~\ref{alg:protocol} accepts one block per round on one map, Algorithm~\ref{alg:best-combination} accepts on average 1.7 blocks in each round that makes progress, and $\kappa^*_k$ averages about 1.2 against $\kappa=3$.

\begin{corollary}[Convergence with realized colors]\label{cor:realized-rate}
Under the assumptions of Theorem~\ref{thm:rate}, suppose that Algorithm~\ref{alg:best-combination} has $\kappa^*_k\leq\bar\kappa$ in every epoch, for some $\bar\kappa\geq1$. Then its iterates satisfy~\eqref{eq:rate} with $\kappa$ replaced by $\bar\kappa$ in $\alpha$ and $B$ of~\eqref{eq:constants}.
\end{corollary}
\begin{proof}
By~\eqref{eq:realized-colors}, $F(x)-F(x^+)\geq\sum_bg_b/\kappa^*_k\geq\bar\kappa^{-1}\sum_bg_b$. The first relation is an equality unless no block improves, in which case $\sum_bg_b=0$ and the merge never raises the loss; the second holds because the gains are nonnegative. This is~\eqref{eq:merge} with $\bar\kappa$ in place of $\kappa$, and the proof of Theorem~\ref{thm:rate} uses the coloring only through~\eqref{eq:merge}.
\end{proof}

\paragraph{Effective number of colors.} Corollary~\ref{cor:realized-rate} holds for any $F$, but $\kappa^*_k$ can only be measured after the epoch. We can, however, bound it from above in advance. Let us assume that blocks interact only in pairs, that is, every factor of $F$ in~\eqref{eq:factor-definition} depends on at most two blocks. This is the case whenever the coupling consists of pairwise conflicts, as in the routing task of Section~\ref{sec:main-routing}, where each collision term involves one pair of robots. For a base $x$ with proposals $U_b$ and $b\neq b'$, the interaction
\begin{equation}\label{eq:merge-interaction}
\xi_{bb'}\eqdef F(x_{-\{b,b'\}},U_b,U_{b'})-F(x)-f_b(U_b;x)-f_{b'}(U_{b'};x)
\end{equation}
is zero unless $b$ and $b'$ share a factor, and expanding the pairwise factors gives, for every set $S$ of blocks,
\begin{equation}\label{eq:pairwise-expansion}
F(x_{-S},U_S)-F(x)=\sum_{b\in S}f_b(U_b;x)+\sum_{\{b,b'\}\subseteq S}\xi_{bb'}.
\end{equation}
Let $I^k\eqdef\{b:g_b>0\}$ be the improving blocks of epoch $k$. Two improving blocks $b$ and $b'$ \emph{conflict} if $\xi_{bb'}>0$, that is, if their proposals together gain less than the sum of their gains, and a set of improving blocks is \emph{conflict-free} if no two of its blocks conflict. The \emph{effective number of colors} $\kappa_k$ is the smallest number of conflict-free sets that cover $I^k$ ($\kappa_k\eqdef1$ if $I^k=\emptyset$), and $\deg_k(b)$ is the number of other improving blocks that conflict with $b$. Whereas $\kappa$ must separate every pair of blocks that could interact for some candidates, $\kappa_k$ counts only the groups that epoch $k$ needs. Two blocks of one color share no factor and therefore never conflict, so the $\kappa$ color classes already split $I^k$ into at most $\kappa$ conflict-free groups; and one group per improving block always works. Hence $\kappa_k\leq\min\{\kappa,|I^k|\}$ whenever $I^k$ is nonempty. Moreover, if every improving block conflicts with at most $d$ others, $d+1$ groups are always enough: assign the blocks to groups one at a time; a block conflicts with at most $d$ blocks, so at least one of $d+1$ groups contains none of them, and the block joins that group. Hence $\kappa_k\leq1+\max_{b\in I^k}\deg_k(b)$ for nonempty $I^k$.

\begin{proposition}[Realized and effective colors]\label{prop:realized-conflicts}
If every factor of $F$ in~\eqref{eq:factor-definition} depends on at most two blocks, then every conflict-free set $S$ of improving blocks satisfies
\begin{equation}\label{eq:realized-merge}
F(x)-F(x_{-S},U_S)\geq\sum_{b\in S}g_b.
\end{equation}
Consequently, Algorithm~\ref{alg:best-combination} with all combinations checked gains at least $\kappa_k^{-1}\sum_bg_b$, that is, $\kappa^*_k\leq\kappa_k$.
\end{proposition}
\begin{proof}
For an improving block, $f_b(U_b;x)=-g_b$. By~\eqref{eq:pairwise-expansion}, accepting a set $S$ therefore changes $F$ by $-\sum_{b\in S}g_b$ plus the interactions $\xi_{bb'}$ of its pairs. If $S$ is conflict-free, all these interactions are at most zero, which proves~\eqref{eq:realized-merge}. Now split the improving blocks into $\kappa_k$ conflict-free groups. Their gains add up to $\sum_bg_b$, so the best group alone gains at least $\kappa_k^{-1}\sum_bg_b$. The merge checks this group along with all other sets and accepts a plan at least as good, so $F(x)-F(x^+)\geq\kappa_k^{-1}\sum_bg_b$, which by~\eqref{eq:realized-colors} is $\kappa^*_k\leq\kappa_k$.
\end{proof}

\paragraph{What this means.} The three numbers of colors are ordered, $\kappa^*_k\leq\kappa_k\leq\kappa$. The number $\kappa$ in Theorem~\ref{thm:rate} is the price of throwing away all colors but one, and it is fixed by how the objective could couple the blocks. Corollary~\ref{cor:realized-rate} replaces it by an upper bound $\bar\kappa$ on $\kappa^*_k$, which reflects how well the proposals of each epoch actually combine; the error floor $B/\alpha$ stays the same. The effective number $\kappa_k$ gives a guarantee that can be stated in advance: if no improving proposal conflicts with more than $d$ others, then $\kappa^*_k\leq\kappa_k\leq d+1$, and Corollary~\ref{cor:realized-rate} holds with $\bar\kappa=d+1$. This matters most for large teams: on a densely coupled task $\kappa$ grows with the number of agents, whereas $d+1$ need not. The realized number can be smaller still, because accepting conflicting proposals together counts whenever it lowers $F$ most.

\paragraph{Cost of checking and limits.} Checking all $2^m$ combinations is affordable for a few agents, as in our experiments with three. For more agents, pairwise interactions help: by~\eqref{eq:pairwise-expansion}, the loss of every combination of improving blocks follows from one check per improving block and one per pair of them, so about $|I^k|^2/2$ checks replace $2^m$. Appendix~\ref{app:routing-merge-theory} gives a simple rule that picks a combination from these checks and keeps Corollary~\ref{cor:realized-rate} with $\bar\kappa=d+1$. Combining proposals cannot, however, remove the need for Assumption~\ref{ass:eb}. At a single-block trap with exact corrections no agent finds an improvement, so every agent returns its current block and no combination of the returned proposals changes anything; Appendix~\ref{app:routing-merge-theory} discusses ways around this.

\section{Learning corrections}
\label{sec:main-learning}
Before every epoch, Algorithm~\ref{alg:protocol} updates the correction with verified values (line~2). This can happen in two ways. First, agents can receive the verifier's reports in their prompts; in the routing task of Section~\ref{sec:main-routing}, for example, every collision that the verifier reports becomes a known pair whose penalty the correction adds.

Second, when the hidden part of the objective has a known structure, the correction can be learned numerically. Suppose that $F(x)$ equals $G(x)+\langle\vartheta,\phi(x)\rangle$ up to a constant, with a known function $G$ and known features $\phi$ but unknown weights $\vartheta$. Every evaluated trial solution $z$ then gives one linear equation, $F(z)-F(x)-[G(z)-G(x)]=\langle\phi(z)-\phi(x),\vartheta\rangle$. The estimate $\widehat\vartheta$ moves by the smallest amount that satisfies it, a sketch-and-project step \citep{strohmer2009,gower2015}. Updating the estimate in this way before an epoch is called \emph{calibration}. Agents keep the learned weights across epochs and recompute the known terms when the base changes. If the trial solutions explore every unknown direction on average, the solution and the correction contract jointly at a linear rate set by the slower of search and learning, with no error floor (Theorem~\ref{dev:thm:joint} in Appendix~\ref{sec:learn}). Already evaluated proposals, including rejected ones, can be reused as measurements (Proposition~\ref{prop:reuse}). Without such structure, sparse evaluations cannot determine the values of unqueried candidates (Appendix~\ref{dev:sec:structure}). Appendix~\ref{app:synthetic-learning} tests numerical learning on the synthetic task.

\paragraph{The cost of learning.}
Learning repays its queries when the needed information is sparse and reused; otherwise measuring the alternatives in advance is cheaper (Appendix~\ref{sec:certificate} accounts for the cost of calibration). If the unknown terms are linear in individual binary variables, one extra verifier query identifies one coefficient exactly, so at most one query per variable is ever needed, and a computable bound on the remaining correction error contracts linearly together with the gap (Proposition~\ref{prop:radius}, Corollary~\ref{cor:observable}).

\section{Private work and synchronization}
\label{sec:main-cost}
Longer private search makes each epoch more productive but also slower, whereas frequent synchronization spends time on verification and context transfer. Let one parallel private step cost $c>0$, and let each synchronization, including correction construction, context transfer and verification, cost $C\geq0$. With exact corrections, Theorem~\ref{thm:rate} shows that $K_\varepsilon(H)=\lceil\kappa\log(\Delta(x^0)/\varepsilon)/(\mu q(H))\rceil$ epochs reach expected gap $\varepsilon$, at cost at most $(cH+C)K_\varepsilon(H)$, where $\lceil\cdot\rceil$ rounds up to an integer. With $\lambda=-\log(1-\min_b\rho_b)$, dropping the ceiling and the constant factor $\kappa\log(\Delta(x^0)/\varepsilon)/\mu$ and using $q(H)=1-e^{-\lambda H}$ gives the relaxed cost $J(H)\eqdef(cH+C)/(1-e^{-\lambda H})$ for $H>0$, whose unique minimizer for $C>0$ satisfies (Proposition~\ref{prop:horizon})
\begin{equation}\label{eq:stationary}
e^{\lambda H_*}=1+\lambda H_*+\lambda C/c.
\end{equation}
Costlier synchronization therefore calls for longer private work: $H_*\approx\sqrt{2C/(c\lambda)}$ when $\lambda C/c$ is small. Section~\ref{sec:synthetic-horizon} compares this choice with observed completion costs.

\section{Experiments}
\label{sec:experiments}

We test the theory in three settings. The main study, robot routing with LLM agents (Section~\ref{sec:main-routing}), asks three questions. (Q1) Can decomposition with corrections let LLM agents solve a coupled task that a single agent struggles with? (Q2) How much does the team gain by merging the best checked combination of proposals (Section~\ref{sec:main-combination}) instead of a single color? (Q3) Does the bound of Theorem~\ref{thm:rate}, evaluated round by round with constants measured on the runs, hold for the team's progress, and how closely? Two smaller studies test the remaining parts of the theory. Query planning in SQLite (Section~\ref{sec:main-sqlite}) asks (Q4) when acquiring correction information pays off (Section~\ref{sec:main-learning}). A synthetic task with certified constants (Section~\ref{sec:main-synthetic}) asks (Q5) how tight the bound is, whether a persistent correction error leaves the predicted floor, and how long agents should search between synchronizations (Section~\ref{sec:main-cost}).

\subsection{Robot routing with LLM agents}
\label{sec:main-routing}
Each robot must reach its goal while avoiding walls and other robots, a multi-agent pathfinding (MAPF) task \citep{stern2019mapf}. The rules are simple and visual, yet the task remains hard for LLMs \citep{chen2024mapfllm}. Specialized solvers such as conflict-based search \citep{sharon2012cbs} can solve this problem exactly, and we use them only to certify the optimum. Most importantly, coupling is built in: robots of different agents can collide, so the best routes for one agent's robots can block another agent's.

\begin{figure*}[t]
\centering
\includegraphics[width=.92\textwidth]{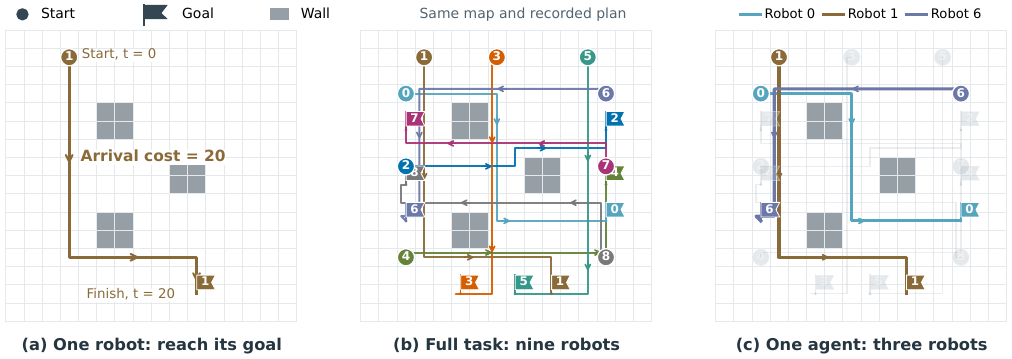}
\Description{Three views of the same recorded plan, which is optimal. Left: robot one goes down from its start, then right and down to its goal, with Start at tick zero and Finish at tick twenty, and arrival cost twenty. Center: all nine robot routes, with coincident segments separated within their cells. Right: one agent's robots zero, one and six highlighted, with the six other robots faded. Numbered filled markers show starts; matching flags show goals. Gray cells are walls.}
\caption{Recorded plan at the optimum (first map, team seed 19 after round 4): nine delivered, zero collisions, $F=F^*=144$. (a) Robot 1: arrival at tick 20, cost 20. (b) All nine routes. (c) Block $\{r_0,r_1,r_6\}$ with peers faded. Markers: starts; flags: goals. Overlapping segments are offset within cells; spatial crossings need not coincide in time.}
\label{fig:main-routing-repair}
\end{figure*}

Software was our first target, and the introduction uses it as an example. In code, however, strong coupling between the parts that different agents would own appears mainly in large systems, because good design keeps modules apart \citep{parnas1972}. Experiments at that scale are hard to control and to analyze: the best program is unknown, so one cannot say how far a team is from it or which interactions slowed it down. Routing, on the other hand, provides a certified optimum, lower bounds and exact block minima, so every constant of the theory can be measured. This small, fully analyzable setting showed us what actually happens during a run. Observations made in it inspired two results of this paper: the realized number of colors (Section~\ref{sec:main-combination}) and the error condition~\eqref{eq:error}, which concerns only two candidates per agent instead of all of them (Section~\ref{sec:main-convergence}). Nothing in the protocol is specific to routing, so it applies to code as well, at a scale where the parts are strongly coupled (Appendix~\ref{app:task-choice}). Below we describe the task and the agents, compare the team with single agents (Q1), test the merge of Section~\ref{sec:main-combination} (Q2) and finally confront the runs with the theory (Q3).

\paragraph{Rules.}
Agents write explicit command lists for nine robots on a $16\times16$ grid with walls, over 48 time steps, which we call ticks. At each tick a robot moves one cell \texttt{up}, \texttt{down}, \texttt{left} or \texttt{right}, or stays in place with \texttt{wait}; directions refer to the map, not to the robot. A route may not enter a wall, and a robot that reaches its goal stays there. The pictures below show, from left to right, the commands, a blocked move, a legal route that arrives at tick 4, and a swap.
\par\nobreak\smallskip
\noindent\makebox[\columnwidth]{\includegraphics[width=\columnwidth]{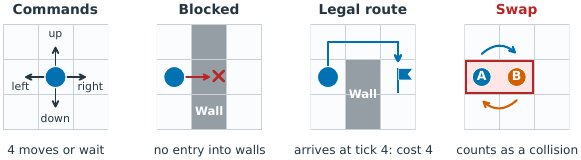}}
\par\smallskip
\noindent Collisions couple the robots. Two robots collide when they occupy the same cell at the same tick. They also collide when they exchange cells during one tick, as in the rightmost picture above, since they would have to pass through each other. Two routes can each avoid every wall and still collide. In the example below, if A and B both move, they meet in the center at tick 1. If B waits one tick instead, A passes first and both arrive safely.
\par\nobreak\smallskip
\noindent\makebox[\columnwidth]{\includegraphics[width=\columnwidth]{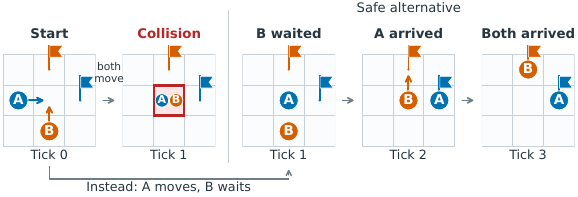}}
\par\smallskip

\begin{figure*}[t]
\centering
\includegraphics[width=\textwidth]{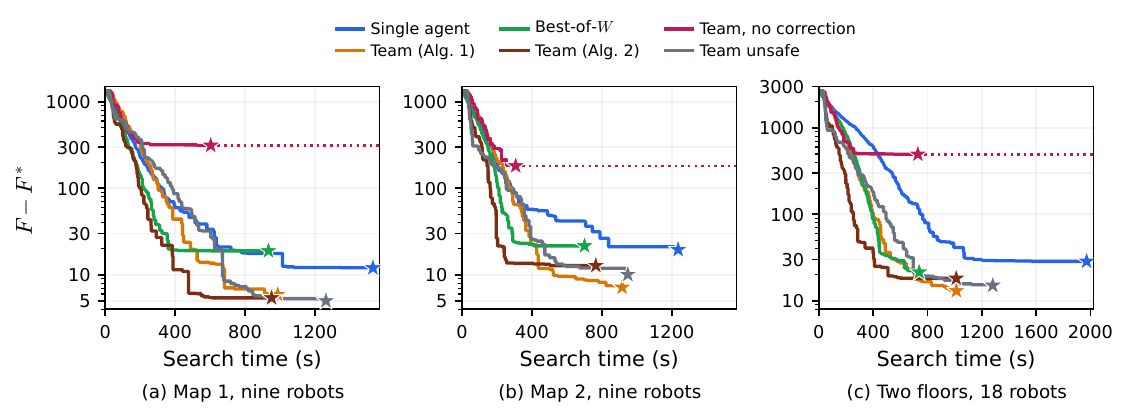}
\Description{Three panels show the mean gap to the optimum against search time, with the gap on a logarithmic axis; the panels for map 1 and map 2 share a time axis up to about 1,570 seconds, and the two-floor panel extends to about 2,000 seconds. On map 1, the team with best combination falls fastest and ends near 5 after about 950 seconds; the team with Algorithm 1 ends near 6 after about 990 seconds; the team unsafe falls more slowly and ends near 5 after about 1,260 seconds; the best-of-three falls fast at first and levels off near 19; the single agent ends near 12; the team without correction stalls near 310. On map 2, the team with best combination falls fastest but levels off near 13, above Algorithm 1, which ends near 7; the team unsafe ends near 10; the best-of-three near 22 and the single agent near 20; the team without correction stalls near 180. On two floors, the team with best combination falls fastest and ends near 18 after about 1,010 seconds, Algorithm 1 near 13 after about 1,010 seconds, the team unsafe near 15 after about 1,280 seconds, the best-of-six near 21 and the single agent near 29 after about 1,970 seconds; the team without correction stalls near 495. Dotted lines continue the final values of the team without correction. Stars mark curve endpoints.}
\caption{Mean gap $F-F^*$ against search time (logarithmic gap axis); 20 runs per curve, 10 without correction. (a, b) Maps 1 and 2: nine robots, three team agents; (c) both maps as two floors: 18 robots, six team agents. Best-of-$W$: $W$ parallel single agents, as many as team agents. Dotted: final mean of the team without correction. Variants: Section~\ref{sec:main-routing}.}
\label{fig:routing-main-progress}
\end{figure*}

\paragraph{Loss, local objectives and corrections.}
The delivery cost of robot $i$ is $c_i(x)=\tau_i$, where $\tau_i$ is the tick at which it reaches its goal. If the robot does not arrive within the horizon, its cost is $c_i(x)=100+48+d_i$ instead, where $d_i$ is its remaining shortest-path distance to the goal. Here 100 is the penalty for non-delivery; we also add the horizon 48, so that an undelivered robot always costs at least 100 more than a delivered one, and $d_i$ rewards getting closer. Each collision adds a penalty of 100:
\begin{equation}\label{eq:routing-objective}
F(x)=\sum_{i=0}^{8}c_i(x)+100\sum_{0\leq i<j\leq 8}C_{ij}(x),
\end{equation}
where $C_{ij}(x)$ counts the ticks at which robots $i$ and $j$ collide. In the example above, the safe plan costs $2+3=5$, since A arrives at tick 2 and B at tick 3. If both move at once, each arrives at tick 2, but the collision raises the loss to 104. We call a plan \emph{safe} if it delivers every robot without collisions.

Agent $b$ controls a set $\mathcal I_b$ of three robots, and its block $u$ lists their commands. Its local objective keeps what it controls, the delivery costs of its robots and the collisions among them:
\begin{equation}\label{eq:routing-local-main}
L_b(u)=\sum_{i\in\mathcal I_b}c_i(u_i)+100\sum_{\substack{i<j\\ i,j\in\mathcal I_b}}C_{ij}(u_i,u_j).
\end{equation}
Collisions between robots of different agents are the reason why the blocks are coupled. The correction accounts for them, but only for the pairs that the agent already knows about. Let $\mathcal P^k$ contain the pairs of robots of different agents whose collision the verifier has found before round $k$. With every other robot $j$ kept on its current route $x_j^k$, agent $b$'s correction is the change in the penalty for these pairs:
\begin{equation}\label{eq:routing-correction-main}
\widehat R_b(u;x^k)=100\sum_{\substack{\{i,j\}\in\mathcal P^k\\ i\in\mathcal I_b,\ j\notin\mathcal I_b}}\big[C_{ij}(u_i,x_j^k)-C_{ij}(x_i^k,x_j^k)\big].
\end{equation}
The agent learns these pairs through its prompt: the verifier reports every collision it finds, and from then on the agent sees the other robot's current route. The LLM never computes~\eqref{eq:routing-local-main} or~\eqref{eq:routing-correction-main}; it searches with what its prompt shows. The system scores each of its attempts by $L_b+\widehat R_b$ and keeps the best one as the proposal $U_b$. Any two agents' robots may collide, so the dependency graph is complete, $\kappa=3$, and Algorithm~\ref{alg:protocol} accepts one block per round.

\paragraph{Maps.}
Figure~\ref{fig:main-routing-repair} shows the first map, with three $2\times2$ walls: (a) one robot's route and cost, (b) an optimal plan found by the team, and (c) the three robots of one agent. Every run starts from the plan in which all robots wait at their starts, which costs $F=1476$ on this map. No plan can cost less than the sum of the robots' shortest-path distances, 144 here, and collision-free plans with exactly this cost exist, such as the one in Figure~\ref{fig:main-routing-repair}(b), so the optimum is $F^*=144$. A plan with a collision or an undelivered robot costs at least $F^*+100$, so any plan within 100 of the optimum is safe. We did not rely on a single map. We repeated every experiment on a second map, on which the robots start from different positions, head to different goals and the walls are placed differently (Figure~\ref{fig:routing-maps} in Appendix~\ref{app:routing-task}). Its optimum is $F^*=147$, and we report the results on both maps.

\paragraph{Agents.}
All agents are the open-weight model GPT-OSS-120B \citep{openai2025gptoss} with medium reasoning, temperature 0.6 and top-$p=0.95$, served by Groq through OpenRouter (Appendix~\ref{app:routing-measurement}). Every prompt states the rules once, then shows what the agent can see of the current plan: a labeled map, the routes and costs of its own robots, a timeline of every robot it can see, drawings of the collisions it knows about, and the attempts already checked. A team agent sees its own robots and the robots in its correction~\eqref{eq:routing-correction-main}. The prompt gives no strategy advice. We tried several prompts and found that the way the plan is presented matters a great deal; we use the best one we found (Appendix~\ref{app:routing-development}). The team has three agents, each owning three robots (Figure~\ref{fig:main-routing-repair}(c)). Each round is one epoch of Algorithm~\ref{alg:protocol}: every agent gives up to $H=3$ answers on the frozen plan, and the coordinator accepts the best checked block.

\paragraph{What the correction does.}
Recall robots A and B from the collision example, and suppose they belong to different agents. Once B's agent knows about A, its correction shows that moving now costs a collision, 100, while waiting costs only one tick. The optimal plan in Figure~\ref{fig:main-routing-repair}(b) came from such a repair. In team seed 19 on the first map, robot $r_1$ swapped cells at tick 11 with robot $r_8$ of another agent, a pair that an earlier check had already reported. In round 4, the agent owning $r_1$ moved it to another shortest route with the same arrival tick 20. This removed the collision and lowered $F$ from 244 to the optimum 144 (Appendix~\ref{sec:routing-recorded-repairs}). The agent's local loss did not change; only the correction, built from $r_8$'s route, made the new route better.

\begin{table*}[t]
\centering\footnotesize
\setlength{\tabcolsep}{2.8pt}
\begin{tabular}{@{}l *{3}{c r@{${}\pm{}$}l r@{${}\pm{}$}l}@{}}\toprule
& \multicolumn{5}{c}{Map 1} & \multicolumn{5}{c}{Map 2} & \multicolumn{5}{c}{Two floors}\\
\cmidrule(lr){2-6}\cmidrule(lr){7-11}\cmidrule(l){12-16}
Variant & Safe runs $\uparrow$ & \multicolumn{2}{c}{$F-F^*$ $\downarrow$} & \multicolumn{2}{c}{Search (s) $\downarrow$} & Safe runs $\uparrow$ & \multicolumn{2}{c}{$F-F^*$ $\downarrow$} & \multicolumn{2}{c}{Search (s) $\downarrow$} & Safe runs $\uparrow$ & \multicolumn{2}{c}{$F-F^*$ $\downarrow$} & \multicolumn{2}{c}{Search (s) $\downarrow$}\\\midrule
Single agent & \textbf{100\%} & 12.0 & 10.5 & 613 & 398 & 95\% & 19.5 & 44.4 & 538 & 366 & \textbf{100\%} & 28.6 & 23.3 & 1149 & 420\\
Best-of-$W$ & 95\% & 18.9 & 48.8 & \textbf{326} & 198 & 90\% & 21.6 & 33.9 & 326 & \textbf{147} & \textbf{100\%} & 21.4 & 16.3 & 517 & \textbf{124}\\
Team (Alg.~\ref{alg:protocol}) & \textbf{100\%} & 5.9 & 2.7 & 481 & 227 & \textbf{100\%} & \textbf{7.2} & \textbf{12.6} & 482 & 202 & \textbf{100\%} & \textbf{13.1} & \textbf{12.7} & 656 & 206\\
Team (Alg.~\ref{alg:best-combination}) & \textbf{100\%} & 5.4 & 3.2 & 340 & \textbf{197} & \textbf{100\%} & 12.8 & 17.2 & \textbf{310} & 204 & \textbf{100\%} & 18.2 & 17.4 & \textbf{452} & 227\\
Team unsafe & \textbf{100\%} & \textbf{5.0} & \textbf{2.6} & 601 & 284 & 95\% & 10.1 & 22.3 & 424 & 232 & 95\% & 15.1 & 22.5 & 717 & 263\\
Team, no correction & 0\% & 313.1 & 109.2 & 326 & 172 & 20\% & 181.4 & 153.4 & 178 & 87 & 0\% & 494.5 & 181.9 & 382 & 176\\\bottomrule
\end{tabular}
\caption{All outcomes per variant and task (20 runs, 10 without correction); mean $\pm$ sample SD. Safe runs: share of runs whose final plan delivers every robot without collisions. $F-F^*$: final gap to the optimum ($F^*=144$, 147, 291). Bold: best per column (mean and SD separately), excluding the team without correction. Best-of-$W$: $W=3$, or 6 on two floors.}
\label{tab:routing-main-results}
\end{table*}

\paragraph{Baselines.}
We compare the team with two baselines that use the same model and the same kind of prompt. The \emph{single agent} sees and scores all nine robots. Each answer replaces the route of one robot of its choice; the agent accepts every improvement at once and stops after 15 answers in a row without improvement. Of the single-agent protocols we tried, this one was the most reliable (Appendix~\ref{app:routing-development}). The \emph{best-of-$W$} runs $W$ copies of the single agent in parallel on the same plan and accepts the best improving answer of each round. With $W$ equal to the number of team agents, three on one map, it matches the team's parallelism. It stops after five rounds without improvement. So does the team; since each round gives every agent three answers, a team agent, like the single agent, has 15 answers to find an improvement before the run stops. Every variant also stops once the plan is safe and within 9 of the optimum ($F\leq153$ on the first map and $F\leq156$ on the second). This threshold only ends a run early, and every run is judged by its final plan. Search time adds up the latencies of the model's answers, network overhead included; in a parallel round only the slowest agent counts (Appendix~\ref{app:routing-measurement}). Unless stated otherwise, each variant has 20 runs per map, which differ in the model's sampling seed.

\paragraph{The team finds the best plans.}
Table~\ref{tab:routing-main-results} lists the outcomes of every variant, and Figure~\ref{fig:routing-main-progress}(a,b) shows how the mean gap $F-F^*$ falls with search time on the two maps. Both maps tell the same story. All three protocols reach near-optimal plans, but the team's plans are the best and the most consistent. The team is safe in all 40 runs and meets the threshold in 39 of them, whereas the single agent meets it in 32 and the best-of-three in 31. The team also ends with the smallest gap on both maps: its mean is lower on map 1 (permutation tests, $p<0.001$ against the single agent and $p=0.011$ against the best-of-three) and its median on map 2 ($p\leq0.002$), where a few baseline runs that end with collisions make the difference in means less clear ($p=0.26$ and $p=0.08$). Of the three, only the team's mean gap falls below 10. The single agent and the team need similar time in median, but the single agent is less predictable: on the first map, seven of its runs took longer than 700 seconds, whereas only two team runs did. Three single agents in parallel are the fastest, with median search times of about four and a half minutes, against seven for the team ($p=0.011$ on map 1 and $p=0.022$ on map 2). They pay for it with worse plans, and 3 of their 40 runs end unsafe. At nine robots, parallel search thus buys speed, whereas decomposition with corrections buys the best plans. Part of the team's own speed comes from parallelism: played one after another, its recorded answers would take 1.8 times as long on the first map (Appendix~\ref{app:routing-decomposition}). The team also asks the model more often than the single agent, 47 answers against 28 on the first map, and costs USD 0.23 per run against 0.16 (Table~\ref{tab:routing-variants}).

\paragraph{Is the correction needed?}
If the agents' parts were independent, the correction would not matter. To test this, we removed it: agents are told nothing about other agents' robots, so they receive neither the collision reports nor the peers' routes that carry the correction, while verification and merging stay unchanged (Appendices~\ref{app:routing-revised} and~\ref{app:routing-no-correction}). We ran this variant 10 times per map, and only 2 of the 20 runs end safe (Table~\ref{tab:routing-main-results}). Every run on the first map delivers all nine robots but keeps one to five collisions, all between robots of different agents. Without the collision penalties its final gap would average 3.1, close to the corrected team's 5.9: each agent solves its own part well, and the remaining loss comes entirely from interactions between the parts. In most runs, the search also ends early, but not because the agents run out of answers. In 6 of 10 runs on the first map, every agent's robots soon take shortest routes without colliding with each other. Each block is then optimal for its own objective, so no agent can improve it, and the plan freezes with its collisions; 7 of 10 runs on the second map end this way. The coupling is real, and the correction is what resolves it.

\begin{figure*}[t]
\centering
\includegraphics[width=\textwidth]{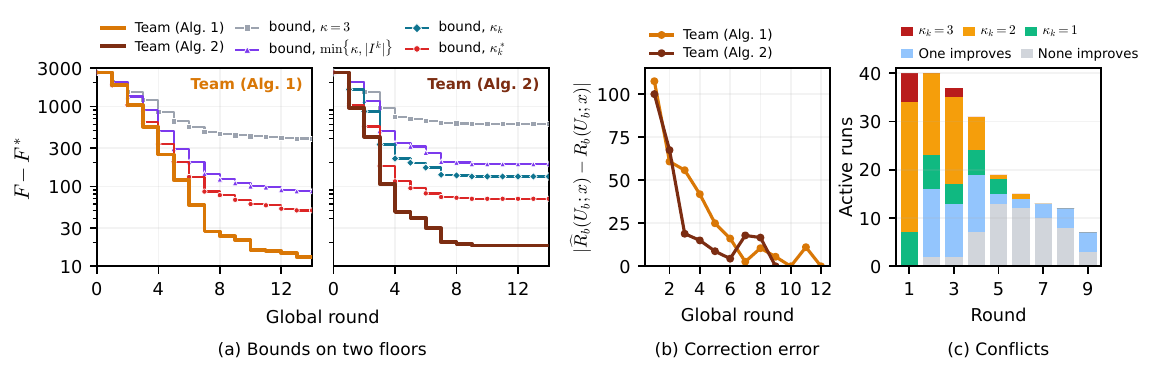}
\Description{Three panels. Left, in two halves with a shared logarithmic gap axis: on two floors, the observed mean gap of the team with Algorithm 1 (orange) ends near 13 and its bounds end near 396 with kappa equal to 3, 87 with the smaller of kappa and the number of improving blocks, and 50 with the realized number of colors; the observed mean gap of the team with Algorithm 2 (brown) ends near 18 and its bounds end near 605, 191, 133 with the effective number of colors, and 70. Middle: the mean correction error at the agents' returned proposals on maps 1 and 2 falls from about 100 in round 1 to a few units by rounds 6 and 7 for both teams, with small rises when new collision pairs appear. Right: stacked bars per round for the 40 runs of the team with Algorithm 2, with bar height equal to the active runs; in round 1 most runs have conflicts that need two colors and six need three, rounds 2 to 4 mix rounds with conflicts, rounds where several proposals improve without conflict and rounds where one improves, and in rounds 5 to 8 most remaining runs have no improving proposal.}
\caption{(a) Two floors: mean gap of each team and its bounds with the divisors of Section~\ref{sec:main-combination}. (b) Mean correction error at the agents' proposals, maps 1 and 2; 100 is one collision unknown to the correction. (c) Rounds of Team (Alg.~\ref{alg:best-combination}) with several improving proposals, by their $\kappa_k$, or with one or none; bar height: active runs. All tasks: Figure~\ref{fig:routing-theory-maps}.}
\label{fig:routing-theory}
\end{figure*}

\paragraph{A larger task: two floors.}
To test a larger task, we stack both maps as the two floors of one building. Each floor keeps its own nine robots, so the task has 18 robots, loss $F=F_1+F_2$ and optimum $F^*=144+147=291$. Robots on different floors never meet. The team has six agents, three per floor, with the same blocks as before. Its dependency graph consists of two separate triangles, so it still needs $\kappa=3$ colors and can accept one block on each floor per round. Since each team agent sees only its own floor, a two-floor team run is simply a run on each floor, played side by side round by round; each round lasts as long as its slowest agent on either floor. We therefore build the two-floor team runs from the recorded runs of both maps, pairing seed $s$ of the first map with seed $s$ of the second, without new model calls. The single agent, in contrast, sees both floors in one prompt and changes one route on either floor per answer. The parallel baseline now uses six such agents, as many as the team has (Appendix~\ref{app:routing-two-floors}).

With 18 robots the team's advantage grows (Table~\ref{tab:routing-main-results}, Figure~\ref{fig:routing-main-progress}(c)). All three protocols stay safe in every run, but the team meets both thresholds in 19 runs, against 9 for the single agent and 16 for the best-of-six. The team also ends with the best plans: its median final gap is 11, against 22.5 for the single agent and 15.5 for the best-of-six ($p\leq0.004$). The single agent now struggles with the size of the task. Still changing one route per answer, it needs 63 answers on average, more than twice as many as on one map, and its median search takes 1200 seconds, twice the team's 595 ($p<0.001$). The team's time grows less, because both floors are searched at the same time. The best-of-six is faster than Team (Alg.~\ref{alg:protocol}) on average, though not significantly in median (means 517 and 656 seconds, $p=0.01$; medians 511 and 595, $p=0.19$), and it ends with worse plans. Without correction, no run is safe.

\paragraph{Keeping the discarded work.}
Because the dependency graph is complete, Algorithm~\ref{alg:protocol} accepts one block per round and throws the other agents' checked proposals away. In the runs above this happened in most rounds that made progress: in 79 of 121 such rounds on the first map and in 64 of 115 on the second, at least one discarded proposal would also have lowered the loss. On average, the improving proposals of a round, each counted on its own, gained 1.5 times as much as the accepted one. This waste led us to Algorithm~\ref{alg:best-combination}. In \emph{Team (Alg.~\ref{alg:best-combination})}, the coordinator checks every nonempty combination of the changed blocks' proposals, at most $2^3-1=7$, and accepts the one with the lowest loss; with three agents, checking them all is cheap. Theorem~\ref{thm:rate} still applies, because the best combination is never worse than the best color. For comparison, \emph{Team unsafe} merges all changed blocks without any check and keeps the best plan it has seen. Both variants differ from Team (Alg.~\ref{alg:protocol}) only in the merge rule and in the one prompt sentence that describes it (Appendix~\ref{app:routing-merge}).

Keeping the work pays off (Table~\ref{tab:routing-main-results}, Figure~\ref{fig:routing-main-progress}). In rounds that make progress, Team (Alg.~\ref{alg:best-combination}) updates 1.7 blocks on average, whereas Algorithm~\ref{alg:protocol} updates only one. This makes it about as fast as the best-of-$W$, the fastest protocol so far, on map 1 and faster on map 2 and two floors. Its mean gap lies below that of the best-of-$W$ almost throughout the search, so it reaches almost every level of loss sooner, and it ends lower on every task (Figure~\ref{fig:routing-main-progress}). Against Team (Alg.~\ref{alg:protocol}), its mean gap falls faster on every task as well. It is also the first to reach a safe plan, with a median under four minutes on every task. That is 40--50\% sooner than Team (Alg.~\ref{alg:protocol}) ($p\leq0.007$) and 22--39\% sooner than the best-of-$W$ ($p<0.001$ on map 2 and two floors, $p=0.27$ on map 1). Its whole search is about 30\% shorter in mean than with Algorithm~\ref{alg:protocol} ($p<0.05$ on every task), with fewer rounds and a lower cost. The final plans stay close: all are safe, and their median gaps are within 2 of those of Algorithm~\ref{alg:protocol} on every task, with no significant difference.

Two caveats remain. First, on map 2 the mean gap of Team (Alg.~\ref{alg:best-combination}) levels off near 13, above that of Team (Alg.~\ref{alg:protocol}). Six of its runs stop above the threshold there, against one with Algorithm~\ref{alg:protocol} ($p=0.09$); the same runs raise its final gap on two floors. These stops were not caused by the merge discarding improvements. In each of these runs some block could still be improved, but for five rounds its agent either found nothing or proposed a change that collided with a robot missing from its correction, and the check rejected that change. Any improvement that passed the check would have been accepted (Appendix~\ref{app:routing-merge}). Second, merging without a check does not help. Team unsafe is not significantly faster than Team (Alg.~\ref{alg:protocol}), and it is slower in mean on map 1 and two floors. Its plan gets worse in 32 of 185 rounds on map 1 and in 37 of 148 on map 2, and one of its final plans keeps a collision. What makes merging pay off is the check: every merged plan is evaluated before it is accepted. The check need not cover every combination, since the greedy version of Algorithm~\ref{alg:best-combination}, which needs far fewer checks when there are many agents, keeps the guarantee of Theorem~\ref{thm:rate} (Section~\ref{sec:main-combination}).

\begin{figure*}[t]
\centering
\includegraphics[width=.94\textwidth]{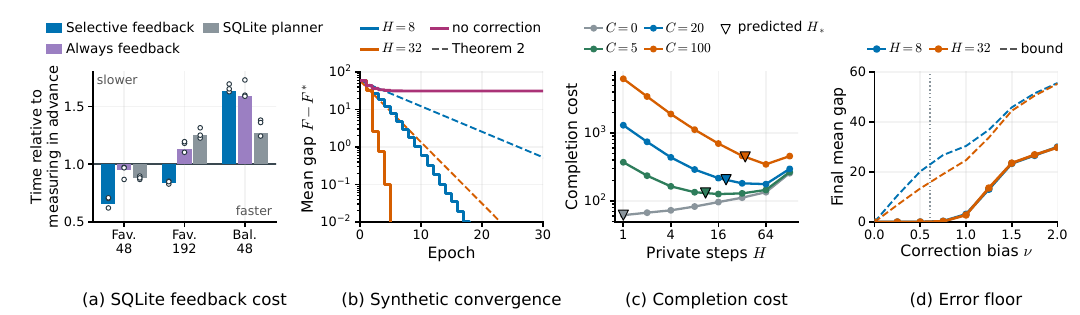}
\Description{Four panels. First: SQLite controller time relative to measuring the alternatives in advance, as bars that grow up or down from one; selective feedback is faster on the two favorable workloads and slower on the balanced one, with always-on feedback and the native planner for comparison and white dots for individual databases. Second: mean gap to the optimum on a logarithmic axis for the synthetic task; corrected runs with eight and thirty-two private steps fall to zero, dashed theoretical bounds lie above them, and runs without correction stall near thirty. Third: modeled completion cost against the number of private steps for four synchronization costs, U-shaped for positive cost, with triangles on each curve at the horizon predicted by the theory, left of the observed minimum for positive synchronization cost. Fourth: final mean gap against a persistent bias on the correction; it stays at zero up to about 0.6 and then rises to about thirty, nearly identically for eight and thirty-two private steps, with dashed upper bounds above it.}
\caption{(a) SQLite controller time relative to measuring in advance (below 1: faster) on favorable (Fav.) and balanced (Bal.) workloads; points: databases. (b--d) Synthetic task, 2,000 runs per curve or point: (b) mean gap and Theorem~\ref{thm:rate} bound (dashed); (c) modeled cost, triangles at $H_*$ (Section~\ref{sec:main-cost}); (d) final gap under a correction bias $\nu$ and the bound (dashed).}
\label{fig:main-secondary-results}
\label{fig:main-sqlite-time}
\label{fig:main-synthetic-known}
\end{figure*}

\paragraph{Measured constants.}\label{sec:routing-assumptions}
To compare the runs with the theory, we audited every search of Team (Alg.~\ref{alg:protocol}) offline, computing the exact block minima, with and without the correction, at every recorded base (Appendix~\ref{app:routing-assumptions}). The audit first confirms that the coupling is strong: about a third of the proposals that deliver all three of an agent's robots without collisions among them still hit another agent's robot (30\% on map 1, 32\% on map 2). It then measures the constants of Theorem~\ref{thm:rate}. The block error bound (Assumption~\ref{ass:eb}) holds with $\mu=0.43$ on map 1 and $0.70$ on map 2 at every state that starts a round. Six final states, all within 3 of the optimum, are single-block traps, where only a joint change of two blocks helps. A search captures on average about two thirds of the available corrected improvement (0.66 and 0.65), which plays the role of $q_b(H)$. The correction error is zero at the true block minimizer of every recorded base and averages about 40 at the proposals the agents return (45.5 and 36.5), whereas a uniform bound over all routes would exceed $10^4$. This gap is what the two-point condition~\eqref{eq:error} exploits.

\paragraph{The bound, round by round.}
We then apply the recurrence $\Delta_{k+1}\leq(1-\alpha_k)\Delta_k+B_k$ behind Theorem~\ref{thm:rate} to each run, computing $\alpha_k$ and $B_k$ from that round's measurements, and average the resulting bounds over runs (Appendix~\ref{app:routing-empirical-rate}, Proposition~\ref{dev:prop:recorded}). Nothing is fitted to the loss curve. The bound holds in every round of every run with either merge rule, on all three tasks. It also carries over between runs: on both maps, coefficients averaged over 10 runs of Team (Alg.~\ref{alg:protocol}) bound the mean gap of the other 10 in every round. As the agents learn more collision pairs, the correction becomes accurate (Figure~\ref{fig:routing-theory}(b)). The error at the true block minimizers is zero, so $B_k$ comes only from the error at the returned proposals. That error averages about 100 in round 1, one collision the correction did not know about, and falls to a few units within six or seven rounds, an empirical counterpart of Theorem~\ref{dev:thm:joint}. It falls faster with Algorithm~\ref{alg:best-combination}, to 19 by round 3 against 56 with Algorithm~\ref{alg:protocol}: its coordinator checks combinations of proposals, and every collision found in these checks also becomes known to the correction.

With $\kappa=3$, however, the bound is loose late in a run. On two floors it ends at 396 for Algorithm~\ref{alg:protocol} and 605 for Algorithm~\ref{alg:best-combination} (gray curves in Figure~\ref{fig:routing-theory}(a)), while the observed mean gaps are 13 and 18. The cause is the discarded work seen above: with $\kappa=3$ the bound credits a round with only a third of the blocks' gains, whereas Algorithm~\ref{alg:protocol} accepts the best block and Algorithm~\ref{alg:best-combination} often more. The smaller divisors of Section~\ref{sec:main-combination} tighten the bound step by step. Dividing by $\min\{\kappa,|I^k|\}$, which bounds the number of colors that contain an improving block, and by 1 in rounds without one, lowers the final bounds to 87 and 191. The effective number of colors $\kappa_k$, which bounds the realized one by Proposition~\ref{prop:realized-conflicts}, needs checked combinations and therefore applies only to Algorithm~\ref{alg:best-combination}; it lowers that bound further to 133. The realized number of colors $\kappa^*_k$ (Corollary~\ref{cor:realized-rate}), which averages about 1.5 for Algorithm~\ref{alg:protocol} and 1.2 for Algorithm~\ref{alg:best-combination}, gives the tightest bounds, 50 and 70. On all three tasks, the bound with $\kappa^*_k$ ends within a factor of 3 to 7 of the observed mean gap, instead of 30 to 60 with $\kappa=3$ (Appendix~\ref{app:routing-merge-theory}).

\paragraph{Coupling during a run.}
Why is the realized number of colors so much smaller than $\kappa=3$? Because the coupling changes as the plan improves. Team (Alg.~\ref{alg:best-combination}) checks every combination, so its runs show which proposals actually conflict (Table~\ref{tab:routing-conflict-phases}). Figure~\ref{fig:routing-theory}(c) sorts its rounds by the effective number of colors $\kappa_k$ that their improving proposals need; rounds in which only one proposal improves, or none, are shown apart, since they leave nothing to combine. The worst case $\kappa_k=3$ is rare: it occurs in 8 of the 105 rounds with several improving proposals, 6 of them in round 1. A run passes through three phases. At the start, nearly every agent improves its block, and the new routes often collide: in round 1, 34 of 58 pairs of proposals conflict on map 1 and 19 of 56 on map 2, and most rounds need two colors. Conflicting proposals are still worth combining, because delivering robots gains far more than a collision costs; the best combination accepts a conflicting pair in more than half of the first rounds. The plan therefore goes from no collisions to 2.3 and 3.3 on average after round 1. In the repair phase that follows, while the plan still has a collision or an undelivered robot, fewer agents improve per round. Their proposals still conflict often, in half of the pairs on map 1 and in more than two thirds on map 2, frequently because both remove the same collision. Yet in rounds 2 to 4, 16 of the 60 rounds with several improving proposals have no conflict at all, and the best combination accepts all of them ($\kappa_k=1$). Once the plan is safe, the task is nearly decoupled: in most rounds at most one agent improves, and $\kappa_k$ averages 1.03 and 1.04; from round 5 on, nearly every round has one improving proposal or none. Coupling is thus strongest in the first rounds, when most of the loss is removed, and fades near the optimum, whereas the coloring, which must hold for every candidate, assumes three colors throughout.

\paragraph{In short.}
Decomposition with corrections gives the most reliable and the best plans, and its advantage grows with the size of the task (Q1). Checking combinations of proposals keeps the agents' work, so the team reaches small gaps sooner than parallel single agents and is the first to reach a safe plan on every task (Q2). The bound of Theorem~\ref{thm:rate}, in its round-by-round form (Proposition~\ref{dev:prop:recorded}), holds in every round of every run with a checked merge, and with realized colors it ends within a factor of 3 to 7 of the observed gap (Q3).

\subsection{Execution feedback in SQLite}
\label{sec:main-sqlite}

This study isolates the acquisition trade-off of Section~\ref{sec:main-learning} on a real database engine; the controller is a deterministic program, not an LLM. A query joins two groups of rows through a large edge table, and SQLite can start from either group; the cheaper side depends on the groups' edge counts, which are unknown in advance. The controller's correction is a table of learned edge counts: after a query runs, SQLite reports how many edges it visited \citep{sqliteScan}, and the controller reuses this count for later queries on the same group, as execution feedback does in LEO and Bao \citep{stillger2001leo,marcus2021bao}. The alternative is to count the candidate groups in advance, which costs extra queries but avoids mistakes.

Figure~\ref{fig:main-sqlite-time}(a) compares the two on three workloads over three generated databases with 4.9 million edges, charging all measurement time. On a favorable stream of 48 queries, where a few costly groups recur, learning from feedback takes 0.65 of the time of counting in advance (4.92 versus 7.53\,ms). Over 192 queries, amortized advance counting narrows this to 0.84. On a balanced stream, where costly groups are frequent and rarely recur, learning loses and takes 1.64 times as long. Learning a correction thus pays off when the needed information is sparse and reused (Appendix~\ref{app:sqlite-feedback}).

\subsection{Synthetic task with certified constants}
\label{sec:main-synthetic}

\paragraph{Objective.}
Six blocks of eight binary variables, $x_b\in\{0,1\}^8$, interact along a path. With targets $t_{bj}$, unary weights $a_{bj}\in[1,1.5]$ and coupling weight $w=0.2$, the loss is
\begin{equation}\label{eq:main-synthetic-objective}
F(x)=\sum_{b=1}^{6}\sum_{j=1}^{8}a_{bj}|x_{bj}-t_{bj}|+w\sum_{b=1}^{5}\sum_{j=1}^{8}|x_{bj}-x_{b+1,j}|.
\end{equation}
The fixed task starts at $x^0=1-t$; its unique minimizer is $t$, with $F^*=3.4$ and $\Delta_0=58.60$. The private score $L_b(u)=\sum_j a_{bj}|u_j-\tilde t_{bj}|$ uses the wrong target $\tilde t_b=1-t_b$ in blocks 1, 3 and 5, a deliberate local/global mismatch that the exact correction removes.

\paragraph{Search and verified assumptions.}\label{sec:synthetic-assumptions}
Each private step picks a random bit of the block and, with a block-specific probability $\pi_b$ between 0.5 and 1.0, sets it to minimize the corrected score; the two colors are $\{1,3,5\}$ and $\{2,4,6\}$. With exact corrections, $\delta_b^*=\delta_b^U=0$, $\rho_b=\pi_b/8$ and $\kappa=2$, and an exact computation certifies $\mu=0.7193$, rounded down (Appendix~\ref{app:synthetic-correction}). Theorem~\ref{thm:rate} then gives, with no fitted coefficient,
\[
\mathbb E F(x^K)\leq F^*+
\left(1-\frac{\mu q(H)}{2}\right)^K\Delta_0.
\]

\paragraph{Results and the cost of private work.}\label{sec:synthetic-horizon}
For every horizon $H$ from 1 to 128 (powers of two), all 2,000 exact-correction runs reach the optimum, whereas none of 2,000 uncorrected runs at $H=8$ does: they stall at mean gap 31 (Figure~\ref{fig:main-synthetic-known}(b)). The certified bound holds at every epoch and is within a factor of two of the observed mean gap for three epochs at $H=8$; later, convergence is faster than the worst case. Figure~\ref{fig:main-synthetic-known}(c) charges $H+C$ modeled units per epoch, where $C$ is the cost of one synchronization. Its triangles mark the horizon $H_*$ that minimizes $J(H)$ (Section~\ref{sec:main-cost}), computed from the certified constants alone: 1, 11, 20 and 35 for $C=0,5,20,100$. Like the observed minima (1, 16, 64, 64), $H_*$ grows with $C$, and its interpolated cost is at most 1.29 times the best tested one. The observed minima lie further right because long private search solves each block exactly, so runs finish in about two epochs, faster than the geometric rate behind $J$.

\paragraph{Approximate corrections and the floor.}\label{sec:synthetic-floor}
Every exact correction coefficient here receives a fixed bias $\pm\nu$ with random signs, unknown to the agents. Below the smallest coefficient magnitude, 0.605, the bias changes no decision and every run reaches the optimum. Larger biases leave a floor that grows with $\nu$, from 3.08 at $\nu=1$ to 30.0 at $\nu=2$ for $H=8$, and is essentially the same for $H=32$ (Figure~\ref{fig:main-synthetic-known}(d)): longer private search cannot remove it. The recurrence with measured two-point errors bounds the floor in every run (Appendix~\ref{app:synthetic-floor}).

\section{Conclusions and limitations}
\label{sec:conclusions}
Dividing a shared task among agents works when each agent can see, through a correction, how its changes affect the others. We formalized this as corrected block optimization with verified, compatible merging, whose rate and error floor are set by correction accuracy, local search quality and block geometry. A coordinator that accepts the best verified combination of proposals keeps these guarantees, and its rate depends on how well the proposals actually combine. In practice, agents need corrections built from verified interactions, given as numbers or as information in their prompts, proposals verified against a common base, and merges that are either verified as a whole or limited to non-interacting changes.

With GPT-OSS-120B, a team of three agents found collision-free plans in all 40 runs on two maps, better than those of a single agent, which was safe in all but one run; three single agents searching in parallel were faster than this team but ended with worse plans more often. On a larger task made of both maps, a team of six agents kept its better plans and needed half the single agent's median search time. Merging the best verified combination of proposals keeps the agents' compatible work: it cut the team's search time by about 30\%, let it reach low losses sooner than the parallel single agents and made it, in median, the first to reach a safe plan, with the same guarantees, while merging all proposals unchecked did not help.

Coupling also rose and fell during the solve: the plan had no collisions at the start, several after the first round, when nearly all agents changed their parts at once, and none at the end, when at most one agent still improved in most rounds. We expect the same arc in other shared tasks, because parts that are first changed independently must then be reconciled, and near a good solution few changes remain; a merge that adapts to the conflicts that actually occur, as the best combination does, then pays for coupling only while it lasts.

\paragraph{Limitations.}
The block error bound excludes tasks that require coordinated changes of several blocks; in routing it holds at every audited state that starts a round, but a few final states near the optimum are single-block traps. For LLM agents we measure the constants on visited histories but cannot certify the uniform conditional contraction of Assumption~\ref{ass:local}. Our LLM evidence comes from two maps and their two-floor combination, one model and up to 20 runs per variant. On a single map, a single agent with a well-presented state also solves the task and parallel single agents search faster than Team (Alg.~\ref{alg:protocol}). The prompt strongly affects both protocols (Appendix~\ref{app:routing-development}), so prompts must be matched before protocols are compared.

\begin{acks}
The research reported in this publication was supported by funding from King Abdullah University of Science and Technology (KAUST): i)~KAUST Baseline Research Scheme, ii)~CRG Grant ORFS-CRG12-2024-6460, and iii)~Center of Excellence for Generative AI, under award number 5940.
\end{acks}

\bibliographystyle{ACM-Reference-Format}
\bibliography{references}

\clearpage
\onecolumn
\section*{Notation}
\phantomsection\label{supp:notation}
This table collects the notation used in the main text and the appendices. Each entry gives the meaning of a symbol. The same letter sometimes has a different meaning in a proof or experiment; these uses are listed separately. Block subscripts and fixed arguments are occasionally suppressed in proofs, so $f,\widehat f,R,\widehat R,D,g,U,q,\delta,\delta^*,\delta^U,\varepsilon$ then denote the corresponding block quantities. The index $j$ denotes a coordinate within a block, a color class, or a single-agent answer, as stated. A hat denotes an implemented approximation or an estimate, not a different task.

\begingroup
\small
\setlength{\tabcolsep}{3pt}
\renewcommand{\arraystretch}{0.98}
\begin{longtable}{@{}>{\raggedleft\arraybackslash}p{.16\textwidth}>{\raggedright\arraybackslash}p{.82\textwidth}@{}}
\caption{Notation used in the paper.}\label{tab:notation}\\
\toprule
Symbol & Meaning\\\midrule
\endfirsthead
\toprule
Symbol & Meaning (continued)\\\midrule
\endhead
\midrule
\multicolumn{2}{r}{Continued on the next page}\\
\endfoot
\bottomrule
\endlastfoot

\multicolumn{2}{@{}l}{\textbf{Shared task, local objectives and corrections}}\\*\midrule
$m$ & Number of blocks or agents.\\
$b$ & Block or agent index.\\
$\cX_b,\cX$ & Finite block domain and product domain $\cX_1\times\cdots\times\cX_m$.\\
$x,x_b,x_{bj}$ & Shared candidate solution, its block $b$, and coordinate $j$ of that block.\\
$x_{-b},(x_{-b},u)$ & All blocks other than $b$, and the solution with block $b$ replaced by $u$.\\
$x^0,x^k,x^+$ & Initial solution, solution at epoch $k$, and solution after the next commit.\\
$u,U_b,U_{kb}$ & Trial block value, returned block proposal, and the proposal in epoch $k$.\\
$u_b^*,x^*$ & A true conditional block minimizer and a global minimizer.\\
$F(x),F^*$ & Global loss $F:\cX\to\mathbb R$ and its minimum over $\cX$.\\
$\Delta(x),\Delta_0$ & Global objective gap $F(x)-F^*$ and its initial value $\Delta(x^0)$.\\
$L_b(u)$ & Available local objective $L_b:\cX_b\to\mathbb R$; $L$ is its one-block example.\\
$f_b(u;x)$ & Exact conditional global change $F(x_{-b},u)-F(x)$; $f_b:\cX_b\times\cX\to\mathbb R$.\\
$D_b(x)$ & Best true block decrease $-\min_u f_b(u;x)$.\\
$R_b(u;x)$ & Exact anchored correction $f_b(u;x)-[L_b(u)-L_b(x_b)]$; $R_b:\cX_b\times\cX\to\mathbb R$.\\
$\widehat R_b(u;x)$ & Implemented approximation to the exact correction; $\widehat R_b:\cX_b\times\cX\to\mathbb R$.\\
$\widehat f_b(u;x)$ & Corrected local change $L_b(u)-L_b(x_b)+\widehat R_b(u;x)$; $\widehat f_b:\cX_b\times\cX\to\mathbb R$.\\
$\widehat f_b^*$ & Minimum of $\widehat f_b(\cdot;x)$ over $\cX_b$.\\
$\widehat F(x)$ & Learned global surrogate $G(x)+\langle\widehat\vartheta^+,\phi(x)\rangle$.\\
$M,\ell_a,N(a)$ & Number of objective factors, factor $a$, and the blocks on which it depends.\\
$x_{N(a)}$ & Subvector indexed by the factor's block set.\\
$\chi(b),\kappa$ & Color assigned to block $b$ and number of colors.\\
$\mathcal C_j,j^k$ & Color class $\{b:\chi(b)=j\}$, and the color selected in epoch $k$.\\
$g_b$ & Verified nonnegative gain $\max\{0,-f_b(U_b;x)\}$.\\
$S^k$ & Blocks committed in epoch $k$: the improving blocks of the selected color (Algorithm~\ref{alg:protocol}) or the best verified combination (Algorithm~\ref{alg:best-combination}).\\
$(x_{-S},U_S)$ & Solution $x$ with every block $b\in S$ replaced by its proposal $U_b$.\\
$\Phi(S)$ & Loss $F(x_{-S},U_S)$ of accepting the proposals of the blocks in $S$ (Section~\ref{sec:main-combination}).\\
$I^k$ & Improving blocks of epoch $k$, those with $g_b>0$.\\
$\xi_{bb'}$ & Interaction of the proposals of blocks $b$ and $b'$: their joint change minus the sum of their single changes; they conflict if $\xi_{bb'}>0$.\\
$\kappa^*_k$ & Realized number of colors of the best-combination merge in epoch $k$: $\sum_bg_b$ over the realized gain (Section~\ref{sec:main-combination}).\\
$\kappa_k,\deg_k(b)$ & Effective number of colors, the smallest number of conflict-free sets covering $I^k$, and the number of improving blocks that conflict with $b$ (Section~\ref{sec:main-combination}); $\deg(b)$ in the synthetic task is the degree in the dependency graph.\\
$\bar\kappa$ & Bound on $\kappa^*_k$ in every epoch (Corollary~\ref{cor:realized-rate}).\\

\multicolumn{2}{@{}l}{\textbf{Convergence and stopping certificates}}\\*\midrule
$k,K$ & Current epoch index and prescribed number of epochs.\\
$H$ & Number of private search steps per epoch.\\
$\mathcal H,\mathcal H_k$ & History conditioning a local search, or history before epoch $k$'s calibration.\\
$\delta_b^*,\delta_b^U$ & Two-point correction-error bounds (Assumption~\ref{ass:correction}): at a true block minimizer, and in conditional expectation at the returned proposal.\\
$\delta_b$ & Uniform absolute correction-error bound over block candidates.\\
$\varepsilon(u),\varepsilon_b(u)$ & Correction error $\widehat f(u)-f(u)$ of a candidate in the proofs, and of block $b$ in Proposition~\ref{dev:prop:recorded}.\\
$Q_b,e^*_b,e^U_b,\widehat D_b$ & In one recorded epoch (Proposition~\ref{dev:prop:recorded}): captured fraction of the corrected improvement, correction errors at a true block minimizer and at the returned proposal, and the available corrected improvement.\\
$\rho_b,\rho,\rho_{\min}$ & Per-step local contraction parameter and its minimum across blocks.\\
$q_b(H),q(H)$ & Captured-improvement guarantee $1-(1-\rho_b)^H$, and its minimum over blocks.\\
$\mu$ & Block error-bound constant in $\sum_bD_b(x)\geq\mu\Delta(x)$.\\
$\mu_{\rm fin}$ & Capped minimum improvement-to-gap ratio on the finite domain.\\
$\underline f_b$ & Computable lower bound on $\min_u\widehat f_b(u;x)$.\\
$\alpha$ & Descent coefficient $\mu q(H)/\kappa$.\\
$B$ & Correction-error term $\kappa^{-1}\sum_b(q_b(H)\delta_b^*+\delta_b^U)$; at most $\kappa^{-1}\sum_b(1+q_b(H))\delta_b$ under a uniform bound.\\
$B_\delta$ & Uniform-error counterpart $\kappa^{-1}\sum_b(1+q_b(H))\delta_b$ of $B$ (Appendix~\ref{sec:convergence}).\\
$B/\alpha$ & Error floor in the expected-gap bound.\\
$\Delta_k,\alpha_k,B_k$ & Gap after round $k$ and the round's measured coefficients in $\Delta_{k+1}\leq(1-\alpha_k)\Delta_k+B_k$.\\
$\zeta$ & Relative-error fraction in $(1+q_b)\delta_b\leq\zeta q_bD_b$.\\
$\varepsilon$ & Requested objective-gap tolerance; an expected-gap tolerance where expectation is specified.\\

\multicolumn{2}{@{}l}{\textbf{Learning corrections from global evaluations}}\\*\midrule
$G(x)$ & Known part of the structured global objective.\\
$\phi(x),r$ & Known feature vector in $\mathbb R^r$ and its dimension.\\
$\vartheta,\widehat\vartheta$ & Fixed unknown coefficients and their retained estimate.\\
$\widehat\vartheta^k,\widehat\vartheta^+$ & Estimate indexed by epoch and estimate immediately after calibration.\\
$c_0$ & Unknown additive constant in the objective.\\
$\Lambda_b$ & Known bound on $\|\phi(x_{-b},u)-\phi(x)\|_2$.\\
$z,z_b$ & Feasible calibration probe and singleton proposal $(x_{-b},U_b)$.\\
$a,a_b$ & Feature displacement $\phi(z)-\phi(x)$, or that of proposal $z_b$.\\
$y,y_b$ & Residual global difference after subtracting the known-part difference.\\
$y-\langle a,\widehat\vartheta\rangle$ & Prediction residual for the current coefficient estimate.\\
$P$ & Measurement projector $aa^\top/\|a\|_2^2$; zero for a skipped or zero-length probe.\\
$I$ & Identity matrix on the coefficient space.\\
$\sigma$ & Excitation constant in $\mathbb E[P\mid\mathcal H_k]\succeq\sigma I$.\\
$\eta$ & Learning contraction factor $\sqrt{1-\sigma}$.\\
$\widetilde\vartheta,\widetilde\vartheta^+,\widetilde\vartheta^{\rm next}$ & Parameter error $\widehat\vartheta-\vartheta$, after calibration, and after extra reused measurements.\\
$E_k$ & Upper envelope of the gap along a recorded run (Proposition~\ref{dev:prop:recorded}).\\
$\beta$ & Conversion coefficient $\kappa^{-1}\sum_b(1+q_b(H))\Lambda_b$.\\
$\omega$ & Potential weight $2\beta/(1-\eta)$.\\
$\tau$ & Joint rate $\min\{\alpha,(1-\eta)/2\}$.\\
$\mathcal V_k$ & Joint potential $\Delta(x^k)+\omega\|\widehat\vartheta^k-\vartheta\|_2$.\\

\multicolumn{2}{@{}l}{\textbf{Binary calibration and observable uncertainty}}\\*\midrule
$d_b,d_{\max}$ & Number of binary coordinates in block $b$, and the largest block dimension.\\
$\Psi(x)$ & Known coupling function in the coordinate residual model.\\
$\vartheta_{bj},\widehat\vartheta_{bj}$ & True and estimated coefficient of binary coordinate $(b,j)$.\\
$M_{bj}$ & Known initial bound on $|\vartheta_{bj}-\widehat\vartheta^0_{bj}|$.\\
$s$ & Bit displacement: signed one-bit change $z_{bj}-x_{bj}\in\{-1,1\}$.\\
$p$ & Probe probability: probability that a block with unknown coordinates performs calibration.\\
$\mathcal U_b^k,n_b$ & Remaining unknown coordinates before epoch $k$, and their number in a block.\\
$\Gamma_k,\Gamma_{b,k}$ & Radii: sum of remaining bounds $M_{bj}$ over all blocks, or over block $b$.\\
$d_{\rm tot}$ & Query budget: total number of binary coordinates $\sum_b d_b$.\\
$\gamma$ & Radius contraction parameter $p/d_{\max}$.\\
$\beta_0$ & Coordinate-model conversion factor $\kappa^{-1}\max_b(1+q_b(H))$.\\
$\omega_0,\tau_0$ & Radius weight $2\beta_0/\gamma$ and joint rate $\min\{\alpha,\gamma/2\}$.\\
$\mathcal W_k$ & Observable-radius potential $\Delta(x^k)+\omega_0\Gamma_k$.\\
$M$ & Value range: width of a known interval of objective values in the impossibility argument.\\

\multicolumn{2}{@{}l}{\textbf{Private work and cost accounting}}\\*\midrule
$c$ & Private cost: modeled cost of one parallel private-search slot.\\
$C$ & Other per-epoch cost, including synchronization and context preparation; it includes verifier calls in Appendix~\ref{sec:horizon} and excludes them, charged through $v$, in Appendix~\ref{sec:certificate}.\\
$v$ & Cost of one scalar global evaluation.\\
$K_\varepsilon(H)$ & Sufficient epoch count for expected gap at most $\varepsilon$.\\
$T_\varepsilon(H),T_K$ & Sufficient modeled completion cost and cost through $K$ epochs.\\
$\lambda$ & $-\log(1-\rho)$, the exponential form of the worst-block local rate.\\
$J(H)$ & Relaxed cost ratio $(cH+C)/(1-e^{-\lambda H})$.\\
$H_*$ & Positive continuous minimizer of $J(H)$ for $C>0$.\\
$\psi(H)$ & Proof helper: $e^{\lambda H}-1-\lambda H-\lambda C/c$.\\
$N_{\rm q},K_{\rm end}$ & Synthetic experiments: number of counted global evaluations, and the epoch at which a run completes.\\

\multicolumn{2}{@{}l}{\textbf{Robot-routing task and its offline references}}\\*\midrule
$r_i,i$ & Robot label and robot index, $i=0,\ldots,8$.\\
$\mathcal I_b$ & Set of robots owned by team block $b$.\\
$W$ & Number of single agents searching in parallel in the best-of-$W$ baseline, equal to the number of team agents.\\
$x_i,u_i$ & Routes: accepted and candidate command arrays for robot $i$.\\
$(x,y),p$ & Grid coordinates and a free cell in a reference-search state; $x$ increases rightward and $y$ downward.\\
$t$ & Routing tick; the fixed horizon is 48 ticks.\\
$s_i,g_i,p_i(t)$ & Start, goal and position of robot $i$ at tick $t$.\\
$\tau_i$ & First goal-arrival tick of a delivered robot.\\
$d_i(p,g_i)$ & Shortest legal grid distance from cell $p$ to robot $i$'s goal.\\
$c_i(x_i)$ & Delivery cost of robot $i$. It equals the arrival tick for a delivered robot, otherwise $100+48+d_i(p_i(48),g_i)$.\\
$C_{ij}(x_i,x_j)$ & Number of simultaneous vertex and edge-swap conflicts for a robot pair.\\
$\mathcal P^k$ & External robot pairs discovered before round $k$.\\
$\widehat R_b(u;x^k)$ & Routing correction: anchored collision-penalty change over the retained external pairs.\\
$\Delta L_b,\Delta\widehat R_b$ & Change in local loss and in the retained correction for a proposed route block.\\
$A_b^{\mathcal P^k}(u;x^k)$ & Absolute private score using the retained external pairs.\\
$A_b^{\rm all}$ & Absolute block score using every external pair.\\
$\bar d_b$ & Routing bound: sum of shortest distances for robots owned by block $b$.\\
$A_b^{\rm CBS}$ & Reference score: absolute score of a collision-free delivered reference plan in the soft-penalty certificate.\\
$\underline A_i,w_b$ & Single-robot relaxation minimum and joint block witness program.\\
$C_{b,\rm miss}^k(u)$ & Conflicts involving undiscovered external pairs against frozen peers.\\
$\varepsilon_{kb}(u)$ & Realized score error $\widehat f_b(u;x^k)-f_b(u;x^k)$.\\
$Q_{kb},Q_{s,k,b}$ & Fraction of the corrected improvement captured by a returned block; the counterpart of $q_b(H)$.\\
$\mu_{\rm obs},\mu_s$ & Minimum capped improvement-to-gap ratio over recorded states, or within run $s$.\\
$e^*_{s,k,b},e^U_{s,k,b}$ & Measured correction errors at a true block minimizer and at the returned block; realized counterparts of $\delta_b^*,\delta_b^U$.\\
$\Delta_{s,k}$ & Objective gap for run $s$ after round $k$.\\
$\alpha_{s,k},B_{s,k}$ & Round-specific descent and correction-error coefficients computed from all returned blocks.\\
$\bar\alpha,\bar B$ & Pooled arithmetic means of the measured round coefficients.\\
$x_s^k,K_s$ & Accepted state at round $k$ in run $s$, and that run's stopping round.\\
$E_{s,k}$ & Upper envelope for run $s$'s objective gap at round $k$.\\
$\overline E_k,\bar\alpha_k,\bar B_k$ & Envelope propagated with round-$k$ coefficients averaged over the runs of one half (check on the other half).\\
$\overline U_K$ & Mean loss envelope: the optimum plus the mean of $E_{s,\min\{K,K_s\}}$ over runs.\\
$s$ (replicate), $j,k,h,b$ & Run index; single-agent step, team round, private step and block indices in the API-seed scheme.\\
$a_j,a_{kbh}$ & Measured duration of a checkable model answer.\\
$t_{kb},t_{\rm team}$ & Sum of a block's checked durations in a round, and the team's search time.\\
\shortstack[r]{$T_{\rm parallel}$\\$T_{\rm sequential}$} & Search time of the parallel team and of the sequential replay of the same answers.\\
$p$ & Top-$p$: nucleus-sampling parameter, fixed to $0.95$ in routing.\\

\multicolumn{2}{@{}l}{\textbf{Synthetic optimization experiment}}\\*\midrule
$t_{bj},\tilde t_{bj}$ & Desired target bit and private target bit.\\
$a_{bj}$ & Unary mismatch weight for bit $(b,j)$.\\
$w$ & Neighbor-coupling weight, fixed at $0.2$.\\
$\pi_b$ & Activity probability of block $b$'s randomized coordinate solver.\\
$A_{bj}(x)$ & Exact affine coefficient of candidate bit $u_j$ with peers frozen.\\
$h_{bj}(x)$ & Difference between the exact and private affine coefficients.\\
$\widehat A_{bj}$ & Private coefficient plus learned unary correction and refreshed coupling coefficient.\\
$c_b,\deg(b)$ & Total incident coupling weight $w\deg(b)$ and block degree on the path.\\
$I(x)$ & Set of coordinates whose bits differ from the desired target.\\
$\mu_{\rm dom}$ & Unary-dominance certificate $\min_{b,j}(a_{bj}-c_b)/(a_{bj}+c_b)$.\\
$\widehat D(x)$ & Sum of all exact affine surrogate improvements at the committed base.\\
$\Gamma(x)$ & Sum of remaining coordinate uncertainty bounds at the current base.\\
$\overline\Delta(x)$ & Stopping: certified gap bound $(\widehat D(x)+\Gamma(x))/\mu_{\rm dom}$.\\
$\nu,s_{bj}$ & Persistent correction bias and its random sign in the approximate-correction series.\\
$e^*_b,e^U_b,Q_b$ & Realized correction errors at a true block minimizer and at the returned proposal, and the realized captured fraction, in the same series.\\

\multicolumn{2}{@{}l}{\textbf{Database experiment}}\\*\midrule
$s,s_{\rm db}$ & Database source index and database-generation seed.\\
$A_s,B_s,C_s$ & Left endpoint, edge and right endpoint tables for source $s$.\\
$g_s,h_s$ & Selected left and right endpoint-group IDs.\\
$\ell,r$ & Join directions: left-first and right-first traversal choices.\\
$q,q_s,a_s$ & Query/report index, selected predicates for source $s$, and its join-direction choice.\\
$v_s(\ell,g),v_s(r,h)$ & Edge fan-out of a selected left or right group.\\
$J_q(a),a$ & Separable fan-out surrogate and vector of three join directions.\\

\multicolumn{2}{@{}l}{\textbf{Mathematical operations and summary conventions}}\\*\midrule
$\mathbb R,\mathbb R^r$ & Real numbers and real vectors of dimension $r$.\\
$\langle a,a'\rangle,a^\top$ & Euclidean inner product $\sum_j a_ja'_j$ and vector or matrix transpose.\\
$\|\cdot\|_2,\|\cdot\|_2^2$ & Euclidean norm and its square.\\
\shortstack[r]{$\|z-x\|_1$\\$|z_j-x_j|$} & Sum of absolute coordinate differences and scalar absolute value.\\
\shortstack[r]{$\mathbb E[Z]$\\$\mathbb E[Z\mid\mathcal H]$} & Expectation and conditional expectation of a random quantity $Z$.\\
$A\succeq B$ & Positive-semidefinite matrix order: $A-B$ is positive semidefinite.\\
$\eqdef,=,\leq,\geq$ & Definition, equality and scalar inequality relations.\\
$\min,\max,\sup$ & Minimum, maximum and supremum over the stated index or candidate set.\\
$\sum,\times$ & Sum and multiplication or Cartesian product as indicated.\\
$\in,\subseteq,\varnothing$ & Membership, subset relation and empty set.\\
$\log,\exp,\sqrt{\cdot}$ & Natural logarithm, exponential and square root.\\
$\lceil z\rceil$ & Smallest integer greater than or equal to $z$.\\
$\min\{\cdot\},\max\{\cdot\}$ & Minimum or maximum of listed scalar quantities.\\
$\approx,O(n)$ & Approximation and asymptotic upper-bound notation.\\
$n,\pm,\mathrm{SD},\mathrm{SE}$ & Sample count, displayed variation, sample standard deviation and standard error.\\
$+$, $*$ & A plus superscript marks an update; a star marks an optimum. Hats mark approximations or estimates; a tilde marks the private target of the synthetic task and, in Appendix~\ref{sec:learn}, a parameter error; bars and underlines mark averages or bounds, as defined in each entry.\\
\end{longtable}
\endgroup

\twocolumn
\appendix
\numberwithin{equation}{section}
\numberwithin{theorem}{section}
\numberwithin{assumption}{section}
\numberwithin{figure}{section}
\numberwithin{table}{section}
\section{Further theory and examples}\label{app:development}
This appendix extends Sections~\ref{sec:main-setup}--\ref{sec:main-cost} and uses their notation; the table on page~\pageref{supp:notation} lists every symbol. All guarantees concern the specified objective $F$, not requirements that its evaluation omits, and private work means computation without communication, not confidential data. Section~\ref{sec:obstacle} works through small examples of the model. Section~\ref{sec:convergence} treats a uniform correction-error bound, characterizes the block error bound, gives an optimality certificate and proves the bound along a recorded run used in Section~\ref{sec:main-routing}. Section~\ref{sec:horizon} proves the private-work horizon of Section~\ref{sec:main-cost}. Sections~\ref{sec:learn} and~\ref{sec:certificate} develop the learning theory summarized in Section~\ref{sec:main-learning}, with its certificate and cost. The best-combination merge and its guarantees are stated in Section~\ref{sec:main-combination}, with a conflict-greedy variant in Appendix~\ref{app:routing-merge-theory}.

\subsection{Examples}\label{sec:obstacle}
\paragraph{Why local improvement may not help.}
A module can pass its isolated tests while returning a format that another module cannot read. An agent searching only with those tests has no reason to repair the format. We call the difference between what the local and global objectives reward an \emph{objective mismatch}. More local search does not necessarily make this mismatch larger; it may simply solve the wrong local problem more accurately. Checking a proposal globally can reject it, but rejection alone does not tell the agent how to search better.

The following one-decision example isolates this issue. Let $u=0$ mean retaining an incompatible output format and $u=1$ mean repairing it. Let $F$ be the global objective and $L$ the local objective, both minimized:
\begin{equation}\label{eq:obstacle}
F(u)=1-u,\qquad L(u)=u,\qquad u\in\{0,1\}.
\end{equation}
Starting at zero, even an exact minimizer of $L$ keeps the incompatible format, whereas $F$ is minimized at one.

Verification therefore protects the current solution but does not resolve the cause of the failed search. To produce a useful proposal, the agent needs a different criterion for comparing its candidates, and the correction supplies it. At base $u=0$, the local change from choosing one is $+1$, while the global change is $-1$. A correction of $-2$ to that candidate's local change makes the corrected score prefer the repair, and the correction at the unchanged candidate is zero, as anchoring requires.

\paragraph{A two-block example.}
Let $x_1,x_2\in\{0,1\}$ encode two decisions. Each local objective prefers zero, but the global objective prefers one and also penalizes disagreement:
\begin{equation}\label{eq:worked-task}
\begin{gathered}
L_1(u)=L_2(u)=u,\\
F(x)=(1-x_1)+(1-x_2)+\tfrac14(x_1-x_2)^2.
\end{gathered}
\end{equation}
Starting from $x=(0,0)$, the loss is 2. Changing either block to one lowers it to $1.25$, although its local objective rises by 1. Thus the exact anchored correction at that candidate is $-1.75$: adding it to the local change gives the true change $-0.75$. For either block, the corrected score of leaving it unchanged is zero.

\begin{center}
\small\setlength{\tabcolsep}{3pt}
\begin{tabular}{lrrr}
\toprule
Block-1 choice & Local change & Correction & Sum\\
\midrule
Keep $u=0$ & 0 & 0 & 0\\
Choose $u=1$ & 1 & $-1.75$ & $-0.75$\\
\bottomrule
\end{tabular}
\end{center}
Section~\ref{sec:learn} returns to it to show how objective evaluations reveal its unknown correction coefficients.

\paragraph{One update.}
The two blocks share the disagreement factor, so they have different colors. With exact local minimization, both propose one, with singleton gain $0.75$. Either color is best. Break the tie in favor of block 1: the next solution is $(1,0)$ and its loss is $1.25$. At this new base, changing block 2 to one improves the loss by $1.25$, and the next update reaches $(1,1)$ with loss zero. Notice that the conditional score changed when its neighboring block changed; a correction must be evaluated at the current base.

\paragraph{Changes that clash.}
Two blocks that share a factor can each lower the loss alone and yet raise it together. On two binary blocks, let
\[
F(x)=(1-x_1)+(1-x_2)+3x_1x_2.
\]
From $x=(0,0)$, with loss 2, changing either block to one lowers the loss to 1, but changing both raises it to 3, like two robots that take the same shortcut and collide. The factor $3x_1x_2$ joins the two blocks, so they have different colors, and Algorithm~\ref{alg:protocol} accepts only one of the two changes.

\paragraph{A single-block trap.}
Even a perfect correction cannot resolve a task that violates Assumption~\ref{ass:eb}. For example, on two binary blocks let $F(0,0)=1$, $F(1,0)=F(0,1)=2$, and $F(1,1)=0$. At $(0,0)$, neither single-block change helps, so $D_1+D_2=0$ while the global gap is 1. Joint search or acceptance of intermediate worsening is needed. The parameter $\mu$ describes the task, rather than the agent or communication protocol.

\subsection{Further convergence results}\label{sec:convergence}
\paragraph{A uniform correction-error bound.}
A bound on the correction error at every candidate,
\begin{equation}\label{dev:eq:error}
\sup_{u\in\cX_b}|\widehat R_b(u;x)-R_b(u;x)|\leq\delta_b,
\end{equation}
implies Assumption~\ref{ass:correction} with $\delta_b^*=\delta_b^U=\delta_b$. Lemma~\ref{lem:local} then gives
\begin{equation}\label{dev:eq:localgain}
\mathbb E[g_b\mid\mathcal H]\geq q_b(H)D_b(x)-(1+q_b(H))\delta_b,
\end{equation}
and Theorem~\ref{thm:rate} holds with $B=B_\delta\eqdef\kappa^{-1}\sum_b(1+q_b(H))\delta_b$. Such a bound must also hold for poor candidates that no reasonable search returns, so it can be far larger than the two-point constants; in routing it exceeds $10^4$ (Appendix~\ref{app:routing-assumptions}). Lemma~\ref{lem:local} and the first step of the proof of Theorem~\ref{thm:rate} hold epoch by epoch, conditionally on $\mathcal H$, so they remain valid when $\delta_b$ changes between epochs:
\begin{equation}\label{eq:drift}
\mathbb E[F(x)-F(x^+)\mid\mathcal H]\geq\alpha\Delta(x)-B_\delta.
\end{equation}
The learning results of Sections~\ref{sec:learn} and~\ref{sec:certificate} use them in this form.

\paragraph{A relative correction error.}
If instead $q_b(H)\delta_b^*+\delta_b^U\leq\zeta q_b(H)D_b(x)$ for a fixed $\zeta<1$ at every epoch, the proof of Theorem~\ref{thm:rate} gives the contraction factor $1-(1-\zeta)\mu q(H)/\kappa$ and no floor. This relative certificate is useful analytically but is not automatically observable: $D_b$ is usually unknown.

\paragraph{How restrictive is the block error bound?}
On a finite domain, existence of some positive constant has a simple characterization. The size of that constant is a separate question: it may be arbitrarily small. The following observation also explains when a computable stopping certificate is available.

\begin{proposition}[Block geometry and an optimality certificate]\label{prop:geometry}
On the full finite product domain, a constant $\mu\in(0,1]$ satisfying~\eqref{eq:eb} exists if and only if every point with $D_b(x)=0$ for all $b$ is globally optimal. When this local-to-global property holds and there are suboptimal points, one valid choice is
\begin{equation}\label{eq:finite-mu}
\mu_{\rm fin}\eqdef\min\left\{1,\min_{x:\Delta(x)>0}
\frac{\sum_bD_b(x)}{\Delta(x)}\right\}.
\end{equation}
If every point is optimal, take $\mu_{\rm fin}=1$. Moreover, at any base where a known $\mu>0$ satisfies~\eqref{eq:eb}, exact block improvements give
\begin{equation}\label{eq:stop-exact}
\Delta(x)\overset{\eqref{eq:eb}}{\leq}\frac1\mu\sum_bD_b(x).
\end{equation}
More generally, suppose certified bounds $\delta_b$ satisfy~\eqref{dev:eq:error} and computable lower bounds $\underline f_b\leq\min_u \widehat f_b(u;x)$ are available. Then
\begin{equation}\label{eq:stop-inexact}
\Delta(x)\overset{\eqref{eq:eb},\eqref{dev:eq:error},\eqref{eq:correction},\eqref{eq:surrogate}}{\leq}\frac1\mu\sum_b(\delta_b-\underline f_b).
\end{equation}
\end{proposition}
\paragraph{Proof idea.} Finiteness turns strictly positive improvement ratios into a positive minimum. For certification, the error bound converts upper bounds on exact block improvements into an upper bound on the global gap.
\begin{proof}
If~\eqref{eq:eb} holds and all $D_b(x)$ vanish, then $\Delta(x)=0$. Conversely, the stated local-to-global property makes every ratio in~\eqref{eq:finite-mu} positive; the nonempty set of ratios is finite, so its minimum is positive. At a global optimum every $D_b$ is zero, and hence the resulting constant satisfies~\eqref{eq:eb} everywhere. For the approximate certificate, uniform error gives
\begin{equation}\label{eq:stop-block}
\begin{gathered}
\min_u f_b(u;x)
\overset{\eqref{dev:eq:error},\eqref{eq:correction},\eqref{eq:surrogate}}{\geq}
\min_u \widehat f_b(u;x)-\delta_b\geq\underline f_b-\delta_b,\\
D_b(x)\overset{\eqref{eq:gap},\eqref{dev:eq:error},\eqref{eq:correction},\eqref{eq:surrogate}}{\leq}\delta_b-\underline f_b.
\end{gathered}
\end{equation}
Combining these bounds yields
\begin{equation*}
\Delta(x)\overset{\eqref{eq:eb}}{\leq}\frac1\mu\sum_bD_b(x)
\overset{\eqref{eq:stop-block}}{\leq}\frac1\mu\sum_b(\delta_b-\underline f_b),
\end{equation*}
which proves~\eqref{eq:stop-inexact}. The same bound holds with $\delta_b$ replaced by a bound $\delta_b^*$ on the error at a true block minimizer.
\end{proof}
Neither enumerating~\eqref{eq:finite-mu} nor solving all block subproblems is automatically cheap. Approximate proposals give lower bounds on available improvement, whereas stopping needs upper bounds. An exact block solve or a certified lower bound on each surrogate minimum supplies the missing direction. A known valid $\mu$ is also required. Thus the generic local solver does not automatically provide a stopping test, but the theory does permit one when this additional information is available.

Local-to-global results in discrete convex analysis depend on the allowed neighborhood: coordinated subset moves or exchanges may be essential \citep{murota2003}. Merely calling an objective discrete convex does not establish~\eqref{eq:eb} for the fixed disjoint blocks used here.

\paragraph{A bound along a recorded run.}
The argument of Lemma~\ref{lem:local} also applies to a single realized epoch, with realized local progress in place of $q_b(H)$ and realized two-point errors in place of $\delta_b^*,\delta_b^U$.
\begin{proposition}[Bound along a recorded run]\label{dev:prop:recorded}
Fix a realized epoch $k$ with base $x=x^k$ and $\Delta(x)>0$. Suppose that every returned block scores no worse than the current block, that is, $\widehat f_b(U_b;x)\leq0$. Write
\[
\varepsilon_b(u)\eqdef\widehat f_b(u;x)-f_b(u;x),\qquad \widehat D_b\eqdef-\min_u\widehat f_b(u;x).
\] Let $Q_b\eqdef-\widehat f_b(U_b;x)/\widehat D_b\in[0,1]$ if $\widehat D_b>0$ and $Q_b\eqdef0$ otherwise, and let $e^*_b\eqdef|\varepsilon_b(u_b^*)|$ for a true block minimizer $u_b^*$ and $e^U_b\eqdef|\varepsilon_b(U_b)|$. Then the merge of Algorithm~\ref{alg:protocol} gives
\begin{equation}\label{dev:eq:recorded}
\Delta(x^+)\leq\min\{\Delta(x),\,(1-\alpha_k)\Delta(x)+B_k\},
\end{equation}
where
\[
\alpha_k\eqdef\min\Bigl\{1,\frac{\sum_bQ_bD_b(x)}{\kappa\Delta(x)}\Bigr\},\qquad B_k\eqdef\frac1\kappa\sum_b(Q_be^*_b+e^U_b).
\] Consequently, if every epoch with $\Delta(x^k)>0$ satisfies these hypotheses, and with $\alpha_k\eqdef B_k\eqdef0$ once $\Delta(x^k)=0$, which then persists because the objective never increases, the sequence $E_0=\Delta(x^0)$, $E_{k+1}=\min\{E_k,(1-\alpha_k)E_k+B_k\}$ satisfies $\Delta(x^k)\leq E_k$ for every $k$.
\end{proposition}
\begin{proof}
If $\widehat D_b>0$, then $-\widehat D_b\leq\widehat f_b(u_b^*;x)=-D_b(x)+\varepsilon_b(u_b^*)$, so $-\widehat f_b(U_b;x)=Q_b\widehat D_b\geq Q_b(D_b(x)-e^*_b)$. Hence
\[
\begin{aligned}
g_b\geq-f_b(U_b;x)&=-\widehat f_b(U_b;x)+\varepsilon_b(U_b)\\
&\geq Q_bD_b(x)-Q_be^*_b-e^U_b.
\end{aligned}
\]
If $\widehat D_b=0$, the right-hand side is $-e^U_b\leq0\leq g_b$. Summing over blocks and using~\eqref{eq:merge}, $\Delta(x)-\Delta(x^+)\geq\kappa^{-1}\sum_bg_b\geq\alpha_k\Delta(x)-B_k$; capping $\alpha_k$ at one only weakens this inequality, and $\Delta(x^+)\leq\Delta(x)$ because the objective never increases. Since $1-\alpha_k\geq0$, induction on~\eqref{dev:eq:recorded} gives $\Delta(x^k)\leq E_k$.
\end{proof}
The proposition uses only quantities realized in one run, so it holds on every recorded trajectory without assumptions on the local solver beyond returning a block that scores no worse than the current one. Under the best-combination merge, $\kappa$ can be replaced by the realized number of colors $\kappa^*_k$ (Corollary~\ref{cor:realized-rate}), and, with pairwise factors and all combinations checked, by the effective number of colors $\kappa_k$ (Proposition~\ref{prop:realized-conflicts}, Appendix~\ref{app:routing-merge-theory}). Appendix~\ref{app:routing-empirical-rate} evaluates it on the recorded LLM runs.

\subsection{How much private work is worth doing}\label{sec:horizon}
Section~\ref{sec:main-cost} charges at most $c>0$ for one parallel private step, which lasts as long as its slowest agent, and at most $C\geq0$ for each synchronization: correction construction, certification, message transfer, loading, singleton verification and applying the selected edits. These bounds are uniform, and $C$ does not grow with $H$. With exact corrections, let $T_\varepsilon(H)$ denote the modeled cost of running enough epochs to make the expected gap at most $\varepsilon$. For $0<\varepsilon<\Delta(x^0)$,
\begin{equation}\label{dev:eq:time}
\begin{gathered}
K_\varepsilon(H)\eqdef\left\lceil\frac{\kappa}{\mu q(H)}
\log\frac{\Delta(x^0)}{\varepsilon}\right\rceil,\\
T_\varepsilon(H)\leq(cH+C)K_\varepsilon(H).
\end{gathered}
\end{equation}
Indeed, using $1-t\leq e^{-t}$,
\begin{equation*}
\begin{array}{rcl}
\mathbb E\Delta(x^{K_\varepsilon})
&\overset{\eqref{eq:rate}}{\leq}&(1-\alpha)^{K_\varepsilon}\Delta(x^0)\\
&\leq&e^{-\alpha K_\varepsilon}\Delta(x^0)
\overset{\eqref{eq:constants},\eqref{dev:eq:time}}{\leq}\varepsilon.
\end{array}
\end{equation*}
If correction size or evaluation cost grows with the horizon, one must replace $C$ by that actual bound before optimizing. Token work also grows with the total number of agents' private steps.

Let $\rho\eqdef\min_b\rho_b$ and $\lambda\eqdef-\log(1-\rho)>0$. Dropping the outer ceiling in~\eqref{dev:eq:time} and allowing a real-valued horizon yields the objective
\begin{equation}\label{eq:ratio}
J(H)\eqdef\frac{cH+C}{1-e^{-\lambda H}},\qquad H>0.
\end{equation}
\begin{proposition}[A finite useful private-work horizon]\label{prop:horizon}
For $C>0$, $J$ has a unique positive minimizer $H_*$, characterized by~\eqref{eq:stationary}. Over integer $H\geq1$, it suffices to compare the neighboring integers around $H_*$, clamped to one. For $C=0$, the integer optimum is $H=1$. These optimize the relaxed bound, not necessarily the ceiled bound or actual runtime.
\end{proposition}
\paragraph{Proof idea.} The derivative has the sign of a single strictly increasing function. Its unique zero balances the extra cost of one more private step against the additional improvement that step buys.
\phantomsection\label{proof:prop-horizon}
\begin{proof}
Differentiating~\eqref{eq:ratio} gives
\begin{equation}\label{eq:derivative}
J'(H)\overset{\eqref{eq:ratio}}{=}
\frac{c[1-e^{-\lambda H}]-\lambda(cH+C)e^{-\lambda H}}
{(1-e^{-\lambda H})^2}.
\end{equation}
Its sign is the sign of $\psi(H)=e^{\lambda H}-1-\lambda H-\lambda C/c$. For $C>0$, $\psi(0)<0$, $\psi'(H)=\lambda(e^{\lambda H}-1)>0$ for $H>0$, and $\psi(H)$ tends to infinity. Hence there is exactly one zero, with $J$ decreasing before and increasing after it. This proves both the continuous and integer claims. For $C=0$, $e^t>1+t$ for $t>0$ gives $J'(H)>0$.
\end{proof}

As $H$ increases, $q(H)$ approaches one: the analysis interpolates between inexact and exact block minimization. It supplies a batching calculation, not ProxSkip-type communication acceleration.

\subsection{Learning the correction}\label{sec:learn}
A global objective evaluation can teach the correction if it reveals something reusable. We obtain this property by assuming that the unknown part of the objective has fixed coefficients multiplying known features. Evaluations then give linear equations for those coefficients. Learning them once can help evaluate many later candidates, even as the solution changes.

In the next assumption, $G(x)$ is the known part of the objective. The vector $\phi(x)$ contains $r$ known numerical features, and the unknown vector $\vartheta$ gives their weights. A \emph{probe} is a feasible candidate evaluated to learn these weights; it need not become the team's next solution. We call these learning evaluations \emph{calibration}.

\begin{assumption}[Static identifiable residual]\label{dev:ass:features}
For a positive integer $r$, there are a known function $G$, known features $\phi:\cX\to\mathbb R^r$, and an unknown fixed vector $\vartheta\in\mathbb R^r$ such that
\begin{equation}\label{dev:eq:features}
F(x)=G(x)+\langle\vartheta,\phi(x)\rangle+c_0.
\end{equation}
The unknown constant $c_0$ cancels in differences. Known terms and features can be evaluated locally after context loading; their costs are charged. Exact scalar verifier queries are permitted on feasible speculative solutions. Feature changes obey known bounds
\begin{equation}\label{dev:eq:featurebound}
\|\phi(x_{-b},u)-\phi(x)\|_2\leq\Lambda_b.
\end{equation}
\end{assumption}
The representation helps only if $r$ and the cost of evaluating $G$ and $\phi$ are small. The unknown residual here means $F-G-c_0$; the measurement residual introduced next instead means an observed value minus its prediction.

\paragraph{An objective difference gives a linear measurement.}
At base $x$, query a feasible probe $z$ and compute
\begin{equation}\label{dev:eq:measurement}
\begin{gathered}
a\eqdef\phi(z)-\phi(x),\\
y\eqdef F(z)-F(x)-[G(z)-G(x)]\\
\overset{\eqref{dev:eq:features}}{=}\langle a,\vartheta\rangle.
\end{gathered}
\end{equation}
Let $\widehat\vartheta$ denote the current estimate of the unknown weights. Its predicted measurement is $\langle a,\widehat\vartheta\rangle$, so $y-\langle a,\widehat\vartheta\rangle$ is the prediction error, or residual. For $a\ne0$, the following update moves the estimate to the nearest vector consistent with the observed measurement. The superscript $+$ denotes the updated estimate, which is retained for later epochs:
\begin{equation}\label{dev:eq:project}
\widehat\vartheta^+\eqdef\widehat\vartheta+
\frac{y-\langle a,\widehat\vartheta\rangle}{\|a\|_2^2}a.
\end{equation}
For intuition, return to the one-bit example and take the known part to be $G(u)=L(u)=u$ and the feature to be $\phi(u)=u$. The unknown coefficient is $\vartheta=-2$ and the constant is $c_0=1$. Evaluating the change from zero to one gives $a=1$ and $y=-2$: the global change is $-1$ and the known local change is $+1$. Starting from an estimate of zero, the update sets it to $-2$. The corrected local objective then prefers one. This example requires only one measurement because it has one unknown coefficient; larger models need enough informative directions.

\paragraph{Learning the two-block example.}
The task~\eqref{eq:worked-task} has known part and features
\begin{equation}\label{eq:worked-features}
\begin{gathered}
G(x)=x_1+x_2+\tfrac14(x_1-x_2)^2,\quad \phi(x)=(x_1,x_2),\\
\vartheta=(-2,-2),\qquad c_0=2.
\end{gathered}
\end{equation}
The algorithm knows $G$ and $\phi$, but must learn $\vartheta$. Start with $\widehat\vartheta=(0,0)$ at $x=(0,0)$. Evaluating the probe $z=(1,0)$ gives global change $-0.75$ and known-part change $1.25$. Their difference is $y=-2$, while $a=(1,0)$. Update~\eqref{dev:eq:project} therefore sets the first estimated coefficient to $-2$ and leaves the second at zero. The probe is a measurement; it does not itself change the shared solution.

With that estimate, block 1's corrected candidate now has score $-0.75$, so local search can propose it and verification can accept it. After this accepted update, the base is $(1,0)$. A probe at $(1,1)$ gives global change $-1.25$, known-part change $0.75$, and hence the same unknown coefficient value $-2$ for block 2. Its exact correction is now $-2.25$, rather than the value $-1.75$ at the original base. The coefficient is static; the complete correction function is not. Known coupling terms must still be recomputed at the current context.

This two-query sequence illustrates coefficient reuse. It is an explanatory sequence; a general rate needs the excitation assumption below.

The update is the established Kaczmarz/sketch-and-project estimator \citep{strohmer2009,gower2015}. Ordinary verified edits can supply such measurements too, but they need not explore all parameter directions. We therefore permit deliberate calibration probes that are evaluated without being committed.

Measurements must reveal every unknown direction, not repeatedly test the same one. The next assumption expresses this coverage, often called \emph{excitation}. Here $I$ is the identity matrix, $\succeq$ is the positive-semidefinite matrix order, and $\mathcal H_k$ is the history before epoch $k$'s calibration. A larger $\sigma$ means a stronger guaranteed coverage of parameter directions.

\begin{assumption}[Exciting calibration]\label{dev:ass:excitation}
Conditionally on all past information before calibration, the measurement projector $P=aa^\top/\|a\|_2^2$ satisfies
\begin{equation}\label{dev:eq:excitation}
\mathbb E[P\mid\mathcal H_k]\succeq\sigma I,\qquad\sigma\in(0,1/r].
\end{equation}
A skipped probe has $P=0$ and leaves the estimate unchanged; zero-length measurements are skipped. Their probability is included in $\sigma$.
\end{assumption}
For example, uniformly sampling one of the $r$ coordinate directions gives $\mathbb E[P]=I/r$, so $\sigma=1/r$ is valid. Taking traces also shows that a single rank-one probe per epoch cannot give $\sigma>1/r$. Thus the amount and diversity of measurement information matter. This assumption is a property of an implementable probe distribution, not a consequence of successful task edits. The coordinate specialization below instead samples the remaining uncertain coordinates and uses an observable-radius analysis.

\paragraph{From the estimate to private work.}
After calibration, replace the unknown vector in the exact correction by its estimate:
\begin{equation}\label{eq:learnedcorrection}
\begin{gathered}
\widehat R_b(u;x)\eqdef G(x_{-b},u)-G(x)\\
{}-[L_b(u)-L_b(x_b)]
+\langle\widehat\vartheta^+,\phi(x_{-b},u)-\phi(x)\rangle.
\end{gathered}
\end{equation}
Substituting into the surrogate makes the implemented local score explicit:
\begin{equation}\label{dev:eq:implemented-score}
\begin{gathered}
\widehat f_b(u;x)\overset{\eqref{eq:surrogate},\eqref{eq:learnedcorrection}}{=}
G(x_{-b},u)-G(x)\\
{}+\langle\widehat\vartheta^+,\phi(x_{-b},u)-\phi(x)\rangle.
\end{gathered}
\end{equation}
Thus the implemented method is block descent on the learned global surrogate $\widehat F(x)=G(x)+\langle\widehat\vartheta^+,\phi(x)\rangle$. The local objective cancels algebraically; in this representation it supplies a modeling decomposition, not an extra source of optimization power. This exposes the actual computational requirement: after receiving the needed context and coefficients, the agent must evaluate the known terms and features cheaply. The saving comes from this representation and reuse, not from subtracting and adding $L_b$. A coefficient vector whose evaluation is as expensive as the verifier would not solve the cost problem.

Then run the same $H$-step private solver, verify singleton gains, and merge the best color. Write $\widehat\vartheta=\widehat\vartheta^k$ for the estimate before the calibration of epoch $k$ and $\widehat\vartheta^+=\widehat\vartheta^{k+1}$ for the estimate after it. With $\widetilde\vartheta\eqdef\widehat\vartheta-\vartheta$ and $\widetilde\vartheta^+\eqdef\widehat\vartheta^+-\vartheta$, the surrogate error is bounded by
\begin{equation}\label{eq:learnederror}
\sup_u|\widehat R_b(u;x)-R_b(u;x)|
\overset{\eqref{eq:correction},\eqref{dev:eq:features},\eqref{dev:eq:featurebound},\eqref{eq:learnedcorrection}}{\leq}
\Lambda_b\|\widetilde\vartheta^+\|_2.
\end{equation}
The unknown error norm is only used in analysis. Section~\ref{sec:certificate} supplies an observable certificate for coordinate residuals.

\begin{lemma}[Learning contracts a persistent error]\label{lem:learning}
Under Assumptions~\ref{dev:ass:features} and~\ref{dev:ass:excitation}, let $\eta\eqdef\sqrt{1-\sigma}<1$. Then
\begin{equation}\label{eq:learncontract}
\mathbb E[\|\widetilde\vartheta^+\|_2\mid\mathcal H_k]\leq\eta\|\widetilde\vartheta\|_2.
\end{equation}
\end{lemma}
\paragraph{Proof idea.} The update removes precisely the component of coefficient error along the measured direction. The excitation assumption ensures that, on average, every direction loses a definite fraction of its squared error.
\phantomsection\label{proof:lem-learning}
\begin{proof}
An active update is an orthogonal projection of the error:
\begin{equation}\label{eq:pythagoras}
\begin{gathered}
\widetilde\vartheta^+\overset{\eqref{dev:eq:measurement},\eqref{dev:eq:project}}{=}(I-P)\widetilde\vartheta,\\
\|\widetilde\vartheta^+\|_2^2\overset{\eqref{dev:eq:measurement},\eqref{dev:eq:project}}{=}\|\widetilde\vartheta\|_2^2-\widetilde\vartheta^\top P\widetilde\vartheta.
\end{gathered}
\end{equation}
The same identities hold for a skipped update with $P=0$. Conditioning and applying the excitation bound gives
\begin{equation}\label{eq:squarecontract}
\mathbb E[\|\widetilde\vartheta^+\|_2^2\mid\mathcal H_k]
\overset{\eqref{eq:pythagoras},\eqref{dev:eq:excitation}}{\leq}(1-\sigma)\|\widetilde\vartheta\|_2^2.
\end{equation}
Jensen's inequality for the square root proves~\eqref{eq:learncontract}. Since $\vartheta$ is fixed, committing a new solution does not change the parameter error. This persistence is the crucial benefit of the representation.
\end{proof}

We can now track both processes together. Searching with an inaccurate correction leaves an additional task-error term proportional to the parameter error; calibration reduces that parameter error. Weighting the two errors appropriately gives one quantity that decreases geometrically. This quantity is an analysis tool: it contains the unknown optimum and unknown true weights, so the algorithm cannot use it as a stopping test.

\paragraph{What is coupled to what?}
The potential $\mathcal V_k$ below adds $\omega$ times the coefficient error $\|\widehat\vartheta^k-\vartheta\|_2$ to the task gap. Calibration reduces its conditional expectation by a factor $\eta<1$. Meanwhile, optimization reduces task error by a factor $1-\alpha$, but adds a penalty proportional to the coefficient error remaining after calibration. The proof combines these two statements. The coefficient $\beta$ converts parameter error into objective-error units; the weight $\omega$ makes their weighted sum decrease. A \emph{potential} is simply this weighted quantity used to track the analysis, not an additional objective that an agent minimizes.

\begin{theorem}[A joint task--correction contraction]\label{dev:thm:joint}
Assume known graph structure, Assumption~\ref{ass:local} for every learned surrogate, Assumption~\ref{ass:eb}, and Assumptions~\ref{dev:ass:features}--\ref{dev:ass:excitation}. Set
\begin{equation}\label{dev:eq:jointconstants}
\begin{gathered}
\alpha\eqdef\mu q(H)/\kappa,\quad
\beta\eqdef\kappa^{-1}\sum_b(1+q_b(H))\Lambda_b>0,\\
\omega\eqdef2\beta/(1-\eta),\quad
\tau\eqdef\min\{\alpha,(1-\eta)/2\},\\
\mathcal V_k\eqdef\Delta(x^k)+\omega\|\widehat\vartheta^k-\vartheta\|_2.
\end{gathered}
\end{equation}
With deterministic initialization,
\begin{equation}\label{dev:eq:joint}
\mathbb E\mathcal V_K\leq(1-\tau)^K\mathcal V_0.
\end{equation}
Thus solution error and correction error vanish together; no externally supplied shrinking-error schedule is needed.
\end{theorem}
\paragraph{Proof idea.} Optimization contracts the task error but pays a penalty proportional to the remaining coefficient error. Calibration contracts that coefficient error. The weight $\omega$ is chosen so that the latter decrease pays for the former penalty.
\phantomsection\label{proof:thm-joint}
\begin{proof}
Condition first on the calibration outcome. The drift bound~\eqref{eq:drift} applies with the current errors bounded by~\eqref{eq:learnederror}. Averaging over calibration yields
\begin{equation}\label{eq:coupled}
\begin{gathered}
\mathbb E[\Delta(x^{k+1})\mid\mathcal H_k]\\
\overset{\eqref{eq:drift},\eqref{eq:learnederror},\eqref{dev:eq:jointconstants}}{\leq}
(1-\alpha)\Delta(x^k)+\beta\mathbb E[\|\widetilde\vartheta^+\|_2\mid\mathcal H_k].
\end{gathered}
\end{equation}
Adding the weighted learning error gives
\begin{equation}\label{eq:jointstep}
\begin{gathered}
\mathbb E[\mathcal V_{k+1}\mid\mathcal H_k]\\
\begin{array}{c@{\;}l}
\overset{\eqref{eq:coupled},\eqref{eq:learncontract}}{\leq}&
(1-\alpha)\Delta(x^k)+(\beta+\omega)\eta\|\widetilde\vartheta\|_2\\
\overset{\eqref{dev:eq:jointconstants}}{\leq}&(1-\tau)\mathcal V_k.
\end{array}
\end{gathered}
\end{equation}
For the second step, $\tau\leq\alpha$ and
$(\beta+\omega)\eta\leq\omega[1-(1-\eta)/2]$; the latter follows because the difference equals $\beta(1-\eta)\geq0$. Taking expectations and induction prove~\eqref{dev:eq:joint}.
\end{proof}

In this theorem, Lemma~\ref{lem:local} is applied epoch by epoch, after conditioning on the calibration result. Its error bound may therefore change from epoch to epoch; no fixed nonzero error floor is imposed. The rate $\tau$ is controlled by the slower of two processes: optimizing the solution with the current correction, and identifying the correction itself. The method retains an estimate of fixed coefficients, not a permanently valid table of candidate values: after $x$ changes, known features and coupling terms must still be evaluated at the new context. This is the precise sense in which the correction is reusable.

\paragraph{Can already-paid verification teach the correction?}
Yes. Every proposal verified in line~6 of Algorithm~\ref{alg:protocol}, including a rejected one, supplies a measurement at the common base. With $z_b=(x_{-b},U_b)$, define
\begin{equation}\label{eq:proposal-measurement}
\begin{gathered}
a_b\eqdef\phi(z_b)-\phi(x),\\
y_b\eqdef F(z_b)-F(x)-[G(z_b)-G(x)]
\overset{\eqref{dev:eq:features}}{=}\langle a_b,\vartheta\rangle.
\end{gathered}
\end{equation}
One can apply~\eqref{dev:eq:project} to these measurements after all proposals and merge decisions have been fixed, retaining the resulting estimate for the next epoch. Each equation must use the base at which its two objective values were measured. Reuse needs no additional verifier call; feature evaluation, projection work, and communicating the updated estimate still cost work.

\begin{proposition}[Safe reuse of proposal measurements]\label{prop:reuse}
Under the hypotheses of Theorem~\ref{dev:thm:joint}, keep its deliberate calibration and add any finite sequence of exact projection updates from the already-evaluated proposals, after the current search and merge decision. The bound~\eqref{dev:eq:joint} remains valid for the estimate after these extra updates. No excitation or independence assumption is needed for the extra measurements.
\end{proposition}
\paragraph{Proof idea.} Extra projections cannot increase parameter error. They do not change the proposals used to prove this epoch's task-error bound.
\begin{proof}
Let $\widetilde\vartheta^+$ be the error just after deliberate calibration, and let $\widetilde\vartheta^{\rm next}$ be its value after all proposal measurements have been processed. Every nonzero measurement gives an orthogonal projection; a zero measurement is skipped. Repeated use of the pathwise identity~\eqref{eq:pythagoras} gives
\begin{equation}\label{eq:reuse-norm}
\|\widetilde\vartheta^{\rm next}\|_2\overset{\eqref{eq:pythagoras}}{\leq}\|\widetilde\vartheta^+\|_2.
\end{equation}
The task update still obeys~\eqref{eq:coupled}. Consequently
\begin{equation*}
\begin{gathered}
\mathbb E[\mathcal V_{k+1}\mid\mathcal H_k]\\
\begin{array}{c@{\;}l}
\overset{\eqref{eq:coupled},\eqref{eq:reuse-norm},\eqref{eq:learncontract}}{\leq}&
(1-\alpha)\Delta(x^k)+(\beta+\omega)\eta\|\widetilde\vartheta\|_2\\
\overset{\eqref{dev:eq:jointconstants}}{\leq}&(1-\tau)\mathcal V_k.
\end{array}
\end{gathered}
\end{equation*}
Induction proves the claim. The next epoch's hypotheses must continue to hold for the augmented history, as required by their uniform conditional formulation.
\end{proof}
This extension justifies learning from verification, but does not justify removing deliberate calibration: proposals can all have zero measurement vectors or fail to explore an unknown direction.

The joint theorem shows that learning and optimization can converge together. Yet its estimation error involves the unknown true coefficients, so the algorithm cannot read that error from its current state. Nor does a rate in epochs alone reveal the price of the measurements. A more concrete specialization lets us address both issues.

\subsection{Certifying and paying for learning}\label{sec:certificate}
For binary blocks with unknown individual-bit penalties, each targeted objective evaluation identifies one coefficient exactly. This gives both a computable bound on the remaining correction error and a finite total number of learning queries. We now derive those guarantees. Unlike the preceding parameter-error norm, the remaining uncertainty can be tracked without knowing the true weights. For this specialization, assume $\cX_b=\{0,1\}^{d_b}$ with positive integer $d_b$. Thus every one-bit calibration probe is feasible; a constrained subset of the binary cube need not have this property. Let $\Psi$ be a known function describing interactions between blocks. Assume all remaining unknown requirements are linear penalties on individual bits:
\begin{equation}\label{dev:eq:coordinate-model}
F(x)=\sum_bL_b(x_b)+\Psi(x)+\sum_b\langle\vartheta_b,x_b\rangle+c_0.
\end{equation}
The unknown $c_0$ cancels in differences. Assume known coordinate bounds $|\vartheta_{bj}-\widehat\vartheta^0_{bj}|\leq M_{bj}$. A probe flips bit $j$ of block $b$ at the current base, producing $z$ and $s=z_{bj}-x_{bj}\in\{-1,1\}$. Its residual is
\begin{equation}\label{eq:bitprobe}
\begin{gathered}
y\eqdef F(z)-F(x)-[L_b(z_b)-L_b(x_b)]\\
{}-[\Psi(z)-\Psi(x)]
\overset{\eqref{dev:eq:coordinate-model}}{=}s\vartheta_{bj},\\
\widehat\vartheta_{bj}^+
\eqdef\widehat\vartheta_{bj}+(y-s\widehat\vartheta_{bj})/s.
\end{gathered}
\end{equation}
This identifies one coefficient exactly from one extra scalar query when $F(x)$ is cached. Keep it forever. Conditionally on the current history, each block with unobserved coefficients probes with probability $p\in(0,1]$, choosing uniformly among its unobserved coordinates. Independence between blocks is not required. Probes are not committed.

Let $\mathcal U_b^k$ be the remaining unobserved coordinates before epoch $k$. Until identified, each coefficient estimate is kept at its initial value and retains its error bound $M_{bj}$, while a measured coefficient has zero error. Summing these remaining bounds gives a computable uncertainty radius:
\begin{equation}\label{eq:radius}
\Gamma_{b,k}\eqdef\sum_{j\in\mathcal U_b^k}M_{bj},\qquad\Gamma_k\eqdef\sum_b\Gamma_{b,k},\qquad d_{\max}\eqdef\max_b d_b.
\end{equation}
This sampling rule stops probing already identified coefficients, so it need not satisfy uniform excitation over all original parameter directions. Its analysis instead uses the uncertainty remaining in $\Gamma_k$. Known coupling terms are recomputed at the current context; only the static unknown coefficients are retained. The anchored correction error for block $b$ is at most $\Gamma_{b,k+1}$, its radius after this epoch's probes, because each binary coordinate displacement has absolute value at most one.

\begin{proposition}[Observable uncertainty and bounded calibration]\label{prop:radius}
Write $\gamma\eqdef p/d_{\max}$. The above calibration satisfies
\begin{equation}\label{eq:radiuscontract}
\mathbb E[\Gamma_{k+1}\mid\mathcal H_k]\leq(1-\gamma)\Gamma_k.
\end{equation}
There are at most $d_{\rm tot}\eqdef\sum_b d_b$ additional calibration queries over the entire run. After $K$ epochs their expected number is at most $\min\{d_{\rm tot},mpK\}$.
\end{proposition}
\paragraph{Proof idea.} Each successful probe removes one previously unmeasured coefficient from the uncertainty sum. Bounding the expected removed uncertainty gives contraction; counting distinct coefficients gives the finite query budget.
\phantomsection\label{proof:prop-radius}
\begin{proof}
If a block has $n_b>0$ unobserved coefficients and remaining radius $\Gamma_{b,k}$, its expected removed radius is $p\Gamma_{b,k}/n_b\geq p\Gamma_{b,k}/d_{\max}$. Empty sets contribute zero. Summing gives~\eqref{eq:radiuscontract}; solution updates do not change the learned static parameters or their uncertainty. Every calibration identifies a previously unobserved coefficient, so its count is at most $d_{\rm tot}$. At most $mp$ queries occur in expectation per epoch, proving both bounds.
\end{proof}

\begin{corollary}[Joint contraction with observable correction uncertainty]\label{cor:observable}
Assume the full binary domains, coordinate model~\eqref{dev:eq:coordinate-model}, known initial coefficient bounds, and exact calibration protocol above. Assume a valid dependency-graph coloring, Assumptions~\ref{ass:local} and~\ref{ass:eb} for the resulting surrogates and iterates, and deterministic initialization. Define
\begin{equation}\label{eq:observableconstants}
\begin{gathered}
\beta_0\eqdef\kappa^{-1}\max_b(1+q_b(H)),\quad
\omega_0\eqdef2\beta_0/\gamma,\\
\tau_0\eqdef\min\{\mu q(H)/\kappa,\gamma/2\},\quad
\mathcal W_k\eqdef\Delta(x^k)+\omega_0\Gamma_k.
\end{gathered}
\end{equation}
Then $\mathbb E\mathcal W_K\leq(1-\tau_0)^K\mathcal W_0$ for every integer $K\geq0$. The radius $\Gamma_k$ is observable, although $\Delta(x^k)$ need not be. In particular, $\Gamma_k=0$ certifies that the unknown coefficients have been learned, not that the current solution is globally optimal.
\end{corollary}
\paragraph{Proof idea.} Repeat the joint-potential argument with the computable uncertainty radius replacing the unknown coefficient-error norm. The same weighting principle turns two coupled inequalities into one contraction.
\phantomsection\label{proof:cor-observable}
\begin{proof}
Set $\alpha=\mu q(H)/\kappa$. Here $\Gamma_{k+1}$ is the radius after this epoch's deliberate probes; the ensuing solution update does not change it. Conditional on those probes, each block's correction error is at most its remaining coordinate-radius sum. The gain and merge bounds therefore give
\begin{equation}\label{eq:radius-gap}
\begin{gathered}
\mathbb E[\Delta(x^{k+1})\mid\mathcal H_k]\\
\overset{\eqref{dev:eq:localgain},\eqref{eq:merge},\eqref{eq:eb},\eqref{eq:observableconstants}}{\leq}
(1-\alpha)\Delta(x^k)+\beta_0\mathbb E[\Gamma_{k+1}\mid\mathcal H_k].
\end{gathered}
\end{equation}
Adding the weighted radius yields
\begin{equation}\label{eq:radius-joint}
\begin{gathered}
\mathbb E[\mathcal W_{k+1}\mid\mathcal H_k]\\
\begin{array}{c@{\;}l}
\overset{\eqref{eq:radius-gap},\eqref{eq:radiuscontract},\eqref{eq:observableconstants}}{\leq}&
(1-\alpha)\Delta(x^k)\\
&{}+(\beta_0+\omega_0)(1-\gamma)\Gamma_k\\
\overset{\eqref{eq:observableconstants}}{\leq}&(1-\tau_0)\mathcal W_k.
\end{array}
\end{gathered}
\end{equation}
For the last step, $\tau_0\leq\alpha,\gamma/2$ and the coefficient margin is
\begin{equation*}
\begin{gathered}
\omega_0(1-\gamma/2)-(\beta_0+\omega_0)(1-\gamma)\\
\overset{\eqref{eq:observableconstants}}{=}\beta_0\gamma\geq0.
\end{gathered}
\end{equation*}
Total expectation and induction establish the stated bound.
\end{proof}

\paragraph{Reusing verification while preserving the radius.}
For a binary proposal, subtract the contributions of all already identified coefficients from its residual. If its displacement contains exactly one still-unknown coefficient, the remaining residual identifies that coefficient by division by its nonzero bit displacement. Mark it identified, just as after a deliberate bit probe. Doing this after proposal generation only removes extra nonnegative terms from the final radius. In the proof, retain the radius just after deliberate probing for the task-error bound, and bound the final radius by that intermediate radius in the potential. The same coefficient inequality proves Corollary~\ref{cor:observable}; Proposition~\ref{prop:radius} also remains valid. Local contraction must hold for the augmented histories. A general multi-coordinate projection need not preserve the individual bounds $M_{bj}$; it belongs to Proposition~\ref{prop:reuse}'s norm analysis instead.

\subsubsection{The cost of acquiring the correction}
Let a scalar verifier call cost $v$, a parallel private slot cost $c$, and other per-epoch overhead cost $C$. Here $C$ excludes verifier calls, which are charged separately through $v$; in the exact-correction model of Section~\ref{sec:horizon}, they were included in its aggregate overhead. All terms below use the same modeled cost units. Evaluating each agent's proposal separately costs at most $m$ calls; color additivity supplies the new base value without an extra union query. The bound below is for Algorithm~\ref{alg:protocol}; the exact best-combination merge adds up to $2^m-m-1$ calls per epoch, and with pairwise factors $|I^k|(|I^k|-1)/2$ calls find the best combination of improving blocks (Appendix~\ref{app:routing-merge-theory}). For a prescribed $K$,
\begin{equation}\label{dev:eq:learncost}
\mathbb E T_K\leq K(C+cH)+v[1+mK+\min\{d_{\rm tot},mpK\}].
\end{equation}
The four terms count initial evaluation, local computation and synchronization, proposal evaluations, and calibration queries. The initial base query is charged once. Feature evaluation, coupling-context refresh, and correction-message transfer belong in $C$ or $c$; they are not free. A sufficient horizon for $\mathbb E\Delta\leq\varepsilon$ is
\begin{equation*}
K=\lceil\tau_0^{-1}\log(\mathcal W_0/\varepsilon)\rceil,
\qquad 0<\varepsilon<\mathcal W_0.
\end{equation*}
This is a fixed-horizon guarantee. Since $\tau_0\leq p/(2d_{\max})$, larger $H$ cannot speed it up beyond the calibration rate; with $p=1$ all coefficients are identified within $d_{\max}$ epochs, after which Section~\ref{sec:horizon} applies.

\subsubsection{Why arbitrary objectives need structure}\label{dev:sec:structure}
Without structure, evaluations at previously visited candidates need not reveal anything about an unvisited one.
On an unrestricted finite candidate set, suppose some candidate $u$ has never been queried. Two objectives can agree on every observed value and differ arbitrarily at $u$. They generate the same transcript and correction estimate, but its error at $u$ cannot be bounded uniformly for both objectives. If values are known only to lie in an interval of width $M$, the same argument precludes a guaranteed radius below $M/2$ there. Adaptive queries do not change this indistinguishability argument. To infer unmeasured candidate values, the method needs a low-dimensional identifiable model, another regularity assumption, or enough queries to cover the candidates. Sparse evaluations alone cannot provide a uniformly accurate correction for an arbitrary objective.

\section{Detailed related work}\label{app:related}

This appendix complements the related work of the main paper. It discusses corrected local computation, distributed optimization, surrogate learning, the organization of LLM collaboration and LLM planning for several robots.

\paragraph{Relation to local training.}
ProxSkip uses a persistent control variate to counter the drift caused by skipping an expensive proximal operation \citep{mishchenko2022}. SCAFFOLD uses control variates to address client drift in federated learning \citep{karimireddy2020}. We borrow the question these mechanisms answer: what makes private steps point toward the shared objective? Here the answer is a correction to candidate values, not a gradient vector or an averaging rule. Agents own disjoint blocks, external context is frozen within an epoch, and synchronization combines accepted block replacements instead of averaging full models. Our theorem is therefore a block-descent result; whether ProxSkip-type acceleration carries over remains open.

Table~\ref{tab:proxskip-comparison} makes the distinction explicit. Appendix~\ref{sec:learn} supplies a persistent correction learned from objective-evaluation residuals and a joint error analysis for a restricted identifiable model. Appendix~\ref{sec:certificate} makes its uncertainty observable and charges calibration explicitly.

\begin{table*}[htbp]
\caption{Relation to ProxSkip \citep{mishchenko2022}. Our correction can be learned from verifier feedback; its update and rate differ from ProxSkip's.}
\label{tab:proxskip-comparison}
\centering
\begin{tabular}{@{}p{0.18\textwidth}p{0.37\textwidth}p{0.39\textwidth}@{}}
\toprule
 & ProxSkip & Present construction\\
\midrule
Correction & Updated recursively through occasional proximal steps & Learned persistently from scalar verifier residuals in an identifiable model\\[4pt]
Private work & Corrected gradient steps & Search within an assigned artifact block\\[4pt]
Analysis & Joint iterate--correction error & Joint task--correction contraction; observable uncertainty for coordinate probes\\[4pt]
Communication benefit & Accelerated communication under its optimization assumptions & Batching and a finite calibration budget; no ProxSkip acceleration claim\\
\bottomrule
\end{tabular}
\end{table*}

\paragraph{Classical optimization and distributed local search.}
Parallel numerical iteration and its communication costs have a long history \citep{bertsekas1989}. Inexact coordinate descent explicitly permits approximate block solves \citep{tappenden2016}; our local contraction is an assumption on such a solve, not a new solver. Theorem~\ref{thm:rate} is an inexact block-descent argument under a quantitative local-to-global condition. Its finite domains and candidate-value errors differ from the smooth convex setting of those coordinate-descent analyses, but the descent principle is shared.

MGM chooses improving agents through neighbor gain comparisons, excluding simultaneous conflicting moves; DSA randomizes local updates \citep{maheswaran2004,zhang2005}. Our coordinator evaluates all proposals at a common base and accepts either the improving proposals of the best color (Algorithm~\ref{alg:protocol}) or the best verified combination of proposals (Algorithm~\ref{alg:best-combination}). The monotonicity principle is closely related, while the selection rule and oracle interface differ. Work on $k$-optimal DCOP solutions enlarges the coordinated neighborhood and studies its solution guarantees \citep{bowring2008}; this is directly relevant when our single-block error bound fails. Assigning every agent utility $-F$ also gives an exact potential game with potential $-F$ \citep{monderer1996}. Unilateral improvement then reaches a coordinatewise equilibrium; global optimality still needs the additional geometry in Proposition~\ref{prop:geometry}.

Expected separable overapproximations in parallel coordinate descent handle interference through randomized update models and upper bounds \citep{richtarik2016}. Our exact-additivity coloring argument is simpler and more conservative. The best-combination merge replaces $1/\kappa$ by the fraction of the verified gain it actually keeps, which is at least one over the number of conflict-free groups for pairwise factors (Section~\ref{sec:main-combination}).

\paragraph{Surrogate learning and measurements.}
Additive Bayesian optimization uses structural decompositions to make expensive objective optimization tractable \citep{kandasamy2015}. Linear-bandit best-arm identification studies which measurements reveal enough about a shared unknown parameter to select a good action \citep{soare2014}. Our representation is a noiseless, known-feature surrogate with deterministic candidate-error bounds and block descent. Learning an objective from scalar evaluations is not new; we add a bound that converts the remaining correction error into lost progress and an accounting of measurement cost (Section~\ref{sec:main-cost}).

\paragraph{Established ingredients and proposed combination.}
Block optimization supplies the local-to-global framework \citep{tseng2001,nesterov2012,richtarik2016}. Distributed constraint optimization studies variables owned by agents, with costs that couple their choices. Its DPOP algorithm communicates tables summarizing the best achievable cost for specified neighboring choices, exposing the importance of message size \citep{petcu2005}. The contribution of this paper is the explicit combination of a candidate-value correction, its approximation error, a verified merge, and a communication-horizon calculation for an agent-artifact model.

\subsection{AI-agent collaboration}

The following work studies how LLM agents divide tasks, communicate and use feedback. We compare these mechanisms with the objective correction and convergence analysis developed here.

\paragraph{Shared artifacts and dialogue.}
MetaGPT and ChatDev organize specialized agents around software-development workflows and intermediate artifacts \citep{hong2024,qian2024}. AutoGen builds applications by connecting agents that exchange messages, while AgentVerse studies collaborative agent groups \citep{wu2024,chen2024}. Multiagent debate improves answers through exchanged critiques \citep{du2024}. These works motivate specialization and information exchange. Our model instead makes a particular optimization interface explicit: owned solution blocks, candidate scores, and an exact scalar task verifier.

\paragraph{Communication topology and cost.}
In a communication graph, nodes represent agents or operations and edges specify who sends information to whom. GPTSwarm optimizes such language-agent graphs; DyLAN adapts the participating agent network; G-Designer selects communication patterns for a given task \citep{zhuge2024,liu2024,gdesigner2025}. AgentPrune removes redundant message-passing edges, AgentDropout eliminates agents dynamically, and DALA allocates opportunities to speak through auctions \citep{agentprune2025,agentdropout2025,fan2026}. Thus, optimizing communication expenditure is an established research direction. Here the question is how omitted global requirements distort private search, and which persistent correction permits useful work between synchronizations. Our dependency graph describes objective factors and compatible edits; it is not a learned message-routing graph.

\paragraph{Feedback and evolving collaboration.}
EvoMAC uses verification-derived textual feedback to evolve agents and their connections \citep{hu2025}. AFlow searches over workflows---programs that specify the sequence and connections of model calls---using feedback from executing them \citep{aflow2025}, and evolving orchestration adapts how agents are selected and coordinated \citep{dang2025}. Test-time communication for discovery is another closely related setting \citep{park2026}. These works already connect feedback, accumulated experience, and collaboration. Our narrower contribution is to estimate the missing part of a numerical objective, bound the estimation error, and prove joint convergence of the solution and the estimate when the unknown coefficients can be identified.

\paragraph{LLM planning for several robots.}
In RoCo, one LLM agent per robot negotiates a joint plan through dialogue and re-plans from motion-planner feedback such as collisions \citep{mandi2024roco}. SMART-LLM decomposes instructions and assigns the subtasks to robots \citep{kannan2024smart}; centralized, decentralized and hybrid LLM planners have been compared as teams grow \citep{chen2024scalable}; and LLMs still fail at multi-agent path finding on maps with obstacles \citep{chen2024mapfllm}. Our routing agents also receive collision feedback, but as a numerical correction to their own objective, computed for the robot pairs whose collisions the exact verifier has reported, from those robots' current routes, and only verified combinations of proposals are accepted.

\paragraph{Calibration, reliability, and scaling.}
CommCP calibrates LLM-generated messages through conformal prediction for robots cooperating to answer questions about their physical surroundings \citep{commcp2026}. Its message-reliability objective differs from our correction requirement, which concerns only a true block minimizer and the returned proposal (Assumption~\ref{ass:correction}). Empirical analyses of multi-agent failures and scaling further motivate checking coordination and verification rather than presuming that more agents or messages help \citep{cemri2025,kim2026}. Our assumptions state what such checks must deliver for parallel work to add up to global progress.

\section{Routing: task, verifier and measurement}\label{app:routing-task}\label{app:robot-ninecars-multiseed}
This appendix defines everything the routing experiments share: why we chose the task, the two maps and their two-floor combination, commands, loss, local objective and correction, how feedback reaches the agents, the protocols, the measurement of search time and the exact references that certify optima and block minima. Appendix~\ref{app:routing-development} explains how we chose the prompt and the protocols, trying two versions of the prompt, and Appendix~\ref{app:routing-revised} reports the results of the main text.

\subsection{Which tasks need corrections}\label{app:task-choice}
\paragraph{Three regimes.}
A team of agents is useful when a task is too large or too hard for one agent, and the parts can be searched in parallel. Whether the team also needs coordination depends on how the parts interact. A simple task, such as changing a light bulb, gains nothing from a team: splitting it only adds overhead. A task whose parts are independent, such as one agent watering a garden while another goes shopping, can be split without coordination; each agent optimizes its own part, and the team finishes sooner. In the language of Section~\ref{sec:main-setup}, such a task is separable: if $F(x)=\sum_bL_b(x_b)$ up to a constant, then the exact correction $R_b$ is zero, the dependency graph has no edges, a single color contains every block, and all improving proposals can be merged in one epoch. Theorem~\ref{thm:rate} then holds with $\delta_b^*=\delta_b^U=0$ and $\kappa=1$, but it says nothing that independent parallel search does not already achieve.

The case that motivates this paper lies between these extremes: tasks whose parts are coupled. Examples include the modules of a large program, the structure, wiring and plumbing of a building, the subsystems of an aircraft, and the routes of robots that share a floor. Each part is manageable alone, but changing it alters what is best for its neighbors, so independent local improvements need not combine into a better whole. Here corrections tell each agent how its candidates affect the others, and coloring decides which improvements can be applied together. The strength of the coupling, not the number of agents, determines how much these mechanisms matter.

\paragraph{Why a routing task.}
We wanted an experiment with few agents in which this coupling is strong, controlled and measurable. Multi-robot routing meets these needs. Its rules are simple: five commands, walls and a fixed horizon, and a reader can check a plan by eye. It is nevertheless hard for current LLMs, which must track coordinates, obstacles and timing; LLMs still fail at multi-agent path finding on maps with obstacles \citep{chen2024mapfllm}, and in our runs with prompt v1 a single model writing all nine routes reached a safe plan in only one run of ten (Appendix~\ref{app:routing-development}). Most importantly, coupling is built into the task: robots owned by different agents can collide, so an agent's best plan for its own robots can harm another agent's robots. The exact verifier measures this coupling directly; in the team runs of the main text, 30\% of the proposals that delivered all three of an agent's robots without internal collisions still collided with another agent's robot on the first map, and 32\% on the second (Appendix~\ref{app:routing-assumptions}). A certified optimum and exact block minima also let us measure every constant of the theory.

\paragraph{Why not software.}
Code generation was the first application we considered, and the introduction uses it as an example. It is harder to use for a controlled study of coupling. Good software is designed to limit coupling between modules \citep{parnas1972}, so small coding tasks rarely exhibit strong interactions between the parts that different agents would own; those interactions mainly appear in large systems. A coding benchmark would also need test suites whose failures can be attributed to particular interactions, and a numerical objective whose optimum is known. Building such a benchmark at a scale where coupling is strong and measurable is a natural next step.

\subsection{Maps, robots and the two-floor task}
\begin{figure*}[t]
\centering
\includegraphics[width=.9\textwidth]{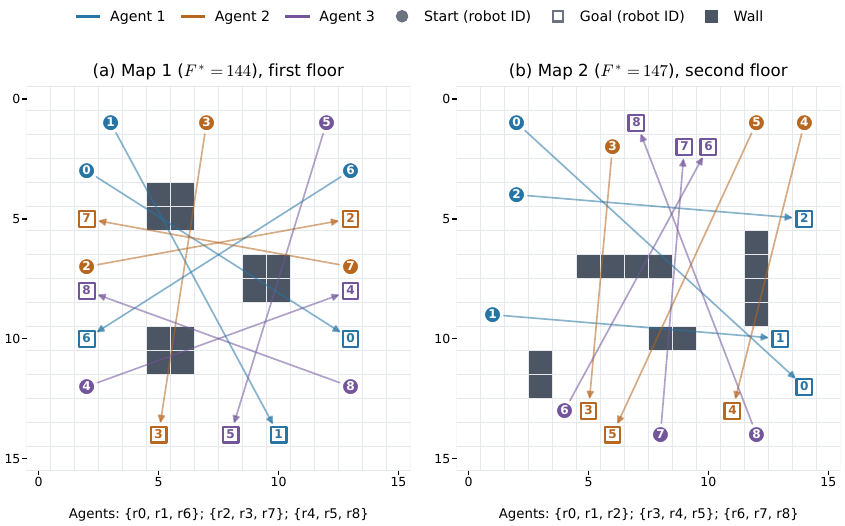}
\Description{Two sixteen-by-sixteen grids side by side. Left, map 1 with three two-by-two obstacles; right, map 2 with a horizontal wall of four cells, a vertical wall of four cells and two pieces of two cells. On each, nine robots have numbered circles at their starts and numbered squares at their goals, colored by the agent that owns them, and a thin arrow points from each start to its goal.}
\caption{The two maps. Circles: starts; squares: goals; numbers: robot IDs; colors: the team agent that owns the robot (the single agent controls all of them); arrows point from start to goal; dark cells are walls. On two floors, map 1 is the first floor and map 2 the second, with its robots named s0--s8.}
\label{fig:routing-maps}
\end{figure*}

The task is discrete multi-agent pathfinding with vertex and edge-swap conflicts and a stay-at-goal convention \citep{stern2019mapf}. Figure~\ref{fig:routing-maps} and Table~\ref{tab:routing-assignments} specify both instances. Coordinates are $(x,y)\in\{0,\ldots,15\}^2$, with $x$ increasing rightward and $y$ downward. On the first map, three $2\times2$ obstacles occupy $\{5,6\}\times\{4,5\}$, $\{9,10\}\times\{7,8\}$ and $\{5,6\}\times\{10,11\}$. On the second map, the twelve wall cells form a horizontal wall of four cells, a vertical wall of four cells and two pieces of two cells. Every other cell is traversable. The simulation horizon is 48 ticks.

A model writes one JSON object with a \texttt{programs} field containing a literal command array for every robot it controls. Allowed commands are \texttt{up}, \texttt{down}, \texttt{left}, \texttt{right}, and \texttt{wait}; each consumes one tick and changes position by $(0,-1)$, $(0,1)$, $(-1,0)$, $(1,0)$, or $(0,0)$, respectively. The model constructs the arrays rather than selecting a precomputed route. Arrays have at most 48 commands. After an array ends, the robot waits in its last cell through tick 48. Upon its first arrival at its goal, the robot must remain there. Every evaluation restarts all robots at their original starts at tick zero; optimization iterations do not advance a physical journey.

\begin{table*}[t]
\centering\footnotesize
\begin{tabular}{@{}lrrrr lrrrr@{}}\toprule
\multicolumn{5}{c}{Map 1 (first floor)} & \multicolumn{5}{c}{Map 2 (second floor, s0--s8 on two floors)}\\
\cmidrule(r){1-5}\cmidrule(l){6-10}
Robot & Agent & Start & Goal & Dist. & Robot & Agent & Start & Goal & Dist.\\\midrule
r0 & 1 & $(2,3)$ & $(13,10)$ & 18 & r0 & 1 & $(2,1)$ & $(14,12)$ & 23\\
r1 & 1 & $(3,1)$ & $(10,14)$ & 20 & r1 & 1 & $(1,9)$ & $(13,10)$ & 13\\
r2 & 2 & $(2,7)$ & $(13,5)$ & 13 & r2 & 1 & $(2,4)$ & $(14,5)$ & 13\\
r3 & 2 & $(7,1)$ & $(5,14)$ & 15 & r3 & 2 & $(6,2)$ & $(5,13)$ & 14\\
r4 & 3 & $(2,12)$ & $(13,8)$ & 15 & r4 & 2 & $(14,1)$ & $(11,13)$ & 15\\
r5 & 3 & $(12,1)$ & $(8,14)$ & 17 & r5 & 2 & $(12,1)$ & $(6,14)$ & 19\\
r6 & 1 & $(13,3)$ & $(2,10)$ & 18 & r6 & 3 & $(4,13)$ & $(10,2)$ & 17\\
r7 & 2 & $(13,7)$ & $(2,5)$ & 13 & r7 & 3 & $(8,14)$ & $(9,2)$ & 15\\
r8 & 3 & $(13,12)$ & $(2,8)$ & 15 & r8 & 3 & $(12,14)$ & $(7,1)$ & 18\\
\midrule Sum &&&& 144 & Sum &&&& 147\\\bottomrule
\end{tabular}

\caption{Assignments and shortest legal grid distances on both maps. On the first map the blocks are $\{r_0,r_1,r_6\}$, $\{r_2,r_3,r_7\}$ and $\{r_4,r_5,r_8\}$; on the second, $\{r_0,r_1,r_2\}$, $\{r_3,r_4,r_5\}$ and $\{r_6,r_7,r_8\}$. The sums are the optima.}
\label{tab:routing-assignments}
\end{table*}

The second map has the same size, horizon, loss and number of wall cells as the first. Starts and goals were drawn at random, and the map was kept because its statistics match the first map's: the sum of shortest distances is 147 against 144, and letting every robot take a shortest path produces five collisions against four. On it, the first agent's three robots move from left to right, the second's from top to bottom and the third's from bottom to top. The two-floor task places both maps on the two floors of one building: eighteen robots, $F=F_1+F_2$ with optimum $291$ and all-waiting value 2955; robots on different floors never collide, and a plan meets the threshold when both floors are safe and within 9 of their optima.

\subsection{Global loss, local loss and correction}
The loss, the local objective and the correction are~\eqref{eq:routing-objective}--\eqref{eq:routing-correction-main} of Section~\ref{sec:main-routing}; this subsection gives their exact conventions. Let $p_i(t)$ be robot $i$'s position at tick $t$, and let $\tau_i\leq48$ denote its first arrival at its goal $g_i$ if it arrives. For a robot that does not arrive, $d_i$ is the shortest four-neighbor grid distance from $p_i(48)$ to $g_i$, respecting walls. Arrival time, including waits, costs one unit per tick. The horizon term applies to every undelivered robot even when its submitted array is short.

For each pair $i<j$, $C_{ij}$ counts a vertex conflict whenever $p_i(t)=p_j(t)$, and an edge-swap conflict whenever the robots exchange distinct neighboring cells during the same tick. Conflicts are checked over ticks zero through 48; repeated co-location contributes one event per pair and tick. Geometric path intersection at different ticks is safe. The collision weight is $100$ for both conflict types. A conflict adds a penalty but does not stop or redirect either submitted trajectory. Walls, in contrast, are hard constraints. They restrict only each robot's own route, so they define the block domains $\cX_b$ rather than coupling blocks; collisions between robots are penalized instead, because they are the coupling under study. Empty arrays initialize the search with all robots stationary: $F(x^0)=9(100+48)+144=1476$ on the first map and $1479$ on the second. The sums of individual shortest-path distances are 144 and 147. Separately computed collision-free witnesses attain these sums, certifying the global optima $F^*=144$ and $F^*=147$. Appendix~\ref{app:routing-references} describes the reference algorithm and its certificates.

The local loss~\eqref{eq:routing-local-main} of a team block counts its three robots' delivery costs and all conflicts among them, and the correction~\eqref{eq:routing-correction-main} covers the external pairs in $\mathcal P^k$, those discovered before global round $k$, with the other blocks frozen at $x^k$. The checked private score change is $L_b(u)-L_b(x_b^k)+\widehat R_b(u;x^k)$. A discovered pair remains known, but the peer's time-indexed positions are refreshed from the current accepted plan at the next round. If every external pair is known, this score change equals the conditional global loss change; before then it can miss conflicts. The experiment learns which pairs to check; the penalty coefficients are known.

\subsection{How feedback works}
The prompts state the rules and show the checked state visible to the agent; their two versions are described in Appendix~\ref{app:routing-development}. The following mechanics are common to both.

Initially $\mathcal P^0=\varnothing$. If global verification discovers an external collision, the following round supplies that pair's event information and the peer's refreshed trajectory. Unrelated robots remain hidden. A spatial line alone would not suffice: the correction uses simultaneous occupancy and movement at specific ticks. Agents receive their previous proposal even if it was not accepted, and the prompt distinguishes it from the current incumbent. Repairs must be new complete arrays from the original start; appending a wait after an earlier collision cannot repair that collision.

\paragraph{Format failure versus route failure.}
A missing or malformed JSON answer, wrong set of IDs, non-array output, or unknown command token is not a checked search answer. It is retried with the identical prompt and settings and a deterministic new retry seed. Format failures are re-asked at most three times per logical attempt, and transport failures at most three times with prompt v1 and twenty times with prompt v2; exhaustion pauses execution for diagnosis. In contrast, a readable array that enters a wall, leaves the grid, exceeds 48 commands, or moves away after arrival is a checked answer. Any such error rejects the entire owned proposal, retains the incumbent and returns route feedback; its candidate global loss is not evaluated. Otherwise, unfinished deliveries and collisions yield a finite score under~\eqref{eq:routing-objective}. These answers count toward search attempts and time even when rejected as worse.

\subsection{Protocols}
All protocols start from the same empty plan and accept only improvements of the global loss.

\paragraph{Team.}
A \emph{global round} is one batch of three concurrent block searches (six on two floors) followed by verification and a commit: one block per round on one map with Algorithm~\ref{alg:protocol}, one per floor on two floors, and the accepted combination with the other merge rules. All searches use the same frozen $x^k$ and $\mathcal P^k$.
\begin{enumerate}
\item Each agent proposes arrays for its three robots, receives private feedback, and may make up to $H=3$ checked answers. Python retains its best strictly improving proposal for $L_b+\widehat R_b$, including the unchanged block as an option.
\item Check the three proposals globally: substitute one proposed block into $x^k$ while keeping both other blocks unchanged. Compare these three full-plan losses with $F(x^k)$.
\item Accept only the proposal with the largest positive global decrease. Ties favor the lower block index; if none improves, retain $x^k$. All block pairs may interact, so a conservative complete dependency graph has three colors, each containing one block.
\item Retain all external collision pairs found during global checks, including those in unselected proposals. Peer trajectories for the next round come from the new accepted plan, rather than a discarded proposal.
\end{enumerate}
This is parallel proposal generation with serial block commitment. Within each private search, the correction and peer routes stay frozen. A deterministic routine computes a certified minimum or a valid lower bound for the available three-robot objective. Search stops early only when a submitted score equals that bound; otherwise it uses its remaining $H$ attempts. Appendix~\ref{app:routing-references} describes this routine and the independent offline certificates. The team's stopping rule differs between the two prompt versions (Appendices~\ref{app:routing-development} and~\ref{app:routing-revised}); restarting a run never resets a counter.

\paragraph{Single agent and best-of-$W$.}
A single agent controls all robots, sees and scores the full plan and receives full-plan feedback; every checked answer is evaluated immediately, and any strictly lower-loss legal plan is accepted at once. There is no grouping into rounds and no hidden-pair correction. The answer scope (all routes, up to $K$ routes or exactly one route per answer) and the stopping rules are given with each variant. The best-of-$W$ runs $W$ copies of the one-route single agent in parallel on the same accepted plan and prompt, with their own seeds; the best strictly improving answer of a round is accepted, its round time is the slowest of the $W$ answers, and $W$ equals the number of team agents.

\subsection{Model, seeds and measurement}\label{app:routing-measurement}
The model is \texttt{openai/gpt-oss-120b}, an open-weight mixture-of-experts reasoning model described in the OpenAI model card \citep{openai2025gptoss}. All routing runs, with both prompt versions, use the same OpenRouter/Groq endpoint, pinned without fallback, with reasoning effort \texttt{medium}, temperature $0.6$ and top-$p=0.95$. No \texttt{max\_tokens}, \texttt{max\_completion\_tokens}, or reasoning-token cap is sent. The network timeout is 3600 seconds; there is no absolute client deadline or automatic reasoning escalation. The API does not expose a controlled accelerator allocation.

\paragraph{Inference service and hardware.}
OpenRouter supplies the API and routes requests to the selected inference provider \citep{openrouter2026routing}; Groq performs inference. The saved payloads restrict routing to \texttt{groq} with \texttt{allow\_fallbacks=false}, and returned provider metadata identifies Groq. Groq documents GPT-OSS-120B serving through its LPU (Language Processing Unit) inference platform \citep{groq2025lpu,groq2026gptoss}. This identifies the accelerator family; the chip generation, accelerator count, location, batching and service load are not exposed by the API.

The client measures each request with a monotonic timer around the HTTP call, response download and response processing. Consequently, the durations below include network, routing and service latency, as well as model generation; these components are not separated. Search time aggregates only checkable-answer durations according to~\eqref{eq:routing-clock}; elapsed time and failed-request costs are recorded separately. All routing variants use the same provider restriction.

For replicate $s$, the base API seed is $4{,}000{,}000+100{,}000s$. Single-agent checked step $j$, starting at zero, adds $j$ with prompt v1 and $1000j$ with prompt v2 (copy $c$ of a best-of-$W$ adds $1000j+c$); team round $k$, private step $h$ and block $b$, also zero-indexed, add $1000k+100h+b$. Retry seeds are deterministic hashes of the logical seed and retry index. Seed settings are recorded, but provider reproducibility is best effort and the protocols induce different request streams. Map, settings, thresholds and limits were fixed before the runs of each campaign, and failed seeds are not rerolled.

Let $a_j$ be the latency of a returned answer whose command object can be checked. The single-agent search time is $\sum_j a_j$. If agent $b$ has checked latencies $a_{kbh}$ in round $k$, the team search time is
\begin{equation}\label{eq:routing-clock}
t_{\rm team}=\sum_k\max_{b\in\{0,1,2\}}\sum_h a_{kbh}.
\end{equation}
We call this the \emph{search time}, as in the main text. The three agents' times are not added as elapsed parallel time. Geometric errors and collisions count; missing or malformed answers and transport failures do not count toward search time. Actual elapsed time, preparation, stopping checks, verification, all physical requests, usage and reported charges are recorded separately. The simulator is inexpensive and has no artificial evaluation delay; this study does not measure a saving from expensive global evaluations.

\subsection{Reference routes and certificates}
\label{app:routing-references}
Reference computations certify the global optimum on the loss curves and the block minima used in the local-quality and block-geometry diagnostics. They use the model evaluator's commands, horizon, goal convention and weights.

\paragraph{Global optimum.}
Breadth-first search on the obstacle grid computes each robot's shortest distance $d_i(s_i,g_i)$, as listed in Table~\ref{tab:routing-assignments}. On the first map, their sum is a lower bound on every full-plan loss:
\begin{equation}\label{eq:routing-reference-global-bound}
F(x)\geq\sum_{i=0}^{8}d_i(s_i,g_i)=144.
\end{equation}
A delivered robot cannot arrive before its grid distance. An undelivered robot costs at least 148, exceeding every individual distance (at most 20 on the first map and 23 on the second), and conflict penalties are nonnegative. Conflict-based search (CBS) \citep{sharon2012cbs} finds a nine-robot plan with all deliveries, no conflicts and total arrival cost 144. Independent replay confirms that it attains the bound, proving $F^*=144$ even when unfinished or colliding plans are considered. The saved search expands seven high-level CBS nodes. The same argument certifies $F^*=147$ on the second map, where CBS again finds a delivered, collision-free plan whose arrival cost equals the sum of shortest distances, and $F^*=144+147=291$ on two floors, whose floors do not interact.

\paragraph{Search algorithms.}
For one robot, breadth-first search uses states $(t,p)$: tick $t\in\{0,\ldots,48\}$ and free cell $p$. A move or wait advances time by one; each transition checks obstacles, bounds and imposed vertex or directed-edge constraints. Goal arrival is accepted only if remaining there through tick 48 respects the constraints. Each state has one predecessor record, merging alternative command prefixes. With 244 free cells on either map, there are at most $49\times244=11{,}956$ states per single-robot search.

CBS starts with independently shortest constrained routes and orders its high-level nodes by total arrival cost. The earliest vertex conflict creates two branches, forbidding that cell at that tick for one robot or the other. An edge swap similarly branches on the directed transitions. Only the affected route is recomputed. The cheapest conflict-free node gives the minimum cost among delivered, conflict-free plans satisfying the constraints. CBS avoids enumerating all joint command arrays, but its tree can grow exponentially; computation-budget exhaustion certifies neither optimality nor impossibility.

\paragraph{Fixed-base block references.}
For a fixed base $x^k$, let $A_b^{\mathcal P^k}(u;x^k)$ be the absolute private score: $L_b(u)$ plus the external conflict penalties for retained pairs $\mathcal P^k$. Then
\begin{equation}\label{eq:routing-audit-score}
\widehat f_b(u;x^k)=A_b^{\mathcal P^k}(u;x^k)-A_b^{\mathcal P^k}(x_b^k;x^k).
\end{equation}
The other six robots' programs remain fixed. Only retained external pairs enter the reference for the agent's available objective; using all hidden peers here would change the denominator of $Q$. Separately, let $A_b^{\rm all}$ include every external pair. Costs involving only the other blocks cancel, so the true conditional global improvement is
\begin{equation}\label{eq:routing-audit-D}
D_b(x^k)=A_b^{\rm all}(x_b^k;x^k)
-\min_u A_b^{\rm all}(u;x^k).
\end{equation}
For either reference, the relevant frozen peer positions become vertex constraints, and peer moves become reverse-edge constraints, for a collision-free witness search. CBS resolves conflicts among the three owned robots. All minima range over legal owned arrays, including waits and unfinished deliveries, rather than only a route catalogue.

\paragraph{Certifying the soft-penalty optimum.}
Finding the best collision-free delivered plan alone need not minimize the soft loss. Let $\bar d_b=\sum_{i\in\mathcal I_b}d_i(s_i,g_i)$ and let $A_b^{\rm CBS}$ be that plan's absolute score. Any plan with a penalized conflict costs at least $\bar d_b+100$, and any unfinished plan costs at least
\[
\bar d_b+\min_{i\in\mathcal I_b}\{148-d_i(s_i,g_i)\}.
\]
The online routine certifies its CBS value only when $A_b^{\rm CBS}$ does not exceed either bound. Otherwise it uses a valid single-robot relaxation bound for the early-stop test, without claiming that the CBS value is the soft minimum.

The offline audit additionally uses independently implemented dynamic programming over $(t,p)$. For each owned robot it minimizes delivery cost plus the relevant frozen-peer conflict penalties, including conflicts after arrival, and retains the non-delivery terminal cost of~\eqref{eq:routing-objective}. Here conflicts remain penalized rather than forbidden. Dropping only conflicts among owned robots relaxes the joint block objective. If $\underline A_i$ is this single-robot minimum and $w_b$ is the joint witness, then, for either choice of external pairs,
\begin{equation}\label{eq:routing-reference-sandwich}
\sum_{i\in\mathcal I_b}\underline A_i
\;\leq\;\min_u A_b(u;x^k)
\;\leq\;A_b(w_b;x^k).
\end{equation}
A separate replay checks the witness's legality and score. On the first map both inequalities are equalities in every audited subproblem, for the v1 and v2 team runs. On the second map, where conflicts among owned robots bind, the single-robot bound lies 1 or 2 below the witness in 83 of the 1,074 subproblems of the v2 team runs (40 of 786 and 81 of 1,008 in the runs with best combination and unsafe merging); their minima are certified instead by the CBS optimum together with the penalty and non-delivery bounds above. Every block minimum used in the audit of Appendix~\ref{app:routing-revised-audit} is therefore certified. Diagnostic witnesses and lower bounds are saved separately from model answers. These certificates support the assumption checks of Appendix~\ref{app:routing-assumptions}; they do not supply routes to agents or alter any recorded decision.

\section{Routing: prompt and protocol design}\label{app:routing-development}
We wrote two versions of the prompt. With prompt v1 we tried several protocols for the single agent and the team on the first map, ten runs each. Where these runs stalled, the model had to simulate the other robots' commands itself; prompt v2 presents the full checked state instead and made both protocols faster. For the main text we chose the combination that worked best: prompt v2, a single agent that replaces one route per answer, the best-of-$W$ such agents as the parallel control, and the team that rewrites its three routes. Task, loss, correction, protocols and measurement are those of Appendix~\ref{app:routing-task}. Figure~\ref{fig:routing-development} shows the mean curves of all variants and Table~\ref{tab:routing-variants} their outcomes.

\begin{figure*}[t]
\centering
\includegraphics[width=.98\textwidth]{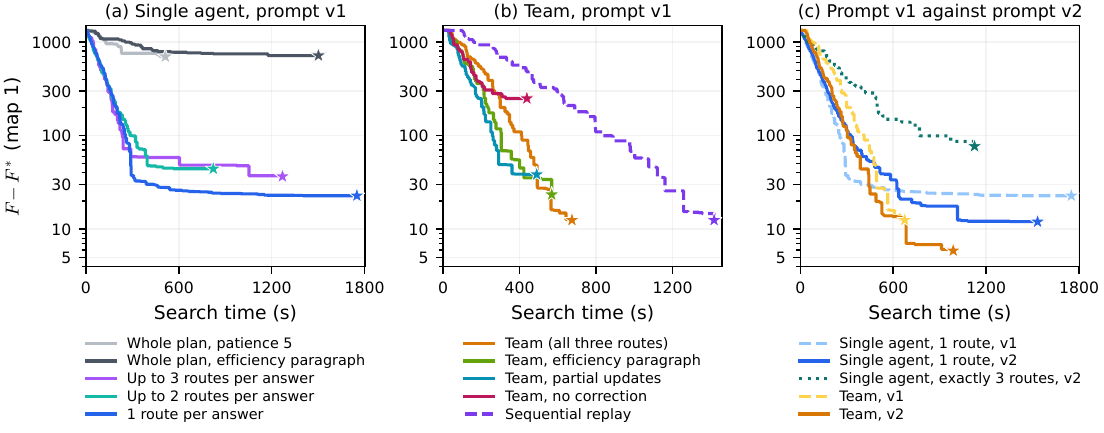}
\Description{Three panels of mean gap to the optimum against search time on logarithmic axes, all on the first map. (a) Single-agent protocols with prompt v1: the two whole-plan protocols stall near a gap of 700; answers with up to three or two routes stall near 37 and 44; one route per answer continues lower and ends near 23. (b) Team protocols with prompt v1: the team ends near 12, the team with the efficiency paragraph near 23, the team with partial updates near 38, the team without correction stalls near 250, and the sequential replay of the team reaches the team's values more slowly. (c) Prompt v1 against prompt v2: with v2 the one-route single agent ends near 12 instead of 23 and the team near 6 instead of 12; the single agent also ends sooner; the single agent with exactly three routes per answer and prompt v2 ends near 77.}
\caption{How the protocols and the prompt were chosen (first map; mean gap $F-F^*$ against search time). (a) Single-agent protocols with prompt v1, ten runs each. (b) Team protocols with prompt v1, ten runs each; the sequential replay plays the team's answers one agent at a time. (c) Prompt v1 against prompt v2: twenty runs for the v2 one-route single agent and team, ten for the other curves. Stars mark the end of recorded support.}
\label{fig:routing-development}
\end{figure*}

\subsection{Prompt v1}\label{app:routing-prompt-v1}
An initial prompt gives the command rules, coordinate orientation, horizon, loss weights, an ASCII obstacle map, owned robot IDs and start/goal coordinates, and their current command arrays. Team agents initially do not receive other blocks' assignments or routes. The single agent receives all nine assignments and checks every pair internally. Each request states that the output must contain complete literal arrays for exactly the owned IDs.

After a checked answer, Python reports its score and per-robot arrival or remaining distance. For an illegal route it identifies the first rejected command, its tick, the previous and attempted cells, and legal directions from that cell. Feedback includes the legal position prefix and an ASCII route drawing for a failed or unfinished route. For visible conflicts it gives the robot pair, tick, vertex cell or swapped edge, and time-labeled map frames. A safe arrived route does not need a repeated full drawing. A supplied external peer is accompanied by its complete time-indexed positions, with only a constant trailing wait suffix compressed; the last listed position persists thereafter. At most three illustrative collision frames are shown, while complete event lists and peer motion remain available.

Format failures were re-asked at most three times and transport failures at most three times. The team stopped upon reaching the threshold ($F\leq153$ with a safe plan), after five rounds without an improving commit, or at a ceiling of eight global rounds.

\subsection{Protocols tried with prompt v1}
\paragraph{Team and whole-plan single agent.}
The team is the protocol of Appendix~\ref{app:routing-task}. The first single agent rewrote all nine routes in every answer. It controls all nine arrays and receives full-plan feedback. Each checked answer is evaluated immediately; any strictly lower-loss legal plan becomes the incumbent immediately. There is no grouping into three-answer batches, no external hidden-pair correction, and no global-round counter. The chain stops upon reaching the stopping threshold, after five consecutive checked answers without an accepted improvement, or after sixty checked answers. A route-invalid answer contributes to that five-answer streak. Actual chains used 5--13 checked answers; sixty was a safety ceiling, not the number performed. Both variants also stop early once all nine robots are delivered without conflicts and $F\leq153$; this rule only ends a run.

The twenty runs of these two protocols complete under their stopping rules. The team safely delivers every robot in $10/10$ runs, with final losses 149--167; the whole-plan single agent does so in $1/10$. Table~\ref{tab:routing-variants} reports final loss, search time and reported cost for these and all other variants. Table~\ref{tab:routing-extra-accounting} reports elapsed time and answer counts; Table~\ref{tab:routing-seeds} retains every seed.

\begin{table}[t]
\centering\footnotesize
\begin{tabular}{lrr}\toprule
Metric & \makecell[rb]{Single agent\\(whole plan)} & Team\\\midrule
Actual elapsed (s) & $315.25\pm132.36$ & $1089.55\pm896.47$\\
Checked answers & $8.30\pm2.71$ & $59.30\pm7.92$\\
Safe delivery of all nine & 1/10 & 10/10\\\bottomrule
\end{tabular}
\caption{All ten outcomes per variant; mean $\pm$ sample SD where applicable. Team answer counts sum across agents, whereas its search time uses the per-round maximum. Actual elapsed time includes failed requests and recovery pauses.}
\label{tab:routing-extra-accounting}
\end{table}

\begin{table*}[t]
\centering\footnotesize
\begin{tabular}{rrrrrrrrr}\toprule
 & \multicolumn{4}{c}{Single agent, whole plan} & \multicolumn{4}{c}{Team (Alg.~\ref{alg:protocol})}\\
Seed & Final $F$ & Delivered/conflicts & Answers & Time (s) & Final $F$ & Rounds & Answers & Time (s)\\\midrule
0 & 1476 & 0/0 & 5 & 165.6 & 152 & 7 & 45 & 608.0\\
1 & 1476 & 0/0 & 5 & 135.4 & 158 & 8 & 58 & 650.2\\
2 & 455 & 9/3 & 8 & 264.4 & 152 & 7 & 57 & 494.2\\
3 & 490 & 8/2 & 10 & 372.6 & 155 & 8 & 59 & 673.6\\
4 & 722 & 9/5 & 9 & 253.7 & 155 & 8 & 66 & 578.9\\
5 & 959 & 9/7 & 8 & 307.7 & 159 & 8 & 59 & 633.3\\
6 & 1476 & 0/0 & 5 & 176.1 & 164 & 8 & 58 & 582.3\\
7 & 609 & 8/3 & 9 & 294.0 & 149 & 8 & 51 & 486.0\\
8 & 150 & 9/0 & 11 & 513.3 & 167 & 8 & 72 & 544.5\\
9 & 597 & 9/4 & 13 & 330.9 & 154 & 8 & 68 & 642.9\\
\bottomrule\end{tabular}

\caption{Every completed seed, without selection. Time is the search time defined above. All ten team outcomes deliver 9/9 without conflicts. Single-agent delivered/conflict counts refer to its final accepted plan, including an unchanged initial plan.}
\label{tab:routing-seeds}
\end{table*}

Figure~\ref{fig:routing-development}(a, b) plots the mean gap $F-F^*$ over all ten runs of each variant; an accepted plan is held until the next accepted change, and a stopped run keeps its final loss. The team and its sequential replay end at mean $F=156.50$, versus $841.00$ for the whole-plan single agent.

The journals contain 676 returned, checkable model responses and 142 HTTP 429 failures, for 818 physical requests. All returned responses include API-returned reasoning. Checked responses include 120 route-invalid proposals and 437 legal candidate proposals with collisions. Across both variants, the provider reports USD 3.553312 in charges; charges for the 142 HTTP failures are unknown, so reported costs are not a certified total bill. Diagnosed rate-limit interruptions resume after cooldowns by reusing saved answers and identical prompts; they do not restart a seed or change the model, map, loss or reasoning. Elapsed time includes recovery pauses. Independent replay checks every returned command, frozen base, local score, correction, block selection and final plan.

\paragraph{A flat curve can conceal attempted repairs.}
Single-agent seed 0 stops at the initial loss 1476 after five checked answers. It does not simply return stationary plans: the first answer has three illegal routes, the third has one, the fourth is legal but worse with $F=2092$, and the fifth has one illegal route. Entire-plan rejection retains the incumbent, while the next prompt still contains attempted arrays and offending commands. With the efficiency paragraph and patience fifteen (below), seed 0 again ends at $F=1476$.

The nine failed runs each end with five nonimproving answers, which take $162.98\pm24.31$ seconds per run (mean $\pm$ sample SD): 34 of these 45 answers are route-invalid. All error durations already count toward search time. The last response, at $513.28$ seconds, belongs to the successful seed; no later attempts were measured. Counts and clocks are verified by independent replay.

\paragraph{Team without correction.}\label{app:routing-no-correction}
This ablation disables only the cross-block correction channel of the team; Appendix~\ref{app:routing-revised} repeats it with prompt v2. Agents receive no acquired external pairs, no peer timelines and no global collision feedback in their local searches; feedback about their own robots, including internal collisions, is unchanged. The coordinator still evaluates the full objective, logs collisions and commits the single best improving block. Model, provider, sampling parameters, prompt, $H=3$, the eight-round ceiling, the five-round plateau rule, the safe target and the ten prescribed seed streams are those of the v1 team. The first request payloads are byte-identical, so the two variants diverge only when the corrected controller would first have revealed a conflict. Local early stopping is unchanged: a block that is optimal for its local objective may stop requesting answers while the joint plan still contains external collisions.

\begin{table}[!h]
\centering\footnotesize
\setlength{\tabcolsep}{2.6pt}
\begin{tabular}{lrrrrrrrrrr}\toprule
Seed & 0 & 1 & 2 & 3 & 4 & 5 & 6 & 7 & 8 & 9\\\midrule
Final $F$ & 452 & 444 & 152 & 451 & 448 & 294 & 448 & 444 & 548 & 248\\
Collisions & 3 & 3 & 0 & 3 & 3 & 1 & 3 & 3 & 4 & 1\\
Rounds & 8 & 8 & 4 & 8 & 8 & 8 & 8 & 8 & 8 & 8\\\bottomrule
\end{tabular}
\caption{All ten no-correction team outcomes. Every final plan delivers all nine robots; collisions are cross-block events in the final plan.}
\label{tab:routing-no-correction}
\end{table}

All ten runs complete, and every final plan delivers all nine robots (Table~\ref{tab:routing-no-correction}). Nine retain one to four cross-block collisions, 2.4 on average over all ten runs; only seed 2 reaches a safe plan, with $F=152$. Final loss is $392.90\pm120.66$, search time $337.97\pm95.34$ seconds, with $39.8\pm10.4$ checked answers over $7.6\pm1.3$ rounds (mean $\pm$ sample SD). Subtracting the collision penalty of 100 per event leaves a mean of 152.9, close to the corrected v1 team's $156.50$. Delivery quality within blocks is therefore comparable, and the remaining loss consists of exactly the cross-block interactions that the uncorrected local objective omits. Because only improving blocks are committed, the uncorrected team never increases $F$, but its agents cannot tell which of their candidates would remove an external collision, so search stalls once every block is locally good. Reported charges total USD 1.32 over 476 requests, 78 of which return no charge. Independent replay checks every route, acceptance decision and outcome.

\paragraph{Whole-plan single agent with an efficiency paragraph.}\label{app:routing-efficient-p15}
This variant adds to the prompt of the whole-plan single agent the paragraph
\begin{quote}\small Work through the routes efficiently, avoiding repeated calculations or restating the same plan. Once you have a concrete candidate, return its JSON so the checker can provide feedback for the next improvement.\end{quote}
and raises its patience to fifteen answers without improvement, with a ceiling of sixty answers. All ten runs end normally and are independently replayed. One run reaches a safe plan, with $F=154$; five deliver all nine robots but keep one to five collisions, and four never accept a route. The mean final loss is $860.00\pm555.40$ and the search time $728.52\pm366.76$ seconds, with $26.4\pm15.5$ answers per run (Table~\ref{tab:routing-variants}). Table~\ref{tab:routing-single-seeds} lists every run, next to the agents that replace only some routes (below).

\begin{table*}[t]\centering\footnotesize
\setlength{\tabcolsep}{3pt}
\begin{tabular}{r@{\hspace{8pt}}rrrr@{\hspace{10pt}}rrrr@{\hspace{10pt}}rrrr@{\hspace{10pt}}rrrr}\toprule
& \multicolumn{4}{c}{1 route per answer} & \multicolumn{4}{c}{Up to 2 routes} & \multicolumn{4}{c}{Up to 3 routes} & \multicolumn{4}{c}{All nine routes}\\
\cmidrule(lr){2-5}\cmidrule(lr){6-9}\cmidrule(lr){10-13}\cmidrule(l){14-17}
Seed & $F$ & Coll. & Ans. & Search (s) & $F$ & Coll. & Ans. & Search (s) & $F$ & Coll. & Ans. & Search (s) & $F$ & Coll. & Ans. & Search (s)\\\midrule
0 & 152 & 0 & 16 & 241.8 & 152 & 0 & 18 & 298.3 & 160 & 0 & 27 & 637.7 & 1476$^\ast$ & 0 & 15 & 403.4\\
1 & 161 & 0 & 35 & 742.6 & 267 & 1 & 25 & 448.3 & 153 & 0 & 19 & 402.9 & 259 & 1 & 60 & 1502.2\\
2 & 152 & 0 & 32 & 577.1 & 170 & 0 & 25 & 778.1 & 152 & 0 & 17 & 331.2 & 1476$^\ast$ & 0 & 15 & 395.3\\
3 & 150 & 0 & 43 & 814.7 & 152 & 0 & 22 & 396.1 & 152 & 0 & 9 & 161.9 & 1476$^\ast$ & 0 & 15 & 469.2\\
4 & 151 & 0 & 51 & 1153.7 & 152 & 0 & 13 & 194.5 & 152 & 0 & 10 & 168.6 & 691 & 5 & 18 & 395.7\\
5 & 156 & 0 & 91 & 1750.3 & 277 & 1 & 41 & 822.9 & 150 & 0 & 47 & 1269.7 & 365 & 2 & 29 & 917.9\\
6 & 272 & 1 & 26 & 372.1 & 150 & 0 & 22 & 384.4 & 164 & 0 & 21 & 682.0 & 656 & 5 & 22 & 755.4\\
7 & 149 & 0 & 26 & 289.0 & 153 & 0 & 12 & 255.0 & 263 & 1 & 31 & 745.0 & 154 & 0 & 47 & 1120.0\\
8 & 172 & 0 & 37 & 676.1 & 152 & 0 & 26 & 565.8 & 196 & 0 & 32 & 631.8 & 1476$^\ast$ & 0 & 15 & 566.0\\
9 & 153 & 0 & 46 & 999.7 & 257 & 1 & 42 & 751.3 & 264 & 1 & 27 & 560.0 & 571 & 4 & 28 & 760.0\\\bottomrule

\end{tabular}
\caption{Single agents with the efficiency paragraph and patience fifteen, by the number of routes an answer may replace: independently replayed final plans of all forty runs. Coll.: collisions in the final plan; Ans.: checked answers. $^\ast$: no route was ever accepted, so the plan is the initial one and delivers no robot; every other final plan delivers all nine. Runs with a safe plan and $F\leq153$ stopped early, seed 1 of the whole-plan agent at its ceiling of sixty answers, and all others after fifteen answers without improvement. The whole-plan agent's elapsed time, which adds retries, exceeds its search time by less than 5\,s in every run except seed 2 (75\,s).}
\label{tab:routing-single-seeds}
\end{table*}

Reported costs total USD 2.22; one request returned no charge.

With patience five and no efficiency paragraph, the whole-plan agent ends at $841.00\pm483.14$ after $281.36\pm111.67$ seconds, also with one safe run. The longer patience therefore buys more search time but not a better final plan: writing all nine routes in one answer remains the bottleneck. Letting the agent change only some routes per answer, with the same prompt and patience, raises safe completion to seven or eight runs of ten with up to two or three routes, and to nine with one route per answer (below).

\paragraph{Single agents that replace some routes.}\label{app:routing-partial-three}
These agents still observe and score the complete nine-robot problem but replace at most $K\in\{1,2,3\}$ routes per answer and keep the others unchanged. At every answer the agent sees all nine starts, goals and accepted arrays, plus the full global route and collision feedback. It chooses up to $K$ distinct robot IDs, independent of the team's fixed groups. Arrays for omitted IDs remain exactly unchanged. The selected replacement arrays are merged literally into the accepted plan and checked together. A geometry error in any selected array rejects the entire proposal. Otherwise, the merged plan is accepted immediately if and only if its full global $F$ strictly decreases. Collisions can occur in accepted intermediate plans if their total loss improves; safe completion always requires zero collisions.

A run stops after fifteen consecutive checkable answers without an accepted improvement, or once its plan is safe with $F\leq153$. The safety ceiling is sixty checked answers for $K=3$ and 120 for $K\in\{1,2\}$, because one route per answer needs more answers; the $K=3$ runs used at most 47, so the ceiling never stopped them. There are no client output or reasoning-token caps. Geometry-invalid but schema-valid answers consume checked attempts and search time; malformed answers and transport failures are retried with the same prompt and do not.

The shared coordinate, map, loss, collision and global feedback instructions are those of the team prompt, and the efficiency paragraph quoted above is included.
The scope instruction reads:
\begin{quote}\small
You are one agent controlling all nine robots in \texttt{robots}. Propose a partial update to their delivery programs. \dots{} All nine robots, their starts, goals and current programs are visible to you.
\end{quote}
Only the following sentences depend on $K$. For $K=3$ they are
\begin{quote}\small
Choose from one to three distinct robot IDs in each response. Return complete replacement arrays ONLY for those chosen IDs. \dots{} Final answer: exactly one JSON object with key programs mapping one to three selected robot IDs to their new complete literal arrays of command names.
\end{quote}
For $K=2$ they read ``Choose one or two distinct robot IDs in each response'' and ``mapping one or two selected robot IDs''. For $K=1$ they read ``Choose exactly one robot ID in each response. Return a complete replacement array ONLY for that chosen ID'' and ``mapping exactly one selected robot ID''. The structured task records \texttt{max\_updated\_robots}$=K$ and the matching \texttt{output\_scope}. Before the $K\in\{1,2\}$ runs, an offline check confirmed that the code produces, for $K=3$, exactly the first prompt recorded in the $K=3$ runs.

Table~\ref{tab:routing-variants} summarizes the five single-agent protocols and Table~\ref{tab:routing-single-seeds} lists every partial-update run.
\begin{itemize}
\item \textbf{One route per answer:} nine of ten runs are safe and six stop early; final loss $166.80\pm37.60$; search time $761.71\pm456.76$ seconds with $40.3\pm20.6$ answers.
\item \textbf{Up to two routes:} seven runs are safe and six stop early; final loss $188.20\pm54.87$; search time $489.47\pm228.26$ seconds with $24.6\pm10.1$ answers.
\item \textbf{Up to three routes:} eight runs are safe and five stop early; final loss $180.60\pm45.75$; search time $559.08\pm325.35$ seconds with $24.0\pm11.4$ answers.
\end{itemize}
All partial-update protocols reduce the gap at a similar rate during the first 300 seconds; afterwards the one-route agent continues to improve, while runs with larger answers more often stall with a collision. The whole-plan agents stall far higher. 

Requiring a complete answer for all nine routes proved difficult: the whole-plan single agent is safe in one run of ten, with or without the efficiency paragraph. Smaller answers make the single agent more reliable: with one route per answer it is safe in nine runs of ten and reaches the lowest mean loss of all single-agent protocols. The price is time: it needs about 40 answers per run, and its median time to a safe plan with $F\leq153$ is 696 seconds, against 341 and 331 seconds with up to two and three routes. We therefore use the one-route protocol as the single-agent baseline, with prompt v2 in the main text. With prompt v1, it reduces the gap faster at first than the v1 team, but the team's mean loss is lower after about 560 seconds; the v1 team is safe in all ten runs, its search takes $589\pm65$ seconds against $762\pm457$, and its final loss varies far less ($156.50\pm5.60$).

\paragraph{Team with the efficiency paragraph.}\label{app:routing-team-efficient}
This variant adds the same paragraph to the team prompt and changes nothing else; without the paragraph, its first three requests are those of the v1 team. All ten runs end normally and are independently replayed. Every final plan delivers all nine robots, and nine are safe; seed 8 keeps one collision. The mean final loss is $167.40\pm34.15$ and the search time $436.65\pm90.79$ seconds, with $51.8\pm10.8$ answers per run (Table~\ref{tab:routing-variants}). Table~\ref{tab:routing-team-variant-seeds} lists every run, together with the team with partial updates (below).

\begin{table*}[t]\centering\footnotesize
\begin{tabular}{r rrrrr@{\hspace{14pt}}rrrrrr}\toprule
& \multicolumn{5}{c}{Efficiency paragraph} & \multicolumn{6}{c}{Partial route updates}\\
\cmidrule(lr){2-6}\cmidrule(l){7-12}
Seed & $F$ & Coll. & Rounds & Search (s) & USD & $F$ & Coll. & Rounds & Ans. & Search (s) & USD\\\midrule
0 & 150 & 0 & 7 & 400.19 & 0.23792 & 152 & 0 & 5 & 43 & 257.44 & 0.15622\\
1 & 149 & 0 & 5 & 276.32 & 0.16257 & 152 & 0 & 6 & 51 & 320.67 & 0.19878\\
2 & 154 & 0 & 8 & 501.97 & 0.31963 & 248 & 1 & 8 & 58 & 489.37 & 0.26790\\
3 & 188 & 0 & 8 & 567.83 & 0.32965 & 150 & 0 & 4 & 33 & 212.23 & 0.12381\\
4 & 161 & 0 & 8 & 429.93 & 0.17183 & 153 & 0 & 6 & 44 & 268.90 & 0.16628\\
5 & 157 & 0 & 8 & 514.01 & 0.24439 & 164 & 0 & 8 & 57 & 373.95 & 0.20821\\
6 & 152 & 0 & 6 & 382.34 & 0.19034 & 148 & 0 & 4 & 27 & 156.81 & 0.08337\\
7 & 152 & 0 & 8 & 515.22 & 0.27382 & 248 & 1 & 8 & 46 & 291.64 & 0.14034\\
8 & 259 & 1 & 8 & 444.37 & 0.19429 & 150 & 0 & 8 & 39 & 360.10 & 0.15127\\
9 & 152 & 0 & 6 & 334.36 & 0.19404 & 259 & 1 & 8 & 49 & 372.92 & 0.19586\\\bottomrule
\end{tabular}
\caption{Two variants of the v1 team: independently replayed final plans of all ten runs of each. Every final plan delivers all nine robots. Coll.: collisions in the final plan; Ans.: checked answers, summed over agents; USD: reported cost. With partial updates, seed 0 is the adopted pilot; the runs with a safe plan and $F\leq153$ stopped early, the others at the eight-round cap, and elapsed time exceeds search time by less than 3\,s in every run except seed 0 (31\,s).}
\label{tab:routing-team-variant-seeds}
\end{table*}

The paragraph is the only difference from the v1 team. It shortens the mean search time from 589 to 437 seconds and lowers the reported cost per run from USD 0.28 to 0.23, while nine of ten runs remain safe; the mean final loss rises from $156.5$ to $167.4$, mostly because of the one unsafe run. The prompt therefore changes the team's results only slightly.

\paragraph{Team with partial route updates.}\label{app:routing-team-partial-three}
Here each agent may replace the routes of one, two or three of its own robots instead of returning all three; arrays for omitted robots are copied exactly from the frozen plan, separate subset proposals are not combined, and correction, verification and the single best commit are unchanged. The original instruction was:
\begin{quote}\small
Return complete arrays for EVERY owned ID, even if unchanged.
\end{quote}
Its replacement is:
\begin{quote}\small
Choose from one to three distinct IDs from owned in each response. Return complete replacement arrays ONLY for those chosen IDs. All omitted owned robots keep their installed current arrays exactly unchanged. You cannot change robots outside owned. Each private attempt starts from the same current arrays shown below.
\end{quote}
The opening instruction becomes:
\begin{quote}\small
Propose a partial update to delivery programs for your owned robots in \texttt{robots}.
\end{quote}
The final-answer instruction becomes:
\begin{quote}\small
Final answer: exactly one JSON object with key programs mapping one to three selected owned IDs to a literal array of command names. No shorthand, loops, code or alternative answers.
\end{quote}
The structured task adds two fields:
\begin{itemize}
\item \texttt{max\_updated\_robots} is 3;
\item \texttt{output\_scope} has the exact value ``One to three selected owned IDs only; omitted arrays remain exactly current.''
\end{itemize}
All other instructions, including ownership, objective and correction, are unchanged.

All ten runs end normally, six by early stop and four at the eight-round cap. All deliver nine robots, but replicates 2, 7 and 9 each retain one collision, giving seven safe runs. Replicates 2 and 7 retain an edge swap between $r_4$ and $r_8$, which belong to the same agent, at $t=6$; replicate 9 retains a collision between $r_6$ and $r_7$, which belong to different agents, at $t=13$. The mean final loss is $182.40\pm48.08$ and the search time $310.40\pm94.67$ seconds (Tables~\ref{tab:routing-variants} and~\ref{tab:routing-team-variant-seeds}).

Partial updates halve the team's mean search time (310 versus 589 seconds). In exchange, three runs end with a collision, and the mean final loss rises from $156.5$ to $182.4$. Partial updates thus trade reliability for speed for the team. Because a team's value lies in solving the task reliably, we keep the team that rewrites all its routes as the main protocol.

\begin{figure*}[t]
\centering
\includegraphics[width=.92\textwidth]{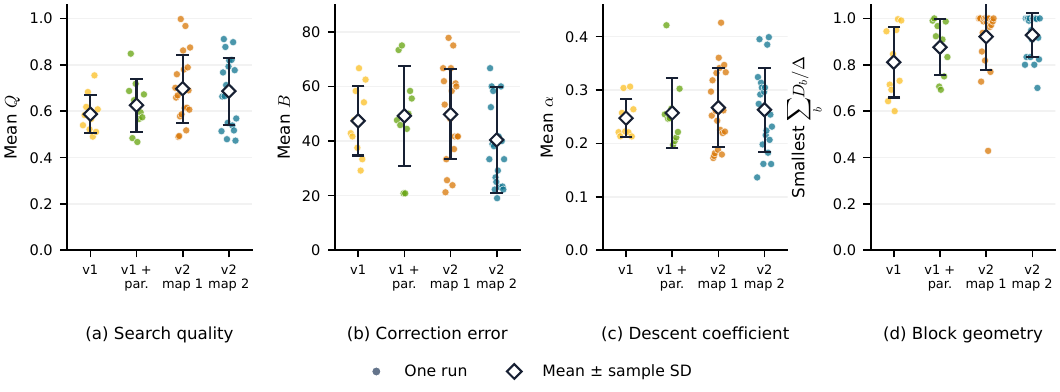}
\Description{Four panels compare four prompt settings using run-level statistics: prompt v1, v1 with the efficiency paragraph, prompt v2 on map 1 and prompt v2 on map 2. Panels show mean realized search quality Q, mean error term B, mean descent coefficient alpha, and the smallest geometry ratio over the states that start a round. Dots are individual runs; white diamonds with bars show mean and sample standard deviation. Search quality is higher with prompt v2; the error term and descent coefficient change little.}
\caption{Measured constants of the team per run for four prompt settings: v1 and v1 with the efficiency paragraph on map 1 (ten runs each), v2 on map 1 and on map 2 (twenty runs each). Panels (a)--(c) average within a run; (d) takes the smallest ratio $\sum_bD_b/\Delta$ over the states of the run that start a round. Diamonds and bars: mean $\pm$ sample SD, not confidence intervals. Since every true-minimizer error is zero, a run's mean $B$ equals its mean error at returned proposals.}
\label{fig:routing-prompt-coefficients}
\end{figure*}

\subsection{Why prompt v1 stalled}\label{app:routing-revision-diagnosis}
With prompt v1, six of the thirty partial-update single-agent runs ended with exactly one collision and fifteen answers without improvement. An offline check with the exact single-robot best response (Appendix~\ref{app:routing-references}) shows that in each of these six final plans one of the two colliding robots could be rerouted, with the other eight routes fixed, removing the collision without delaying that robot's arrival; in three of them the best such route even arrives earlier than the accepted one. Five of the six collisions are head-on: two robots move toward each other along the same row or column, so waiting only moves the point where they meet. The ninety answers given in these stalled phases show two failure modes. First, 56 of the 113 edited routes keep exactly the same cells and only insert or remove waits. Second, the 20 answers that did remove the original collision all created a new one with another robot. Prompt v1 listed every robot's commands but gave tick-by-tick positions only for the robots involved in a reported collision, and drew collision frames only for the previous attempt; to check a detour against the other robots, the model had to simulate their commands itself. The map had no column labels, and the rule that a swap between ticks $t-1$ and $t$ is reported at tick $t$ was not stated.

In prompt v2 we therefore changed the presentation, not the information structure, of both prompts. Asked again in the six stalled states, the single-agent prompt v2 produced an accepted improvement in 21 of 60 answers (ten per state, with and without the list of rejected attempts), including at least one in every state, against none of the ninety recorded answers with prompt v1. The team prompt had the same presentation, and only 22 of the 207 private searches that could improve their block reached the exact corrected block minimum.

\subsection{Prompt v2}\label{app:routing-prompt-v2}
Both prompts state the command, timing, collision and loss rules once, without strategy advice: no route, distance to the optimum or repair rule is supplied. They then present the checked state:
\begin{itemize}
\item the map with column and row labels, and the list of wall cells;
\item every controlled robot's start, goal, current program, arrival tick and cost;
\item a timeline with the position of every visible robot at every tick, marking robots that have arrived;
\item every visible collision of the accepted plan, described in words (for example, ``tick 7: r2 moves [7,6]$\to$[8,6] while r7 moves [8,6]$\to$[7,6]'') and drawn as maps of all visible robots at the tick before and the tick of the collision;
\item the attempts already checked against the accepted plan, with their losses and collisions.
\end{itemize}
The single agent sees all nine robots, the global loss and the stopping threshold, and lists the attempts rejected since the plan was last improved. A team agent sees only its three robots and its known peers, exactly as with prompt v1: robots of other agents whose collisions with its robots the coordinator has already found, with their current fixed timelines (Section~\ref{sec:main-task}). Other robots appear in no table, drawing or collision list. A team agent sees neither the global loss nor the threshold; it lists its earlier attempts in the current round and, if its previous proposal collided in the coordinator's global check, those collisions. Without correction, a team agent sees only its own robots. Answers have the same format as with prompt v1: complete literal command arrays for the chosen robots.

\subsection{Effect of the prompt version}\label{app:routing-prompt-coefficients}
Figure~\ref{fig:routing-development}(c) compares the two prompt versions. For the single agent with one route per answer, prompt v2 raises safe completion from nine of ten runs to all twenty, lowers the final loss from $166.80\pm37.60$ to $156.00\pm10.53$, and shortens the search from $761.7$ to $613.0$ seconds on average with fewer answers (27.8 against 40.3). For the team, it shortens the search from $589.4$ to $481.0$ seconds and lowers the final loss from $156.50\pm5.60$ to $149.90\pm2.67$; both are safe in every run, and the v2 team reaches the threshold in all twenty runs against three of ten. The team's stopping rule changed as well: the v1 team stopped after at most eight rounds, which ended seven of its ten runs before the plan reached $F\leq153$, four of them right after an improving round, whereas the v2 team may run up to thirty rounds; this comparison therefore also reflects that change. Geometrically invalid answers did not become rarer: they are 17\% of the checked answers for both v2 protocols, against 12\% for the single agent and 11\% for the team with v1. The gains come from repairs: the v2 team's searches reach the exact corrected block minimum in 82 of the 343 searches that can improve their block (22 of 207 with v1) and capture on average a fraction 0.66 of the available corrected improvement, pooled over searches (0.58 with v1).

\begin{table}[!htb]
\centering\footnotesize
\setlength{\tabcolsep}{3pt}
\begin{tabular}{@{}lrrrr@{}}\toprule
Per-run statistic & v1 & v1 + par. & v2, map 1 & v2, map 2\\\midrule
Mean $Q$ & $0.59\pm0.08$ & $0.63\pm0.11$ & $0.70\pm0.15$ & $0.69\pm0.14$\\
Mean $B$ & $47.4\pm12.7$ & $49.2\pm18.3$ & $49.7\pm16.5$ & $40.4\pm19.5$\\
Mean $\alpha$ & $0.25\pm0.04$ & $0.26\pm0.07$ & $0.27\pm0.07$ & $0.26\pm0.08$\\
Smallest ratio & $0.81\pm0.15$ & $0.88\pm0.12$ & $0.92\pm0.14$ & $0.93\pm0.10$\\\bottomrule
\end{tabular}
\caption{Mean $\pm$ sample SD over runs (ten for v1 settings, twenty for v2). Smallest ratio: the minimum of $\sum_bD_b/\Delta$ over the states of a run that start a round; its mean is not a certificate over the union of runs.}
\label{tab:routing-prompt-coefficients}
\end{table}

Applying the audit of Appendix~\ref{app:routing-revised-audit} to every team run with a full set of block references shows how the prompt moves the measured constants (Figure~\ref{fig:routing-prompt-coefficients}, Table~\ref{tab:routing-prompt-coefficients}): the v1 team, the v1 team with the efficiency paragraph (72 rounds, 216 returned searches, 518 checked answers) and the v2 team on both maps. Averaged per run, the realized search quality rises from $0.59\pm0.08$ with v1 to $0.63\pm0.11$ with the paragraph and to $0.70\pm0.15$ and $0.69\pm0.14$ with v2 on the two maps, while the error term stays at the same level ($47.4\pm12.7$, $49.2\pm18.3$, $49.7\pm16.5$ and $40.4\pm19.5$) and the descent coefficient changes little ($0.25$, $0.26$, $0.27$ and $0.26$). The error at every true block minimizer is zero for every prompt, so $B$ comes entirely from returned proposals. The smallest geometry ratio at the states that start a round is at least 0.6, 0.69, 0.43 and 0.70 in every run, with means between 0.81 and 0.93. With the efficiency paragraph, the per-round recurrence holds in all 72 rounds and ends at a mean gap bound of $267.46$ against an observed mean gap of $23.40$. A better prompt thus raises search quality and changes which states are visited, while the correction errors stay at the same level.

\begin{table*}[t]
\centering\footnotesize
\begin{tabular}{@{}llrrrrr@{}}\toprule
Variant & Appendix & Safe & Final $F$ & Search time (s) & Answers & Reported USD/run\\\midrule
\multicolumn{7}{@{}l}{\emph{Prompt v2, map 1 (main text)}}\\
Single agent, 1 route per answer & \ref{app:routing-revised} & 20/20 & $156.00\pm10.53$ & $613.01\pm397.82$ & $27.80\pm11.67$ & $0.161\pm0.101$\\
Single agent, exactly 3 routes per answer & \ref{app:routing-revised} & 9/10 & $220.80\pm220.38$ & $561.24\pm253.45$ & $20.40\pm9.89$ & $0.147\pm0.065$\\
Best-of-3 one-route single agents & \ref{app:routing-revised} & 19/20 & $162.90\pm48.82$ & $326.41\pm198.10$ & $39.30\pm11.30$ & $0.188\pm0.108$\\
Team (Alg.~\ref{alg:protocol}) & \ref{app:routing-revised} & 20/20 & $149.90\pm2.67$ & $481.04\pm227.32$ & $47.10\pm16.08$ & $0.228\pm0.106$\\
Team (Alg.~\ref{alg:best-combination}) & \ref{app:routing-merge} & 20/20 & $149.40\pm3.22$ & $339.74\pm196.91$ & $36.20\pm11.88$ & $0.181\pm0.074$\\
Team unsafe & \ref{app:routing-merge} & 20/20 & $149.00\pm2.55$ & $601.43\pm284.49$ & $64.55\pm25.93$ & $0.357\pm0.154$\\
Team without correction & \ref{app:routing-revised} & 0/10 & $457.10\pm109.17$ & $326.19\pm171.96$ & $37.00\pm21.42$ & $0.133\pm0.077$\\
Sequential replay of the team & \ref{app:routing-revised} & 20/20 & $149.90\pm2.67$ & $853.95\pm374.23$ & $47.10\pm16.08$ & ---\\
\midrule
\multicolumn{7}{@{}l}{\emph{Prompt v2, map 2}}\\
Single agent, 1 route per answer & \ref{app:routing-map2} & 19/20 & $166.50\pm44.38$ & $537.91\pm365.71$ & $24.50\pm11.25$ & $0.137\pm0.093$\\
Best-of-3 one-route single agents & \ref{app:routing-map2} & 18/20 & $168.55\pm33.92$ & $326.29\pm147.01$ & $38.70\pm9.58$ & $0.195\pm0.089$\\
Team (Alg.~\ref{alg:protocol}) & \ref{app:routing-map2} & 20/20 & $154.15\pm12.56$ & $482.26\pm201.58$ & $44.75\pm16.06$ & $0.221\pm0.093$\\
Team (Alg.~\ref{alg:best-combination}) & \ref{app:routing-merge} & 20/20 & $159.75\pm17.21$ & $310.44\pm204.06$ & $33.00\pm15.51$ & $0.174\pm0.135$\\
Team unsafe & \ref{app:routing-merge} & 19/20 & $157.05\pm22.26$ & $423.66\pm232.41$ & $47.95\pm22.63$ & $0.235\pm0.130$\\
Team without correction & \ref{app:routing-map2} & 2/10 & $328.40\pm153.41$ & $178.09\pm87.24$ & $19.00\pm7.18$ & $0.069\pm0.031$\\
Sequential replay of the team & \ref{app:routing-map2} & 20/20 & $154.15\pm12.56$ & $778.16\pm304.62$ & $44.75\pm16.06$ & ---\\\midrule
 
\multicolumn{7}{@{}l}{\emph{Prompt v2, two floors (eighteen robots)}}\\
Single agent, 1 route per answer & \ref{app:routing-two-floors} & 20/20 & $319.60\pm23.31$ & $1148.67\pm419.61$ & $63.20\pm16.16$ & $0.365\pm0.127$\\
Best-of-6 one-route single agents & \ref{app:routing-two-floors} & 20/20 & $312.40\pm16.26$ & $516.84\pm123.87$ & $142.50\pm17.07$ & $0.680\pm0.154$\\
Team (Alg.~\ref{alg:protocol}) & \ref{app:routing-two-floors} & 20/20 & $304.05\pm12.73$ & $656.37\pm206.11$ & $91.85\pm20.73$ & $0.449\pm0.141$\\
Team (Alg.~\ref{alg:best-combination}) & \ref{app:routing-merge} & 20/20 & $309.15\pm17.39$ & $451.64\pm227.10$ & $69.20\pm19.77$ & $0.355\pm0.158$\\
Team unsafe & \ref{app:routing-merge} & 19/20 & $306.05\pm22.54$ & $716.82\pm262.97$ & $112.50\pm28.78$ & $0.592\pm0.172$\\
Team without correction & \ref{app:routing-two-floors} & 0/10 & $785.50\pm181.90$ & $382.05\pm176.14$ & $56.00\pm19.90$ & $0.202\pm0.079$\\
Sequential replay of the team & \ref{app:routing-two-floors} & 20/20 & $304.05\pm12.73$ & $1632.10\pm468.52$ & $91.85\pm20.73$ & ---\\\midrule
 
\multicolumn{7}{@{}l}{\emph{Prompt v1, map 1}}\\
Single agent, whole plan, patience 5 & \ref{app:routing-development} & 1/10 & $841.00\pm483.14$ & $281.36\pm111.67$ & $8.30\pm2.71$ & $0.071\pm0.027$\\
Single agent, whole plan, efficiency paragraph & \ref{app:routing-development} & 1/10 & $860.00\pm555.40$ & $728.52\pm366.76$ & $26.40\pm15.54$ & $0.222\pm0.111$\\
Single agent, up to 3 routes per answer & \ref{app:routing-development} & 8/10 & $180.60\pm45.75$ & $559.08\pm325.35$ & $24.00\pm11.37$ & $0.152\pm0.087$\\
Single agent, up to 2 routes per answer & \ref{app:routing-development} & 7/10 & $188.20\pm54.87$ & $489.47\pm228.26$ & $24.60\pm10.13$ & $0.136\pm0.066$\\
Single agent, 1 route per answer & \ref{app:routing-development} & 9/10 & $166.80\pm37.60$ & $761.71\pm456.76$ & $40.30\pm20.63$ & $0.208\pm0.125$\\\midrule
Team (Alg.~\ref{alg:protocol}) & \ref{app:routing-development} & 10/10 & $156.50\pm5.60$ & $589.39\pm64.85$ & $59.30\pm7.92$ & $0.285\pm0.042$\\
Team (Alg.~\ref{alg:protocol}), efficiency paragraph & \ref{app:routing-development} & 9/10 & $167.40\pm34.15$ & $436.65\pm90.79$ & $51.80\pm10.79$ & $0.232\pm0.060$\\
Team (Alg.~\ref{alg:protocol}), partial updates & \ref{app:routing-development} & 7/10 & $182.40\pm48.08$ & $310.40\pm94.67$ & $44.70\pm9.86$ & $0.169\pm0.051$\\
Team without correction & \ref{app:routing-development} & 1/10 & $392.90\pm120.66$ & $337.97\pm95.34$ & $39.80\pm10.37$ & $0.132\pm0.037$\\
Sequential replay of the team & \ref{app:routing-development} & 10/10 & $156.50\pm5.60$ & $1207.14\pm165.61$ & $59.30\pm7.92$ & ---\\\bottomrule
\end{tabular}
\caption{All routing variants, with prompt v2 of the main text and with prompt v1; twenty runs for the v2 one-route single agent, its best-of-three (best-of-six on two floors), the team with each of its three merge rules and its sequential replay on each task, ten for every other variant; mean $\pm$ sample SD over all runs, including unsafe ones. Safe: all robots delivered without collisions in the final plan. Search time ends at the stopping rule; for an unsafe run it is not a time to a solution. Answers counts checked answers per run, summed over agents for teams. Reported costs omit charges that the provider did not return. Each sequential replay reuses its team's recorded answers, so it has no new cost.}
\label{tab:routing-variants}
\end{table*}

\subsection{The protocols of the main text}
Three observations explain the choice of protocols in the main text. First, a single agent that rewrites all nine routes in every answer rarely reaches a safe plan, with or without the efficiency paragraph and longer patience. Letting it update only some robots per answer, while it still sees and scores the whole plan, raises safe completion to seven or eight runs of ten with up to two or three robots, and to nine with exactly one; with prompt v2, one route per answer is safe in all twenty runs and exactly three routes per answer in nine of ten. The one-route agent is therefore the single-agent baseline. Second, with prompt v1 a partial-update option makes the team faster but less reliable, with three runs ending in a collision, two of them between robots of the same agent; the team that rewrites all its routes is kept for the main text. The efficiency paragraph shortens the v1 team's search (437 versus 589 seconds) but leaves one run unsafe and raises its mean loss to 167.4. Third, prompt v2 shortens the search of both the one-route single agent and the team, and the team without correction is rarely safe with either prompt. With prompt v2, three one-route single agents searching in parallel are faster than the team with Algorithm~\ref{alg:protocol} on both maps, while the team ends with the best plans. On two floors, six such agents are faster than the team in mean but not in median, the single agent needs about twice the team's search time, and the team again ends with the best plans. The main text therefore uses prompt v2 with the one-route single agent, its best-of-$W$ as the parallelism-matched control, the team that rewrites all three of an agent's routes, with Algorithm~\ref{alg:protocol} and with the best-combination and unsafe merges, and the team without correction; the sequential replay is reported in Appendix~\ref{app:routing-decomposition}.

\section{Routing results with prompt v2}\label{app:routing-revised}
This appendix reports every run behind Figure~\ref{fig:routing-main-progress} and Table~\ref{tab:routing-main-results}, all with prompt v2 (Appendix~\ref{app:routing-development}): on the first map, twenty runs each for the single agent with one route per answer, the best-of-three such agents and the team, and ten each for the single agent with exactly three routes per answer and the team without correction; on the second map, twenty each for the single agent, the best-of-three and the team, and ten for the team without correction; on two floors, twenty each for the single agent and the best-of-six, with the team variants evaluated on the runs of both maps; and for two other merge rules of the team, best combination and unsafe, twenty runs each on both maps. It then shows recorded collision repairs, audits the assumptions of Theorem~\ref{thm:rate} on the team runs, and evaluates the trajectory-wise bound.

\subsection{Protocols and measurement}
The team protocol is that of Appendix~\ref{app:routing-task}: three agents with fixed blocks, up to $H=3$ answers per agent and round on a frozen base, one accepted block per round, and the certificate-based early stop of a private search. The team stops once the plan is safe and within 9 of the optimum ($F\leq153$ on the first map), after five rounds without an improving commit (fifteen answers per agent, as many as the single agent), or after thirty rounds. The single agent stops once the plan is safe and within the threshold, after fifteen consecutive answers without improvement, or after 120 answers; it either replaces one route per answer or returns complete programs for exactly three robots of its choice per answer, the size of a team agent's answer (a program may be repeated unchanged). An answer with another number of robots is a format failure. The best-of-three runs three copies of the one-route single agent in parallel: every round, all copies receive the same prompt with their own seeds (copy $c$ in round $r$ adds $1000r+c$), the best strictly improving answer is accepted and the other two are discarded; if no answer improves, all three enter the list of rejected attempts. Its round time is the slowest of the three answers, as for the team, and it stops once the plan is safe and within the threshold, after five rounds without improvement (fifteen answers) or after forty rounds.

Model, provider and sampling settings are those of Appendix~\ref{app:routing-measurement}. Replicate $s$ uses base seed $4{,}000{,}000+100{,}000s$; the team adds $1000k+100h+b$, and the single agent adds $1000j$ for its $j$-th answer. Format failures are re-asked with a deterministic new seed, at most three times, and transport failures at most twenty times; neither counts toward search time. Fourteen times in about 5,400 requests on the two maps, the provider stopped an answer at its 65,536-token output limit after the visible answer had degenerated into repetition; each such answer was kept, counted as a format failure, and re-asked once with the next retry seed. Search time is defined by~\eqref{eq:routing-clock}; the sequential replay sums the three agents' times in every round instead of taking the maximum. Every run was replayed independently from the recorded answers: proposals, block selection, accepted plans, losses, known pairs and clocks.

\subsection{Results on the first map}\label{app:routing-revised-results}
Table~\ref{tab:routing-revised-seeds} lists every run; Table~\ref{tab:routing-variants} gives means and costs. Seeds 10--19 of the single agent with one route and of the team were run after seeds 0--9 with byte-identical code and settings; the best-of-three ran all twenty seeds with the same code.
\begin{itemize}
\item \textbf{Single agent, one route per answer:} safe in all twenty runs, sixteen of which stop at $F\leq153$; final loss $156.00\pm10.53$ (median 153); search time $613.0\pm397.8$ seconds (median 448) with $27.8\pm11.7$ answers. Fifteen runs finish within 900 seconds; five need 955 to 1,532 seconds, three of which stop above the threshold, at 157, 160 and 191; the fourth run above the threshold stops at 180 after 855 seconds.
\item \textbf{Best-of-three one-route single agents:} safe in nineteen runs, sixteen of which stop at $F\leq153$; final loss $162.90\pm48.82$ (median 152); search time $326.4\pm198.1$ seconds (median 256) over $13.1\pm3.8$ rounds with $39.3\pm11.3$ answers. Seed 10 stops at $F=370$ with two collisions; seeds 0, 11 and 12 stop at 154 to 158 after 587 to 934 seconds.
\item \textbf{Single agent, exactly three routes per answer (ten runs):} safe in nine runs, all of which stop at $F\leq153$; final loss $220.80\pm220.38$ (median 152); search time $561.2\pm253.5$ seconds with $20.4\pm9.9$ answers. In seed 2 the agent stopped at $F=848$, with four robots undelivered and one collision, after fifteen answers without improvement, eight of them geometrically invalid. Overall 37\% of its answers are invalid, against 17\% for the one-route agent, because one invalid route rejects the whole three-route answer.
\item \textbf{Team:} safe in all twenty runs, all of which stop at $F\leq153$; final loss $149.90\pm2.67$ (median 150.5); search time $481.0\pm227.3$ seconds (median 440; two runs take longer than 700 seconds, at most 987) over $7.3\pm2.7$ rounds with $47.1\pm16.1$ answers. Seed 19 reaches the optimum $F=144$; the two longest runs need 12 and 13 rounds.
\item \textbf{Team without correction (ten runs):} no run is safe; final loss $457.10\pm109.17$, search time $326.2\pm172.0$ seconds over $10.8$ rounds on average. Every run delivers all nine robots and keeps one to five collisions, all between robots of different agents. Six runs end at a fixed point: every agent's own robots take shortest routes without colliding with each other, so all three blocks are certified at the start of a round, no model is called, and every later round would repeat the plan. The other four stop after five rounds without improvement, still answering.
\item \textbf{Sequential replay of the team:} $853.9\pm374.2$ seconds against $481.0\pm227.3$ in parallel, a ratio of means of 1.78; answers, selections and losses are unchanged.
\end{itemize}

\begin{table*}[t]\centering\footnotesize
\setlength{\tabcolsep}{3.4pt}
\begin{tabular}{r rrr rr rrr rrr rr}\toprule
& \multicolumn{3}{c}{Single agent, 1 route} & \multicolumn{2}{c}{Best-of-3} & \multicolumn{3}{c}{Single agent, exactly 3 routes} & \multicolumn{3}{c}{Team (Alg.~\ref{alg:protocol})} & \multicolumn{2}{c}{Team, no correction}\\
\cmidrule(lr){2-4}\cmidrule(lr){5-6}\cmidrule(lr){7-9}\cmidrule(lr){10-12}\cmidrule(l){13-14}
Seed & $F$ & Ans. & Time (s) & $F$ & Time (s) & $F$ & Ans. & Time (s) & $F$ & Rounds & Time (s) & $F$ & Time (s)\\\midrule
0 & 148 & 20 & 318.9 & 157 & 673.7 & 153 & 10 & 280.6 & 145 & 6 & 304.2 & 444$^\dagger$ & 209.3\\
1 & 153 & 19 & 276.9 & 151 & 250.8 & 151 & 26 & 554.3 & 152 & 7 & 542.3 & 452$^\dagger$ & 533.2\\
2 & 153 & 20 & 323.4 & 152 & 195.3 & 848$^\dagger$ & 20 & 448.2 & 152 & 5 & 253.0 & 546$^\dagger$ & 357.1\\
3 & 151 & 14 & 209.8 & 152 & 213.4 & 153 & 14 & 504.3 & 147 & 11 & 666.7 & 444$^\dagger$ & 163.7\\
4 & 150 & 22 & 334.7 & 148 & 156.3 & 148 & 13 & 267.0 & 150 & 12 & 987.2 & 444$^\dagger$ & 270.8\\
5 & 152 & 17 & 260.1 & 153 & 449.7 & 149 & 24 & 477.0 & 145 & 13 & 911.9 & 444$^\dagger$ & 183.0\\
6 & 191 & 48 & 1106.0 & 152 & 190.5 & 152 & 16 & 769.4 & 152 & 6 & 386.5 & 361$^\dagger$ & 290.4\\
7 & 153 & 47 & 1080.7 & 148 & 279.5 & 149 & 43 & 1124.9 & 150 & 3 & 121.5 & 244$^\dagger$ & 125.0\\
8 & 157 & 34 & 955.1 & 152 & 363.3 & 152 & 12 & 478.3 & 149 & 8 & 625.9 & 544$^\dagger$ & 525.2\\
9 & 153 & 26 & 581.9 & 153 & 212.0 & 153 & 26 & 708.4 & 152 & 9 & 673.7 & 648$^\dagger$ & 604.3\\
10 & 152 & 12 & 144.0 & 370$^\dagger$ & 325.6 & --- & --- & --- & 149 & 6 & 366.0 & --- & ---\\
11 & 153 & 22 & 407.3 & 154 & 586.8 & --- & --- & --- & 151 & 8 & 442.7 & --- & ---\\
12 & 153 & 18 & 312.5 & 158 & 934.2 & --- & --- & --- & 152 & 7 & 353.1 & --- & ---\\
13 & 152 & 36 & 810.1 & 153 & 214.2 & --- & --- & --- & 151 & 4 & 302.9 & --- & ---\\
14 & 160 & 41 & 1257.6 & 152 & 360.2 & --- & --- & --- & 153 & 9 & 638.1 & --- & ---\\
15 & 153 & 20 & 382.2 & 153 & 221.3 & --- & --- & --- & 152 & 8 & 596.0 & --- & ---\\
16 & 153 & 50 & 1532.0 & 148 & 188.4 & --- & --- & --- & 152 & 6 & 438.2 & --- & ---\\
17 & 150 & 31 & 623.4 & 149 & 273.8 & --- & --- & --- & 150 & 5 & 323.3 & --- & ---\\
18 & 153 & 25 & 488.2 & 151 & 177.9 & --- & --- & --- & 150 & 8 & 513.0 & --- & ---\\
19 & 180 & 34 & 855.4 & 152 & 261.4 & --- & --- & --- & 144 & 4 & 174.7 & --- & ---\\\bottomrule
\end{tabular}

\caption{Every run on the first map with prompt v2, independently replayed; --- marks a seed that was not run for that variant. $^\dagger$ marks a final plan that is not safe. Ans.: checked answers; team times use the per-round maximum~\eqref{eq:routing-clock}.}
\label{tab:routing-revised-seeds}
\end{table*}

\paragraph{Statistical comparison.}
Two-sided permutation tests over the twenty runs of each protocol compare final losses and search times. The team's final loss is lower than the single agent's (difference of means 6.1, $p<0.001$; of medians 2.5, $p=0.005$); on every seed the single agent ends at least as high as the team, and higher on eighteen. Against the best-of-three, the difference of means is 13.0 ($p=0.011$) and of medians 1.5 ($p=0.09$). The team reaches the threshold in all twenty runs, the two single-agent protocols in sixteen each ($p=0.11$, Fisher's exact test). In search time, the single agent and the team have the same median (difference 7 seconds, $p=0.97$), the single agent's mean is higher by 132 seconds ($p=0.21$) and its spread is larger (standard deviation 1.75 times the team's, $p=0.044$). The best-of-three is faster than the team (difference of medians 184 seconds, $p=0.011$; of means 155 seconds, $p=0.027$), with a similar spread ($p=0.62$); it is faster on fifteen of the twenty seeds.

\subsection{A second map}\label{app:routing-map2}
The second map (Figure~\ref{fig:routing-maps}, Appendix~\ref{app:routing-task}) is run with the code, prompts, model, settings, stopping rules and seed streams of the first map: twenty runs each for the one-route single agent, its best-of-three and the team, and ten for the team without correction. Its threshold is $F\leq156$, nine above the optimum $F^*=147$.

\begin{table*}[t]\centering\footnotesize
\setlength{\tabcolsep}{3.4pt}
\begin{tabular}{r rrr rr rrr rr}\toprule
& \multicolumn{3}{c}{Single agent} & \multicolumn{2}{c}{Best-of-3} & \multicolumn{3}{c}{Team (Alg.~\ref{alg:protocol})} & \multicolumn{2}{c}{Team, no correction}\\
\cmidrule(lr){2-4}\cmidrule(lr){5-6}\cmidrule(lr){7-9}\cmidrule(l){10-11}
Seed & $F$ & Ans. & Time (s) & $F$ & Time (s) & $F$ & Rounds & Time (s) & $F$ & Time (s)\\\midrule
0 & 156 & 20 & 458.2 & 153 & 173.3 & 151 & 12 & 916.2 & 447$^\dagger$ & 93.0\\
1 & 154 & 24 & 526.9 & 152 & 203.3 & 148 & 5 & 341.1 & 247$^\dagger$ & 89.4\\
2 & 155 & 14 & 278.0 & 156 & 409.8 & 155 & 7 & 396.9 & 347$^\dagger$ & 307.1\\
3 & 153 & 45 & 1236.5 & 185 & 476.3 & 151 & 5 & 260.7 & 647$^\dagger$ & 68.0\\
4 & 156 & 20 & 324.7 & 156 & 379.3 & 155 & 6 & 367.6 & 447$^\dagger$ & 194.0\\
5 & 154 & 19 & 293.9 & 150 & 199.0 & 148 & 4 & 180.0 & 152 & 218.2\\
6 & 154 & 15 & 222.6 & 156 & 218.8 & 152 & 7 & 384.1 & 247$^\dagger$ & 114.1\\
7 & 155 & 41 & 746.1 & 156 & 212.0 & 149 & 9 & 434.1 & 251$^\dagger$ & 304.9\\
8 & 161 & 43 & 1224.3 & 149 & 243.4 & 206 & 10 & 625.8 & 347$^\dagger$ & 164.9\\
9 & 156 & 33 & 838.8 & 162 & 701.0 & 148 & 8 & 651.4 & 152 & 227.3\\
10 & 155 & 17 & 322.5 & 152 & 257.8 & 147 & 7 & 310.1 & --- & ---\\
11 & 150 & 20 & 372.1 & 155 & 185.0 & 156 & 8 & 524.2 & --- & ---\\
12 & 154 & 17 & 271.7 & 156 & 197.5 & 151 & 8 & 521.9 & --- & ---\\
13 & 353$^\dagger$ & 31 & 681.4 & 269$^\dagger$ & 452.9 & 151 & 6 & 344.5 & --- & ---\\
14 & 152 & 15 & 258.5 & 153 & 286.3 & 149 & 13 & 823.9 & --- & ---\\
15 & 156 & 12 & 178.7 & 191 & 456.7 & 156 & 5 & 290.9 & --- & ---\\
16 & 167 & 35 & 1178.9 & 155 & 326.7 & 155 & 10 & 638.1 & --- & ---\\
17 & 153 & 14 & 197.6 & 153 & 179.9 & 148 & 10 & 518.9 & --- & ---\\
18 & 156 & 14 & 207.5 & 156 & 433.0 & 154 & 12 & 780.6 & --- & ---\\
19 & 180 & 41 & 939.5 & 256$^\dagger$ & 533.7 & 153 & 7 & 334.2 & --- & ---\\\bottomrule
\end{tabular}

\caption{Every run on the second map with prompt v2, independently replayed. $^\dagger$ marks a final plan that is not safe. Ans.: checked answers; team times use the per-round maximum~\eqref{eq:routing-clock}.}
\label{tab:routing-map2-seeds}
\end{table*}

\begin{table*}[t]\centering\footnotesize
\begin{tabular}{@{}lrrrrrr@{}}\toprule
Variant & Safe & $F\leq156$ & Final $F$ & Search time (s) & Answers & USD/run\\\midrule
Single agent, 1 route per answer & 19/20 & 16/20 & $166.50\pm44.38$ & $537.9\pm365.7$ & $24.5\pm11.2$ & $0.137\pm0.093$\\
Best-of-3 one-route single agents & 18/20 & 15/20 & $168.55\pm33.92$ & $326.3\pm147.0$ & $38.7\pm9.6$ & $0.195\pm0.089$\\
Team (Alg.~\ref{alg:protocol}) & 20/20 & 19/20 & $154.15\pm12.56$ & $482.3\pm201.6$ & $44.8\pm16.1$ & $0.221\pm0.093$\\
Team without correction & 2/10 & 2/10 & $328.40\pm153.41$ & $178.1\pm87.2$ & $19.0\pm7.2$ & $0.069\pm0.031$\\
Sequential replay of the team & 20/20 & 19/20 & $154.15\pm12.56$ & $778.2\pm304.6$ & $44.8\pm16.1$ & ---\\\bottomrule
\end{tabular}
\caption{All runs on the second map; mean $\pm$ sample SD over all runs, including unsafe ones.}
\label{tab:routing-map2}
\end{table*}

Tables~\ref{tab:routing-map2} and~\ref{tab:routing-map2-seeds} and Figure~\ref{fig:routing-main-progress}(b) summarize the runs. The team is safe in all twenty runs and reaches the threshold in nineteen; the remaining run stops at $F=206$, and seed 10 reaches the optimum. Its final loss has median 151, lower than the single agent's 155 and the best-of-three's 156 (permutation tests on the medians, $p=0.002$ and $p=0.001$); on eighteen of twenty seeds it ends at least as low as each of them. The means differ less clearly ($p=0.26$ and $p=0.08$) because one single-agent run stops at $F=353$ with two collisions and two best-of-three runs stop with one collision each. Only the team's mean gap falls below 10, at about 520 seconds. In search time, the best-of-three is fastest (median 272 against 416 seconds for the team, $p=0.02$); the single agent and the team do not differ significantly in median or mean, but the single agent's spread is larger (standard deviation 1.8 times the team's, $p=0.03$), with six runs longer than 700 seconds against three. Played in turn, the team's answers would take $778\pm305$ seconds instead of $482\pm202$. Without correction, eight of ten runs end with collisions between agents. Seven of them end at the fixed point described for the first map, with no model calls in their last five rounds; the eighth also repeats itself, since in each of its last five rounds one agent returns routes that are optimal for its own robots but collide with other agents' robots, and they are rejected. This is why the uncorrected curve stops after about 310 seconds: more rounds would add no answers, or only rejected ones.

\subsection{Two floors}\label{app:routing-two-floors}
The two-floor task (Appendix~\ref{app:routing-task}) places the first map (robots r0--r8) and the second map (robots renamed s0--s8) on the two floors of one building; both instances are copied unchanged from the single-floor campaigns apart from the renaming. Robots on different floors never collide, so $F=F_1+F_2$ with optimum $F^*=291$ and all-waiting value 2955. A plan meets the threshold when both floors are safe with $F_1\leq153$ and $F_2\leq156$.

\paragraph{Team variants.} The team has six agents with the blocks of the single-floor runs. Its dependency graph consists of two disjoint triangles; each color class holds one block of each floor, and the coordinator commits, in each round, the best improving block of each floor, if any. This is Algorithm~\ref{alg:protocol} with the pairing of the two floors' blocks into colors chosen anew in each round: any such pairing is a proper three-coloring, and the committed gain is at least that of the best color of any fixed coloring, so the analysis applies with $\kappa=3$. Each agent's prompt shows only its own floor, exactly as in the single-floor runs. A two-floor team run is therefore a run on each floor played in lockstep: round $k$ lasts $\max(T_{1,k},T_{2,k})$, where $T_{f,k}$ is the slowest agent's search on floor $f$ in that round, and a floor that has met its threshold or gone five rounds without improvement makes no further calls; the run ends when both floors have stopped. The team, its sequential replay, which sums the times of all six agents, and the team without correction are evaluated on the recorded single-floor runs, pairing seed $s$ of the first map with seed $s$ of the second, without new model calls.

\paragraph{Single agent and best-of-six.} The single-agent prompt states the rules once, adding that each floor is its own grid and that robots on different floors never collide, and shows each floor exactly as the single-floor prompt does: the map with its labels and wall list, every robot's program and cost, the timeline and every current collision with its maps. The accepted changes and the rejected attempts are listed once for the whole plan, with the floor of each changed robot; the loss is shown per floor and in total, and the goal names both thresholds. Each answer replaces the route of one robot on either floor, and a change is accepted only if $F$ strictly decreases. The single agent stops once both floors meet their thresholds, after fifteen answers without improvement or after 240 answers; the best-of-six runs six copies of it as the best-of-three does and stops after five rounds without improvement (thirty answers) or after eighty rounds. Neither cap was reached. Seeds, retries and the clock are those of the single-floor runs; the best-of-six ran seeds 0--9 first and seeds 10--19 afterwards with byte-identical code. The provider stopped two of the 4,207 requests at its output limit, and each was re-asked once as before.

\paragraph{Results.} Table~\ref{tab:routing-two-floors} summarizes the variants and Table~\ref{tab:routing-two-floors-seeds} lists every run.
\begin{itemize}
\item \textbf{Single agent:} safe in all twenty runs; nine meet both thresholds, thirteen the first floor's and eleven the second's. Final loss $319.60\pm23.31$ (median 313.5, at most 409); search time $1148.7\pm419.6$ seconds (median 1200; eleven runs take longer than 1000 seconds) with $63.2\pm16.2$ answers. Eleven runs stop after fifteen answers without improvement.
\item \textbf{Best-of-six:} safe in all twenty runs, sixteen of which meet both thresholds; final loss $312.40\pm16.26$ (median 306.5, at most 363); search time $516.8\pm123.9$ seconds (median 511, at most 739) over $23.8$ rounds with $142.5\pm17.1$ answers.
\item \textbf{Team:} safe in all twenty runs, nineteen of which meet both thresholds; final loss $304.05\pm12.73$ (median 302); search time $656.4\pm206.1$ seconds (median 595, at most 1013) over $9.3$ rounds with $91.9\pm20.7$ answers summed over both floors.
\item \textbf{Team without correction (ten runs):} no run is safe; final loss $785.50\pm181.90$, search time $382.1\pm176.1$ seconds.
\item \textbf{Sequential replay of the team:} $1632.1\pm468.5$ seconds against $656.4\pm206.1$ in parallel, a ratio of means of 2.49 (1.78 on the first map, with three agents).
\end{itemize}

\begin{table*}[t]\centering\footnotesize
\begin{tabular}{@{}lrrrrrr@{}}\toprule
Variant & Safe & Both thresholds & Final $F$ & Search time (s) & Answers & USD/run\\\midrule
Single agent, 1 route per answer & 20/20 & 9/20 & $319.60\pm23.31$ & $1148.7\pm419.6$ & $63.2\pm16.2$ & $0.365\pm0.127$\\
Best-of-6 one-route single agents & 20/20 & 16/20 & $312.40\pm16.26$ & $516.8\pm123.9$ & $142.5\pm17.1$ & $0.680\pm0.154$\\
Team (Alg.~\ref{alg:protocol}) & 20/20 & 19/20 & $304.05\pm12.73$ & $656.4\pm206.1$ & $91.9\pm20.7$ & $0.449\pm0.141$\\
Team without correction & 0/10 & 0/10 & $785.50\pm181.90$ & $382.1\pm176.1$ & $56.0\pm19.9$ & $0.202\pm0.079$\\
Sequential replay of the team & 20/20 & 19/20 & $304.05\pm12.73$ & $1632.1\pm468.5$ & $91.9\pm20.7$ & ---\\\bottomrule

\end{tabular}
\caption{All variants on two floors ($F^*=291$); mean $\pm$ sample SD over all runs. Team answers and costs are summed over both floors.}
\label{tab:routing-two-floors}
\end{table*}

\begin{table*}[t]\centering\footnotesize
\setlength{\tabcolsep}{3.4pt}
\begin{tabular}{r rrr rrr rrrr rr}\toprule
& \multicolumn{3}{c}{Single agent} & \multicolumn{3}{c}{Best-of-6} & \multicolumn{4}{c}{Team (Alg.~\ref{alg:protocol})} & \multicolumn{2}{c}{Team, no correction}\\
\cmidrule(lr){2-4}\cmidrule(lr){5-7}\cmidrule(lr){8-11}\cmidrule(l){12-13}
Seed & $F_1{+}F_2$ & Ans. & Time (s) & $F_1{+}F_2$ & Rounds & Time (s) & $F_1{+}F_2$ & Rounds & Time (s) & Seq. (s) & $F_1{+}F_2$ & Time (s)\\\midrule
0 & 150{+}164$^*$ & 69 & 1778.2 & 151{+}178$^*$ & 26 & 676.8 & 145{+}151 & 12 & 916.2 & 1906.9 & 444{+}447$^*$ & 219.5\\
1 & 150{+}155 & 68 & 1518.6 & 151{+}155 & 31 & 739.3 & 152{+}148 & 7 & 578.4 & 1586.1 & 452{+}247$^*$ & 533.2\\
2 & 152{+}165$^*$ & 62 & 1380.5 & 200{+}163$^*$ & 28 & 638.8 & 152{+}155 & 7 & 423.9 & 1186.9 & 546{+}347$^*$ & 435.2\\
3 & 153{+}161$^*$ & 67 & 1356.9 & 152{+}153 & 24 & 491.6 & 147{+}151 & 11 & 723.4 & 1335.1 & 444{+}647$^*$ & 163.7\\
4 & 152{+}156 & 32 & 473.4 & 150{+}153 & 21 & 390.4 & 150{+}155 & 12 & 1012.8 & 2412.9 & 444{+}447$^*$ & 324.2\\
5 & 153{+}165$^*$ & 56 & 1255.8 & 151{+}156 & 21 & 428.9 & 145{+}148 & 13 & 935.2 & 1804.7 & 444{+}152$^*$ & 257.3\\
6 & 152{+}156 & 68 & 912.8 & 148{+}153 & 20 & 319.7 & 152{+}152 & 7 & 452.5 & 1690.2 & 361{+}247$^*$ & 303.9\\
7 & 171{+}157$^*$ & 82 & 1463.3 & 152{+}156 & 26 & 578.7 & 150{+}149 & 9 & 446.9 & 1014.6 & 244{+}251$^*$ & 304.9\\
8 & 153{+}156 & 46 & 659.9 & 152{+}153 & 19 & 344.3 & 149{+}206$^*$ & 10 & 754.6 & 2242.5 & 544{+}347$^*$ & 549.6\\
9 & 151{+}155 & 79 & 1291.7 & 152{+}166$^*$ & 25 & 692.7 & 152{+}148 & 9 & 828.2 & 1957.3 & 648{+}152$^*$ & 729.1\\
10 & 156{+}158$^*$ & 79 & 1576.7 & 152{+}151 & 21 & 381.7 & 149{+}147 & 7 & 423.2 & 1031.3 & --- & ---\\
11 & 151{+}156 & 49 & 789.7 & 153{+}155 & 23 & 496.2 & 151{+}156 & 8 & 570.9 & 1501.8 & --- & ---\\
12 & 158{+}155$^*$ & 91 & 1971.7 & 152{+}155 & 24 & 455.3 & 152{+}151 & 8 & 541.8 & 1462.1 & --- & ---\\
13 & 173{+}162$^*$ & 72 & 1144.8 & 152{+}154 & 21 & 395.8 & 151{+}151 & 6 & 440.5 & 1283.9 & --- & ---\\
14 & 152{+}155 & 44 & 689.9 & 184{+}165$^*$ & 24 & 524.9 & 153{+}149 & 13 & 963.3 & 2131.4 & --- & ---\\
15 & 152{+}156 & 52 & 832.4 & 150{+}155 & 24 & 535.7 & 152{+}156 & 8 & 596.0 & 1693.5 & --- & ---\\
16 & 166{+}174$^*$ & 53 & 903.9 & 152{+}156 & 25 & 655.1 & 152{+}155 & 10 & 731.6 & 1951.8 & --- & ---\\
17 & 225{+}184$^*$ & 49 & 604.1 & 150{+}154 & 24 & 597.5 & 150{+}148 & 10 & 593.4 & 1252.0 & --- & ---\\
18 & 150{+}156 & 55 & 889.0 & 152{+}156 & 23 & 426.5 & 150{+}154 & 12 & 831.6 & 2396.7 & --- & ---\\
19 & 177{+}149$^*$ & 91 & 1480.1 & 152{+}153 & 25 & 566.9 & 144{+}153 & 7 & 362.9 & 800.3 & --- & ---\\\bottomrule
\end{tabular}

\caption{Every run on two floors. $F_1{+}F_2$: final loss of each floor; $^*$ marks a run that misses a threshold. Every final plan of the single agent, the best-of-six and the team is safe; no final plan without correction is. Team runs pair seed $s$ of both maps; Seq.: sequential replay.}
\label{tab:routing-two-floors-seeds}
\end{table*}

\paragraph{Statistical comparison.} We apply the tests used on the first map. Against the single agent, the team's final loss is lower (difference of means 15.6, $p=0.002$; of medians 11.5, $p<0.001$; the single agent ends higher on seventeen seeds and equal on two), the team meets both thresholds more often (19 against 9, $p=0.001$, Fisher's exact test), and it is faster (difference of medians 606 seconds and of means 492 seconds, both $p<0.001$; faster on seventeen seeds), with half the spread (standard deviation ratio 2.04, $p=0.002$). Against the best-of-six, the team's median final loss is lower by 4.5 ($p=0.004$), the means differ by 8.4 ($p=0.09$), and the best-of-six ends higher on sixteen seeds; the threshold counts do not differ significantly (19 against 16, $p=0.34$). The best-of-six is faster in mean (by 140 seconds, $p=0.013$) but not in median (by 84 seconds, $p=0.19$), with a smaller spread (ratio 0.60, $p=0.02$), and it is faster on fifteen seeds.

\paragraph{Bound.} The bound defined in Appendix~\ref{app:routing-empirical-rate} below applies with six blocks and $\kappa=3$: $\alpha_{s,k}$ adds $Q_{s,k,b}D_b$ over all six blocks and divides by $3\Delta_{s,k}$, and $B_{s,k}$ adds the error terms of both floors; a floor that has stopped contributes neither. The one-step inequality~\eqref{eq:routing-recorded-step-rate} holds in all 186 rounds. Propagated per seed from $E_{s,0}=2664$, held after each run stops and averaged, the envelope ends at mean gap 395.9 after thirteen rounds, against 13.1 observed; it stays within a factor of 2.2 of the observed mean gap through round 3 (1.11, 1.46 and 2.15). The pooled means of the coefficients are $\bar\alpha=0.222$ and $\bar B=66.7$.

\subsection{Best-combination and unsafe merges}\label{app:routing-merge}
Algorithm~\ref{alg:protocol} accepts the best color; on one map the dependency graph is complete, so one block per round. Two variants change only this step. With \emph{best combination}, the coordinator checks every combination of the blocks whose proposal differs from the base, at most seven on one map, and accepts the one with the lowest verified loss, or nothing if none improves (the exact branch of Algorithm~\ref{alg:best-combination}). Because the improving blocks of each color are among these combinations, the round gain is at least that of the best color, and Theorem~\ref{thm:rate} applies unchanged (Section~\ref{sec:main-protocol}). The \emph{unsafe} team accepts every changed block together, whatever the loss, and the run returns the best plan it has seen. Everything else is that of the team with Algorithm~\ref{alg:protocol}: code, prompt v2, model, settings, seed streams, the five-round stopping rule (which counts rounds that do not improve the best plan) and the measurement. The prompt states the merge rule in one sentence, so that sentence differs: with Algorithm~\ref{alg:protocol} the coordinator ``installs the best block proposal of the round'', with best combination it ``checks every combination of the block proposals of the round on the full plan and installs the combination that lowers the loss most'', and in the unsafe team it ``installs the block proposals of all agents of the round together''. Collision pairs found in any checked combination become known pairs. Before the runs, an offline test with scripted answers confirmed that the code with the merge rule of Algorithm~\ref{alg:protocol} reproduces the decisions and prompts of the main runs exactly; the independent replay checks every combination, decision and clock. On two floors both variants are composed from the two maps as for Algorithm~\ref{alg:protocol}: the best combination of six blocks on independent floors is the best combination of each floor. The provider stopped 19 of the 3,664 requests at its output limit, eight of them in one run; each was re-asked once.

\begin{table*}[t]\centering\footnotesize
\begin{tabular}{@{}lrrrrrrrr@{}}\toprule
Task and merge rule & Safe & Threshold & Final $F$ & Search time (s) & Sequential (s) & Rounds & Answers & USD/run\\\midrule
Map 1, Team (Alg.~\ref{alg:best-combination}) & 20/20 & 19/20 & $149.40\pm3.22$ & $339.7\pm196.9$ & $601.7\pm255.5$ & 5.7 & $36.2\pm11.9$ & $0.181\pm0.074$\\
Map 1, Team unsafe & 20/20 & 20/20 & $149.00\pm2.55$ & $601.4\pm284.5$ & $1180.2\pm515.9$ & 9.3 & $64.6\pm25.9$ & $0.357\pm0.154$\\
Map 2, Team (Alg.~\ref{alg:best-combination}) & 20/20 & 14/20 & $159.75\pm17.21$ & $310.4\pm204.1$ & $520.0\pm287.0$ & 5.6 & $33.0\pm15.5$ & $0.174\pm0.135$\\
Map 2, Team unsafe & 19/20 & 19/20 & $157.05\pm22.26$ & $423.7\pm232.4$ & $764.2\pm410.3$ & 7.4 & $48.0\pm22.6$ & $0.235\pm0.130$\\
Two floors, Team (Alg.~\ref{alg:best-combination}) & 20/20 & 14/20 & $309.15\pm17.39$ & $451.6\pm227.1$ & $1121.7\pm421.2$ & 7.1 & $69.2\pm19.8$ & $0.355\pm0.158$\\
Two floors, Team unsafe & 19/20 & 19/20 & $306.05\pm22.54$ & $716.8\pm263.0$ & $1944.3\pm562.7$ & 10.6 & $112.5\pm28.8$ & $0.592\pm0.172$\\\bottomrule

\end{tabular}
\caption{The team with the best-combination and the unsafe merge; mean $\pm$ sample SD over twenty runs. The team with Algorithm~\ref{alg:protocol} is in Table~\ref{tab:routing-main-results}. On two floors, answers and costs are summed over both floors.}
\label{tab:routing-merge}
\end{table*}

\begin{table*}[t]\centering\footnotesize
\setlength{\tabcolsep}{3.6pt}
\begin{tabular}{r rrr rr rrr rr}\toprule
& \multicolumn{5}{c}{Map 1} & \multicolumn{5}{c}{Map 2}\\
\cmidrule(lr){2-6}\cmidrule(l){7-11}
& \multicolumn{3}{c}{Team (Alg.~\ref{alg:best-combination})} & \multicolumn{2}{c}{Team unsafe} & \multicolumn{3}{c}{Team (Alg.~\ref{alg:best-combination})} & \multicolumn{2}{c}{Team unsafe}\\
\cmidrule(lr){2-4}\cmidrule(lr){5-6}\cmidrule(lr){7-9}\cmidrule(l){10-11}
Seed & $F$ & Rounds & Time (s) & $F$ & Time (s) & $F$ & Rounds & Time (s) & $F$ & Time (s)\\\midrule
0 & 152 & 3 & 191.1 & 147 & 843.9 & 148 & 2 & 75.3 & 153 & 948.6\\
1 & 152 & 8 & 595.0 & 149 & 646.0 & 157 & 8 & 489.1 & 155 & 319.2\\
2 & 153 & 6 & 385.7 & 150 & 807.0 & 150 & 9 & 502.5 & 153 & 539.6\\
3 & 148 & 5 & 262.6 & 144 & 342.9 & 153 & 4 & 203.9 & 149 & 938.4\\
4 & 148 & 5 & 252.1 & 152 & 420.9 & 152 & 3 & 149.9 & 149 & 365.8\\
5 & 145 & 4 & 197.8 & 150 & 359.3 & 151 & 2 & 75.9 & 149 & 594.7\\
6 & 145 & 5 & 292.5 & 147 & 427.2 & 154 & 4 & 295.7 & 155 & 381.8\\
7 & 144 & 4 & 198.3 & 150 & 507.2 & 153 & 3 & 122.1 & 251$^\dagger$ & 429.5\\
8 & 152 & 3 & 176.6 & 148 & 1263.0 & 192 & 10 & 566.3 & 154 & 157.4\\
9 & 153 & 9 & 565.6 & 148 & 321.4 & 155 & 4 & 248.6 & 152 & 238.9\\
10 & 153 & 4 & 227.7 & 150 & 818.6 & 149 & 2 & 116.2 & 151 & 399.1\\
11 & 149 & 6 & 383.8 & 150 & 670.7 & 155 & 7 & 410.1 & 149 & 181.4\\
12 & 150 & 4 & 230.2 & 146 & 677.8 & 151 & 3 & 159.4 & 155 & 414.5\\
13 & 146 & 4 & 223.7 & 145 & 381.9 & 203 & 13 & 765.8 & 156 & 260.7\\
14 & 154 & 14 & 952.3 & 147 & 352.6 & 161 & 8 & 476.9 & 150 & 531.8\\
15 & 150 & 4 & 216.6 & 148 & 448.0 & 149 & 4 & 149.5 & 152 & 672.2\\
16 & 150 & 9 & 549.0 & 152 & 86.1 & 153 & 5 & 247.5 & 153 & 340.8\\
17 & 144 & 8 & 443.2 & 153 & 978.5 & 149 & 3 & 164.1 & 150 & 427.1\\
18 & 149 & 4 & 262.8 & 151 & 909.5 & 201 & 8 & 344.8 & 156 & 273.0\\
19 & 151 & 4 & 188.2 & 153 & 766.1 & 159 & 9 & 645.2 & 149 & 58.8\\\bottomrule
\end{tabular}

\caption{Every run of the two merge variants, independently replayed. $^\dagger$ marks a final plan that is not safe.}
\label{tab:routing-merge-seeds}
\end{table*}

\paragraph{Results.} Table~\ref{tab:routing-merge} summarizes the runs and Table~\ref{tab:routing-merge-seeds} lists them. With best combination, a round that changes the plan accepts on average 1.67 blocks on map 1 and 1.72 on map 2, and in every run the first round accepts two or three. The team is faster than with Algorithm~\ref{alg:protocol} on every task: on map 1 the search takes $339.7\pm196.9$ instead of $481.0\pm227.3$ seconds (difference of means $p=0.04$, of medians $p=0.03$), on map 2 $310.4\pm204.1$ instead of $482.3\pm201.6$ ($p=0.01$ and $p=0.04$), and on two floors $451.6\pm227.1$ instead of $656.4\pm206.1$ ($p=0.005$ and $p=0.08$), with fewer rounds and answers and USD 0.18 and 0.17 instead of 0.23 and 0.22 per run on the two maps. In median, it also reaches a safe plan first. A collision or an undelivered robot adds at least 100 to the loss, so a gap below 100 implies a safe plan. The median search time until the gap first falls below 100 is 185.1, 154.7 and 223.2 seconds on the three tasks with best combination, against 313.7, 288.1 and 444.7 with Algorithm~\ref{alg:protocol}, 236.1, 207.2 and 362.9 for the best-of-$W$, 329.1, 274.7 and 675.4 for the single agent and 377.6, 256.6 and 425.7 for the unsafe team; a run that never gets there counts as infinitely slow. In permutation tests on the medians, only the best-of-three on map 1 is not significantly slower ($p=0.27$; $p\leq0.007$ for every other comparison). The final losses do not differ significantly from those of Algorithm~\ref{alg:protocol} on any task ($p\geq0.24$), with medians 150, 153 and 302.5 against 150.5, 151 and 302. On map 1 nineteen of twenty runs meet the threshold; on map 2 fourteen do, against nineteen with Algorithm~\ref{alg:protocol} ($p=0.09$, Fisher's exact test), and six runs stop above it, at $F$ between 157 and 203, against one.

\paragraph{Why six runs stall on map 2.} In each of the six, the last five rounds accept nothing although a block can still improve: at the final state the exact corrected block minimum lies 43 to 52 below the current value in three of the runs and 2 to 14 below in the other three, so no run stops at a single-block trap. In five runs the other two blocks are certified optimal for their agents' corrected objectives, so those agents make no calls; in seed 1 two agents keep searching. In four runs (seeds 1, 8, 14 and 18) a searching agent did return a proposal that improved its corrected objective, but the proposal collided with robots of pairs that the correction did not yet know, so verification rejected it and the collisions became known pairs; in seed 14 the agent's next proposal collided with yet another unknown pair, and in seeds 1, 8 and 18 the agent found no further improvement before the stop. In seeds 13 and 19 the agent found no improvement of its corrected objective in fifteen answers. The merge rule would accept any single improving block, so the stalls come from local search with an incomplete correction, not from the merge. These states differ from those of Algorithm~\ref{alg:protocol} because the first rounds differ: after round 1 the plan has 3.3 collisions on average, against 0.1, since two or three agents' routes are accepted together, and it is safe after 2.8 rounds instead of 5.2. Once no collision is left, the realized search quality falls for every merge rule; on map 2 it is 0.28 with best combination (57 of 85 such searches capture nothing) against 0.34 with Algorithm~\ref{alg:protocol} (47 of 87). On map 1 the difference does not occur (one run above the threshold against none).

\paragraph{Unsafe merging.} The unsafe team accepts 2.2 blocks per changing round on map 1 and 2.1 on map 2, but its accepted plan becomes worse in 32 of its 185 rounds on map 1 and 37 of 148 on map 2, when blocks changed together create new collisions; after round 1 its plan has 6.3 collisions on average on map 1 and 3.0 on map 2. Its search time does not differ significantly from that of Algorithm~\ref{alg:protocol} on any task ($601.4$ against $481.0$ seconds on map 1, $423.7$ against $482.3$ on map 2, $716.8$ against $656.4$ on two floors; $p>0.1$), it needs more answers, and its final losses are similar ($p\geq0.18$); one final plan on map 2 keeps a collision. Accepting worse plans did not significantly lower its final loss, and one final state on each map is a single-block trap.

\paragraph{Audit and bound.} The audit of Appendix~\ref{app:routing-assumptions}, defined below, applies to both variants (Table~\ref{tab:routing-constants}). With best combination, the realized search quality averages 0.65 and 0.60 on the two maps, 26\% and 28\% of the locally safe proposals collide with another agent's robot, and the geometry ratio is at least 0.55 and 0.75 at every state that starts a round; four and two final states are single-block traps, all of them final states at which a run had met its threshold and stopped. The one-step inequality~\eqref{eq:routing-recorded-step-rate} holds in all 113 and 111 rounds and in all 142 on two floors, and the averaged envelope ends at mean gap 328.7, 399.4 and 604.7 against 5.4, 12.8 and 18.2 observed (gray curves in Figure~\ref{fig:routing-theory-maps}); coefficients from seeds 0--9 again bound the mean gap of seeds 10--19 in every round, and vice versa. The envelope ends higher than with Algorithm~\ref{alg:protocol} because these runs stop after fewer rounds; Appendix~\ref{app:routing-merge-theory} sharpens it by counting only the conflicts that occur. For the unsafe team the recurrence does not apply, since the loss can increase; its search quality averages 0.63 and 0.68 and its share of colliding locally safe proposals 29\% and 34\%.

\subsection{Recorded collision repairs}\label{sec:routing-recorded-repairs}
\begin{figure*}[t]
\centering
\includegraphics[width=.96\textwidth]{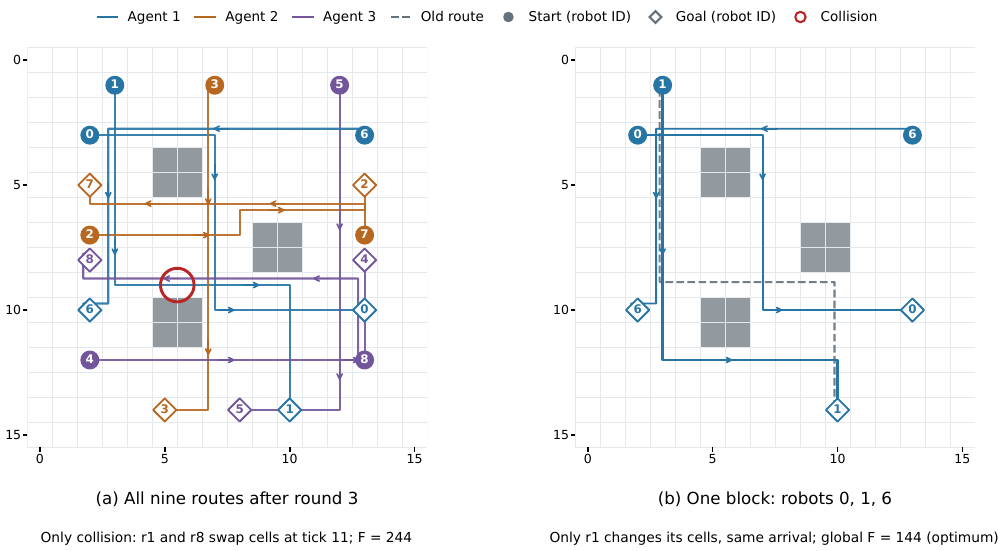}
\Description{Two maps reconstructed from saved commands. Left: all nine routes of team seed nineteen after round three, colored by agent, with a red circle where robots one and eight exchange cells at tick eleven, the only collision of that plan. Right: the three routes owned by the agent that makes the next accepted update; only robot one changes its cells, with its old route dashed in gray and its new route solid, and the same arrival tick.}
\caption{The recorded repair behind Figure~\ref{fig:main-routing-repair} (prompt v2, first map, team seed 19). (a) All nine accepted routes after round 3; colors identify the three agents; filled circles mark starts and open diamonds goals. The red circle marks the only collision of this plan: $r_1$ and $r_8$ exchange cells at tick 11. (b) The block of the agent that owns $r_0,r_1,r_6$ after its accepted answer in round 4: only $r_1$ changes its cells (old route dashed), with the same arrival tick, and the plan reaches the optimum $F=144$. Coincident segments are slightly offset for visibility; all event checks use the original cells and ticks.}
\label{fig:routing-full-plan-detail}
\end{figure*}

The loss curves hide what an accepted change looks like. The plan of Figure~\ref{fig:main-routing-repair} arose from a repair shown in detail in Figure~\ref{fig:routing-full-plan-detail}. In team seed 19 on the first map, the plan accepted after round 3 has $F=244$ and one collision: $r_1$ and $r_8$, which belong to different agents, exchange cells at tick 11. The pair was known from an earlier global check, so in round 4 the agent owning $r_0,r_1,r_6$ saw $r_8$'s current timeline. Its first answer moved $r_1$ to another shortest route with the same arrival tick 20; it also appended waits after arrival to $r_0$ and $r_6$, which leave their cells unchanged. Its local loss did not change, $\Delta L_b=0$, while the correction fell by 100. The verifier accepted the block, and the plan reached $F=144$, the optimum, with all nine robots delivered and no collision. On the first map, 15 of the 121 accepted team rounds remove the last collision between agents in this way, through a pair the agent knew from the correction.

\begin{figure*}[t]
\centering
\includegraphics[width=\textwidth]{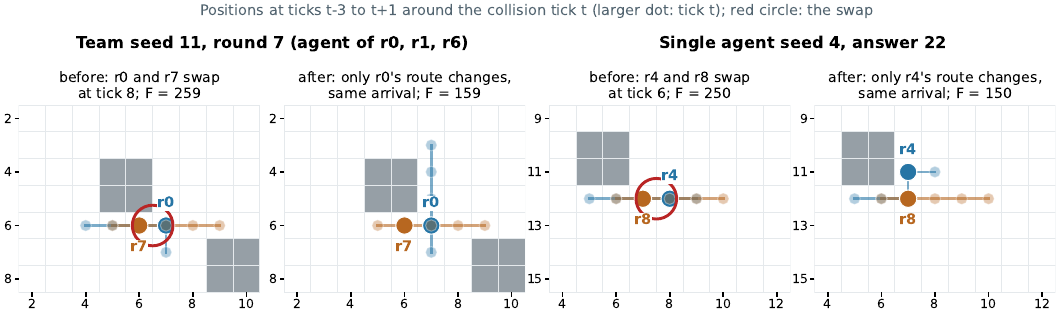}
\Description{Close-up frames around two recorded collisions, before and after the accepted change, showing positions of the two robots over five ticks around the collision. Left: team seed eleven, round seven; robots zero and seven swap cells at tick eight before the change; afterwards robot zero arrives from above and passes behind robot seven. Right: single-agent seed four, answer twenty-two; robots four and eight swap cells at tick six before the change; afterwards robot four turns up and lets robot eight pass.}
\caption{Two more recorded repairs with prompt v2 on the first map, as positions of the two robots at ticks $t-3$ to $t+1$ around the collision tick $t$ (larger dot: tick $t$; red circle: the swap). Left: team seed 11, round 7; the agent of $r_0,r_1,r_6$ knows the swap of $r_0$ with $r_7$ from the correction, and only $r_0$'s route changes, with the same arrival tick ($F$ from 259 to 159). Right: single agent seed 4, answer 22; only $r_4$'s route changes, with the same arrival tick, and the swap with $r_8$ disappears ($F$ from 250 to 150). Python replays the recorded answers without editing them.}
\label{fig:routing-recorded-repairs}
\end{figure*}

Figure~\ref{fig:routing-recorded-repairs} shows two further repairs at the collision tick. In team seed 11, round 7, the agent of $r_0,r_1,r_6$ knew the swap of $r_0$ with $r_7$ at tick 8; its third answer changed only $r_0$'s route, kept its arrival tick 20 (local change 0, correction change $-100$) and removed the plan's last collision ($F$ from 259 to 159; the run ends at 151). The single agent repairs the same kind of conflict without a correction, because it sees every robot: in seed 4, its 22nd answer replaced $r_4$'s route, kept its arrival tick 15 and removed the swap of $r_4$ with $r_8$ at tick 6 ($F$ from 250 to 150, the run's final plan). In all three cases the paths still cross in space but no longer at the same tick, so spatial intersections alone do not determine safety.

\subsection{Audit of the convergence assumptions}\label{app:routing-assumptions}\label{app:routing-revised-audit}
The team runs allow a direct comparison with Assumptions~\ref{ass:correction}--\ref{ass:eb}. Using the exact references of Appendix~\ref{app:routing-references}, an offline audit computes, at every recorded base of every team run, the exact corrected and true block minima. On the first map it covers 145 global rounds, 435 returned block proposals and 942 checked answers (160 of them geometrically invalid) at 165 state occurrences, one of which is optimal; on the second map, 159 rounds, 477 proposals and 895 checked answers (274 invalid) at 179 state occurrences. It includes unselected proposals, searches that return the incumbent and route-invalid answers. Table~\ref{tab:routing-constants} compares the measured constants across prompts and tasks.

\paragraph{Coupling.} Of the 382 returned proposals that deliver all three of an agent's robots without internal collisions, 115 (30\%) still collide with another agent's robot on the first map, and 147 of 461 (32\%) on the second; with prompt v1 the share was 83 of 211 (39\%). Local delivery and internal safety therefore do not suffice for global compatibility. We count simultaneous vertex or edge-swap events, not intersections of spatial route drawings.

\paragraph{Local search quality: Assumption~\ref{ass:local}.}
For each returned block $U_{kb}$ with a positive initial surrogate gap, compute
\begin{equation}\label{eq:routing-audit-Q}
Q_{kb}=\frac{-\widehat f_b(U_{kb};x^k)}{-\min_u \widehat f_b(u;x^k)}\in[0,1].
\end{equation}
The unchanged block is a retained option, so this is the fraction of available corrected local improvement captured by the entire $H=3$ search. It is not the quality of only the winning block or the fraction of the global gap closed. Early termination at a certified optimum can be viewed as leaving that optimum unchanged for the remaining private steps. A route-invalid answer consumes a private step and leaves the best score unchanged; all legal answers are replayed, including worse ones.

\begin{figure*}[t]
\centering
\includegraphics[width=.92\textwidth]{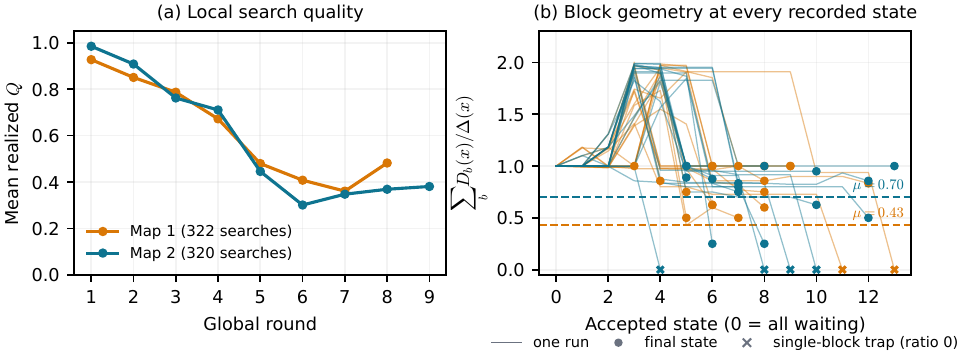}
\Description{Left: mean realized search quality by global round for both maps; it falls from about 0.93 and 0.99 in round one to between 0.3 and 0.5 after round five. Right: the ratio of the summed block improvements to the global gap at every recorded state, one thin line per run and colored by map, with dashed lines at 0.43 for map 1 and 0.70 for map 2; dots mark final states and crosses the six single-block traps at ratio zero.}
\caption{Offline assumption diagnostics for the team with prompt v2. (a) Mean realized $Q$ over searches with a positive corrected gap, in the rounds with at least ten such searches. (b) Exact ratio $\sum_bD_b(x)/\Delta(x)$ at every recorded state, one line per run; dashed lines mark the smallest value over the states that start a round, $\mu=0.43$ on map 1 and $0.70$ on map 2; dots mark final states, crosses the single-block traps (ratio 0).}
\label{fig:routing-assumption-diagnostics}
\end{figure*}

There are 343 searches with a positive corrected gap on the first map and 338 on the second; the others start at a surrogate optimum and are reported separately rather than assigned $Q=1$. The pooled mean is 0.66 on the first map and 0.65 on the second, with 77 and 87 zero-gain searches and 82 and 111 that reach the exact corrected optimum (Table~\ref{tab:routing-audit-Q}); weighting by the initial local gap gives captured fractions of 0.82 and 0.85. The mean falls from 0.93 and 0.99 in round 1 to below 0.5 from round 5 on in the rounds with at least ten searches (Figure~\ref{fig:routing-assumption-diagnostics}(a)). This compares different histories, and later rounds exclude runs that have stopped. Assumption~\ref{ass:local} requires a lower bound on the \emph{conditional expected} fraction at every history, with $q_b(H)=1-(1-\rho_b)^H$, so Appendix~\ref{app:routing-empirical-rate} uses each round's realized $Q$ in the pathwise recurrence rather than a pooled constant.

\begin{table}[t]
\centering\footnotesize
\setlength{\tabcolsep}{2.8pt}
\begin{tabular}{@{}l rrrr rrrr@{}}\toprule
& \multicolumn{4}{c}{Map 1} & \multicolumn{4}{c}{Map 2}\\
\cmidrule(lr){2-5}\cmidrule(l){6-9}
Block & $n$ & Mean $Q$ & $Q=0$ & $Q=1$ & $n$ & Mean $Q$ & $Q=0$ & $Q=1$\\\midrule
Agent 1 & 125 & 0.67 & 26 & 40 & 79 & 0.72 & 18 & 31\\
Agent 2 & 114 & 0.63 & 28 & 16 & 115 & 0.73 & 23 & 58\\
Agent 3 & 104 & 0.68 & 23 & 26 & 144 & 0.56 & 46 & 22\\
\midrule All & 343 & 0.66 & 77 & 82 & 338 & 0.65 & 87 & 111\\\bottomrule
\end{tabular}

\caption{Realized search quality $Q$ after each private search of the team with prompt v2, including unselected proposals; searches with zero initial corrected gap are excluded. Agents 1, 2 and 3 own $\{r_0,r_1,r_6\}$, $\{r_2,r_3,r_7\}$ and $\{r_4,r_5,r_8\}$ on map 1 and $\{r_0,r_1,r_2\}$, $\{r_3,r_4,r_5\}$ and $\{r_6,r_7,r_8\}$ on map 2.}
\label{tab:routing-audit-Q}
\end{table}

\paragraph{Correction accuracy: Assumption~\ref{ass:correction}.}
In the theory, the retained correction~\eqref{eq:routing-correction-main} is $\widehat R_b$; the true correction also includes the undiscovered external pairs. Let $C_{b,\rm miss}^k(u)$ count conflicts with those pairs against the frozen peers. The exact score error is
\begin{equation}\label{eq:routing-audit-error}
\begin{split}
\varepsilon_{kb}(u)&=\widehat f_b(u;x^k)-f_b(u;x^k)\\
&=-100\big[C_{b,\rm miss}^k(u)-C_{b,\rm miss}^k(x_b^k)\big].
\end{split}
\end{equation}
Both corrections are anchored at zero. At every recorded base, the missing-pair conflict count is zero; any existing conflict has already been discovered. Hence a nonzero error at these bases is optimistic: a missing conflict makes the private score understate the true global change.

Among the 782 legal answers on the first map, the mean absolute error is 94.5 and the maximum 3,500; among the 435 returned proposals, 106 have a nonzero error, with mean absolute error 45.5 and maximum 500. On the second map the 621 legal answers have mean absolute error 53.3 (maximum 600), and 114 of the 477 returned proposals have a nonzero error, with mean 36.5 (maximum 600). In 28 and 40 searches the private score improves without a positive verified global gain, which distinguishes local search progress from progress after the correction and the global check.

These observed errors differ from a uniform bound $\delta_b$, which takes a supremum over \emph{all} legal block programs, including poor collision-seeking routes; such a bound implies Assumption~\ref{ass:correction} but is not needed for it. For each recorded state and block we compute that supremum exactly. A time-expanded min/max dynamic program counts undiscovered external conflicts for each owned robot; because those counts add across robots, their extrema add over the product domain. Obstacles and absorbing goals remain enforced. Internal conflicts are penalized in $F$, rather than prohibited in the candidate domain. Independent replay verifies the extremizing joint programs.

Over the recorded states and blocks, these exact suprema average 12,933 on the first map (from 11,400 to 14,000) and 12,518 on the second (from 10,900 to 13,600). A deterministic bound covering \emph{any} history is $\delta_b=86{,}400$: a block has at most $3\cdot6=18$ external pairs, each contributing at most 48 events because starts are distinct, and an anchored count difference is at most $18\cdot48$; multiplying by the collision weight gives the bound. Finite uniform boundedness and anchoring are therefore certified, but these worst-case bounds far exceed the initial global gap of 1332 and give no useful floor in Theorem~\ref{thm:rate}. Assumption~\ref{ass:correction} avoids this obstacle: it needs the error only at a true block minimizer, where it is zero at every audited state, and at the returned proposal, where it averages 45.5 and 36.5.

\paragraph{Block geometry: Assumption~\ref{ass:eb}.}
Using the exact true minima in~\eqref{eq:routing-audit-D}, compute over the recorded states $x$ that start a round
\begin{equation}\label{eq:routing-audit-mu}
\mu_{\rm obs}=\min_{x}\min\left\{1,\frac{\sum_bD_b(x)}{F(x)-F^*}\right\}.
\end{equation}
It equals 0.43 on the first map, where the ratio is below 0.6 at one of the 145 states that start a round, and 0.70 on the second (Figure~\ref{fig:routing-assumption-diagnostics}(b)). Among final states, two on the first map, at $F=147$ and $F=145$, and four on the second, at $F=148$ or $149$, are single-block traps: every $D_b=0$ although the gap is positive, so only a joint change of two blocks can improve them. All six are final states at which the run had met its threshold and stopped; no round starts from a trap, and the recurrence is applied only at states that start a round, where $\mu$ holds. The certificate covers the recorded states, not the full product domain; with prompt v1, $\mu=0.6$ held at all 88 recorded states, including the final ones.

\paragraph{Scope of the connection to the theory.}
The task has a finite product domain of legal command arrays, exact global verification, unary and pairwise factors, and a conservative three-color dependency graph. The frozen parallel searches and best singleton commit implement the theory's merge with $\kappa=3$. The audit certifies the block error bound on the states that start a round and quantifies actual corrected local progress and score error. Assumptions~\ref{ass:correction} and~\ref{ass:local} are statements about conditional expectations; the audit measures their realized counterparts on the recorded states. The two-point errors of Assumption~\ref{ass:correction} are small, whereas a uniform bound over all routes would be far larger, and Appendix~\ref{app:routing-empirical-rate} uses the measured values. Because conflict weights are known, the experiment does not exercise Assumptions~\ref{dev:ass:features}--\ref{dev:ass:excitation}.

\begin{table}[t]\centering\footnotesize
\setlength{\tabcolsep}{3pt}
\begin{tabular}{@{}lrrrrrr@{}}\toprule
& $\mu$ & $\bar Q$ & $\bar\alpha$ & $\bar B$ & Coupled & Gap (bound)\\\midrule
\multicolumn{7}{@{}l}{\emph{Team (Alg.~\ref{alg:protocol})}}\\
Map 1, prompt v1 & 0.60 & 0.58 & 0.247 & 47.4 & 39\% & 12.5 (211.7)\\
Map 1 & 0.43 & 0.66 & 0.246 & 45.5 & 30\% & 5.9 (254.2)\\
Map 2 & 0.70 & 0.65 & 0.242 & 36.5 & 32\% & 7.2 (226.8)\\
Two floors & --- & 0.66 & 0.222 & 66.7 & 31\% & 13.1 (395.9)\\\midrule
\multicolumn{7}{@{}l}{\emph{Team (Alg.~\ref{alg:best-combination})}}\\
Map 1 & 0.55 & 0.65 & 0.244 & 36.3 & 26\% & 5.4 (328.7)\\
Map 2 & 0.75 & 0.60 & 0.222 & 40.2 & 28\% & 12.8 (399.4)\\
Two floors & --- & 0.63 & 0.210 & 60.3 & 27\% & 18.2 (604.7)\\\midrule
\multicolumn{7}{@{}l}{\emph{Team unsafe}}\\
Map 1 & 0.73 & 0.63 & --- & --- & 29\% & 5.0 (---)\\
Map 2 & 0.60 & 0.68 & --- & --- & 34\% & 10.1 (---)\\\bottomrule
\end{tabular}
\caption{Measured constants of the team runs: $\mu$ at states that start a round, mean search quality $\bar Q$, pooled means of the round coefficients, share of locally safe proposals that collide with another agent's robot, and final mean gap with the averaged bound in parentheses. Prompt v1 values use all states and its ten runs; the two-floor rows pool the searches of both maps, and their $\mu$ is not computed. For the unsafe team the round coefficients and the bound are not defined, because its loss can increase. The correction error at true block minimizers is zero in every case, so $\bar B$ is the mean error at returned proposals (summed over both floors on two floors).}
\label{tab:routing-constants}
\end{table}

\subsection{The bound, round by round}\label{app:routing-empirical-rate}
\begin{figure*}[t]
\centering
\includegraphics[width=\textwidth]{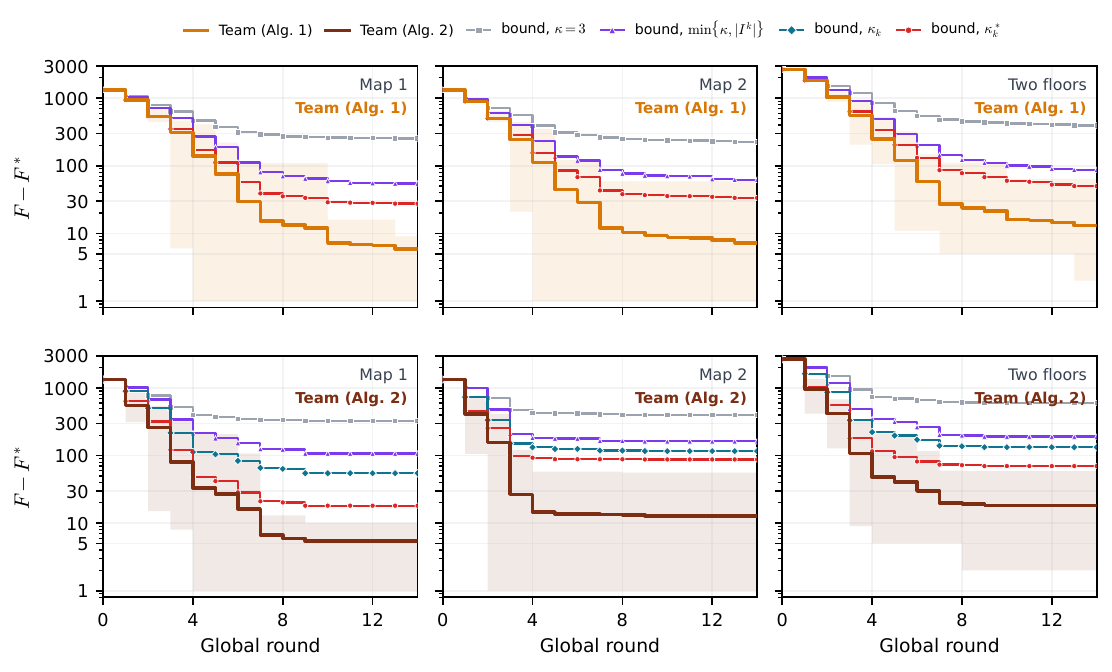}
\Description{Six panels in two rows, for map 1, map 2 and two floors, against the global round, with the gap on a logarithmic axis. Top row: the team with Algorithm 1, its observed mean gap (orange) with a light band between the pointwise minimum and maximum over the twenty runs, and three averaged bounds: with kappa equal to 3 (gray squares), with the smaller of kappa and the number of improving blocks (purple triangles) and with the realized number of colors (red circles). Bottom row: the team with Algorithm 2 (brown), with the same three bounds and the bound with the effective number of colors (teal diamonds). On the two maps the observed bands reach the bottom of the axis where runs reach the optimum.}
\caption{Observed mean gap of each team and its averaged trajectory-wise bound (Appendix~\ref{app:routing-empirical-rate}) for every task, with the divisors of Section~\ref{sec:main-combination}: $\kappa=3$; $\min\{\kappa,|I^k|\}$ (1 if no block improves), which bounds the number of colors that contain an improving block; the effective number of colors $\kappa_k$ (Proposition~\ref{prop:realized-conflicts}), which needs checked combinations and therefore applies only to Team (Alg.~\ref{alg:best-combination}); and the realized number of colors $\kappa^*_k$ (Corollary~\ref{cor:realized-rate}). Bands: pointwise minimum and maximum of the observed gap over the twenty runs, not confidence intervals; a run at the optimum is drawn at gap 1. On two floors each floor uses its own divisor.}
\label{fig:routing-theory-maps}
\label{fig:routing-merge-bound}
\end{figure*}

This subsection defines the retrospective upper bound of Figures~\ref{fig:routing-theory}(a) and~\ref{fig:routing-theory-maps} and evaluates it on every team run. The bound uses actual search quality and correction errors in each round, rather than full-domain worst-case errors or constant coefficients obtained by pooling different histories. For seed $s$ and an active round $k$, define
\begin{equation}\label{eq:routing-empirical-rate-coefficients}
\begin{split}
\alpha_{s,k}&=\frac{\sum_b Q_{s,k,b}D_b(x_s^k)}{3\Delta(x_s^k)},\\
B_{s,k}&=\frac13\sum_b\bigl(Q_{s,k,b}e^*_{s,k,b}+e^U_{s,k,b}\bigr).
\end{split}
\end{equation}
Here $Q$ is the realized corrected-local progress fraction, $D_b$ is the exact available true block improvement, $e^*$ is the absolute anchored score error at a certified true block minimizer, and $e^U$ is that error at the returned proposal. If the initial surrogate gap is zero, $D_b=0$ in these data, so its $QD_b$ contribution is zero. All true-minimizer errors $e^*$ are zero; returned-candidate errors include the actual missing-pair penalties and the zero-error cases.

For each recorded search, the argument of Lemma~\ref{lem:local}, applied to realized rather than expected quantities, gives
\[
g_{s,k,b}\geq Q_{s,k,b}D_b-Q_{s,k,b}e^*_{s,k,b}-e^U_{s,k,b}.
\]
The best of the three singleton colors has at least their mean verified gain. Hence
\begin{equation}\label{eq:routing-recorded-step-rate}
\Delta(x_s^{k+1})\leq(1-\alpha_{s,k})\Delta(x_s^k)+B_{s,k}.
\end{equation}
Independent replay verifies this inequality and $0\leq\alpha_{s,k}\leq1$ in every active round, using all returned block proposals, not just selected blocks. Averaging over blocks is conservative here: with $\kappa=3$ each color is a single block on one map, so the committed gain is the largest singleton gain. The largest per-block lower bound gives the sharper recurrence $\Delta(x_s^{k+1})\leq\max\{0,\Delta(x_s^k)-\max_b(Q_{s,k,b}D_b-Q_{s,k,b}e^*_{s,k,b}-e^U_{s,k,b})\}$, which we also check.

Propagate the bound separately for each seed. Initialize $E_{s,0}=\Delta(x_s^0)$ ($1476-144=1332$ on the first map) and set
\begin{equation}\label{eq:routing-empirical-rate}
E_{s,k+1}=\min\{E_{s,k},(1-\alpha_{s,k})E_{s,k}+B_{s,k}\}.
\end{equation}
The minimum uses the accepted loss's monotonicity. Since $1-\alpha_{s,k}\geq0$, induction with~\eqref{eq:routing-recorded-step-rate} proves $\Delta(x_s^k)\leq E_{s,k}$ at every recorded state. After run $s$ stops at $K_s$, hold both its actual loss and envelope fixed. Only then average the twenty envelopes:
\begin{equation}\label{eq:routing-mean-trajectory-bound}
\overline U_K=F^*+\frac1{20}\sum_{s=0}^{19}E_{s,\min\{K,K_s\}}.
\end{equation}

\paragraph{Results.} On the first map, the one-step inequality~\eqref{eq:routing-recorded-step-rate} and $0\leq\alpha_{s,k}\leq1$ hold in all 145 active rounds. Held after each run stops and averaged over the twenty runs, the envelope~\eqref{eq:routing-mean-trajectory-bound} ends at mean gap 254.2 after thirteen rounds, against 5.9 observed; it stays within a factor of 2.1 of the observed mean gap through round 3 (1.13, 1.49 and 2.03). With the largest per-block lower bound instead of the mean, the final mean gap is 6.8. The mean error term $B_{s,k}$ falls from 137 in round 1 to 0 in round 7, with small values in later rounds over the few runs still active. On the second map the inequality holds in all 159 rounds, and the envelope ends at 226.8 against 7.2 observed, or 8.0 with the largest per-block lower bound. On two floors it holds in all 186 rounds and ends at 395.9 against 13.1 (Appendix~\ref{app:routing-two-floors}). The pooled means of the coefficients are $\bar\alpha=0.246$, $0.242$ and $0.222$ and $\bar B=45.5$, $36.5$ and $66.7$ on the three tasks; they summarize the logs and are not substituted into the recurrence, because keeping each round's coefficients preserves their dependence on the current gap and history. Figure~\ref{fig:routing-theory-maps} shows the pointwise range of the observed gap over runs. With prompt v1 on the first map, the bound held in all 78 rounds of the ten team runs and ended at 211.7 against 12.5, or 15.3 with the largest per-block lower bound.

This is a valid upper bound on the recorded trajectories: the pathwise form of Theorem~\ref{thm:rate}, with realized $Q$ in place of $q_b(H)$ and measured two-point errors in place of $\delta_b^*,\delta_b^U$. It takes no supremum over collision-seeking routes and fits no decay rate to the loss curve. Because its coefficients come from the same runs, it explains the recorded progress rather than predicting fresh runs.

\paragraph{Coefficients from other runs.} The recurrence can also be evaluated with coefficients measured on different runs: average $\alpha_{s,k}$ and $B_{s,k}$ over the runs of one half that are active in round $k$, and propagate $\overline E_{k+1}=\min\{\overline E_k,(1-\bar\alpha_k)\overline E_k+\bar B_k\}$ from $\overline E_0=1332$. On the first map, coefficients from seeds 0--9 give a curve above the mean gap of seeds 10--19 in every round, ending at 141.5 against 6.4, and with the halves exchanged it ends at 167.9 against 5.4, again above in every round; on the second map the curves also lie above in every round in both directions, and with prompt v1, splitting the ten seeds into halves gave the same result.

\subsection{Greedy merging and conflicts}\label{app:routing-merge-theory}
Section~\ref{sec:main-combination} shows that the best-combination merge keeps the guarantees of Theorem~\ref{thm:rate} and that its rate depends on the realized number of colors $\kappa^*_k$, which is at most the effective number of colors $\kappa_k$. This subsection gives a greedy version that needs only quadratically many checks, measures the conflicts and the resulting envelope on the recorded runs, explains why the block error bound is still needed, and lists ways to weaken it.

\paragraph{A greedy rule.} With pairwise factors,~\eqref{eq:pairwise-expansion} determines the loss of every combination from the single checks and one check per pair, so $|I^k|(|I^k|-1)/2$ further checks suffice to find all conflicts and the best combination of improving blocks. When even the search over these combinations is too costly, a greedy rule suffices. Write $\deg'(b)$ for the number of remaining blocks that conflict with $b$; the rule repeatedly keeps the remaining block $v$ with the largest $g_v/(1+\deg'(v))$ and deletes $v$ and the blocks that conflict with it. It returns a conflict-free set with gain at least $\sum_{b\in I^k}g_b/(1+\deg_k(b))$ \citep{sakai2003greedy}. Indeed, each of the $1+\deg'(v)$ blocks $u$ deleted in that step, $v$ included, satisfies $g_u/(1+\deg'(u))\leq g_v/(1+\deg'(v))$, so together they contribute at most $g_v$ to $\sum_ug_u/(1+\deg'(u))$; since $\deg'(u)\leq\deg_k(u)$ and every block is deleted exactly once, the kept gains add up to at least $\sum_{b\in I^k}g_b/(1+\deg_k(b))$. Accepting the better of this set and the best color keeps Theorem~\ref{thm:rate}, and Corollary~\ref{cor:realized-rate} when every improving proposal conflicts with at most $\bar\kappa-1$ others, using a quadratic number of checks.

\paragraph{Recorded runs.} The runs with best combination checked all combinations of changed blocks, which measures every $\xi_{bb'}$. The checks confirm~\eqref{eq:pairwise-expansion} in all 44 and 29 rounds in which all three blocks changed on the two maps, and Proposition~\ref{prop:realized-conflicts}, together with the bound $\sum_bg_b/(1+\deg_k(b))$ of the greedy rule above, in all 113 and 111 rounds. Of the pairs of improving proposals, 71 of 135 conflict on map 1 and 42 of 90 on map 2. In 56 and 28 of these pairs, accepting both proposals creates a collision between their robots that neither proposal causes alone; in the other 15 and 14, both proposals remove the same collision. Another 11 and 15 pairs help each other ($\xi_{bb'}<0$). Rounds with two or more improving proposals have no conflict in 16 of 63 cases on map 1 and 10 of 42 on map 2. With the rounds that have fewer than two improving proposals, $\kappa_k=1$ in 66 of the 113 rounds on map 1 and 79 of the 111 on map 2, $\kappa_k=2$ in 41 and 30, and $\kappa_k=3$ in 6 and 2. The realized number of colors $\kappa^*_k$ averages 1.19 and 1.15 over the rounds that change the plan, never exceeds 1.98 and 2.00, and is below 1 in 8 and 9 rounds. In the rounds with two or more improving proposals it averages 1.29 and 1.26, against 1.84 and 1.81 for $\kappa_k$, and it is smaller than $\kappa_k$ in 49 of 63 and 35 of 42 of them: conflicting proposals are accepted together because that still lowers the loss most. Table~\ref{tab:routing-conflict-phases} splits the rounds into the start, the repair phase, in which the base still has an undelivered robot or a collision, and the refinement of a safe base. At the start every conflict is a new collision between the agents' new routes, yet the best combination often accepts both proposals; during repair many conflicts come from two proposals that remove the same collision; once the base is safe, conflicts are rare because at most one agent improves in most rounds.

\begin{table*}[t]\centering\footnotesize
\begin{tabular}{@{}llrrrrrrr@{}}\toprule
Map & Phase & Rounds & \makecell[rb]{Mean\\$|I^k|$} & \makecell[rb]{Conflicting\\pairs} & \makecell[rb]{New\\collision} & \makecell[rb]{Mean\\$\kappa_k$} & \makecell[rb]{Mean\\accepted} & \makecell[rb]{Accepting a\\conflicting pair}\\\midrule
Map 1 & Start (round 1) & 20 & 2.95 & 34/58 & 34 & 2.20 & 2.35 & 11\\
 & Repair (unsafe base) & 57 & 1.89 & 36/72 & 21 & 1.49 & 1.42 & 2\\
 & Refinement (safe base) & 36 & 0.61 & 1/5 & 1 & 1.03 & 0.61 & 0\\
\midrule
Map 2 & Start (round 1) & 20 & 2.90 & 19/56 & 19 & 1.75 & 2.85 & 13\\
 & Repair (unsafe base) & 35 & 1.63 & 21/30 & 7 & 1.49 & 1.17 & 1\\
 & Refinement (safe base) & 56 & 0.32 & 2/4 & 2 & 1.04 & 0.30 & 0\\
\bottomrule

\end{tabular}
\caption{Realized coupling in the best-combination runs by phase: number of rounds, mean number of improving proposals, conflicting pairs among pairs of improving proposals, conflicts in which the two proposals together create a collision that neither causes alone, mean effective number of colors, mean number of accepted blocks, and rounds whose accepted combination contains a conflicting pair.}
\label{tab:routing-conflict-phases}
\end{table*}

With $g_b\geq Q_bD_b-Q_be^*_b-e^U_b$ (Appendix~\ref{app:routing-empirical-rate}), \eqref{eq:realized-colors} gives $F(x)-F(x^+)=\sum_bg_b/\kappa^*_k\geq\sum_b(Q_bD_b-Q_be^*_b-e^U_b)/\kappa^*_k$, and Proposition~\ref{prop:realized-conflicts} gives the same with $\kappa_k$ in place of $\kappa^*_k$. Hence the recurrence~\eqref{eq:routing-recorded-step-rate} holds with the divisor 3 in~\eqref{eq:routing-empirical-rate-coefficients} replaced by either value in round $k$ of run $s$, with $\alpha_{s,k}$ capped at one; for Algorithm~\ref{alg:protocol} the realized number is $\sum_bg_b/\max_bg_b$. It exceeds one whenever the round discards an improving proposal, which happens in 79 of the 121 rounds of Team (Alg.~\ref{alg:protocol}) that change the plan on map 1 and in 64 of 115 on map 2; over all rounds that change the plan it averages 1.50 and 1.53 and reaches 2.85 and 2.94. The replay checks this inequality in every round, and the envelope~\eqref{eq:routing-empirical-rate} with these coefficients bounds every recorded gap. Figure~\ref{fig:routing-theory-maps} separates three effects. At most $\min\{\kappa,|I^k|\}$ colors contain an improving block, so dividing by this number is valid for any merge that keeps at least the best color. If no block improves, then $|I^k|=0$ and every $g_b$ is zero, so $\sum_b(Q_bD_b-Q_be^*_b-e^U_b)\leq0$: the gain bound holds with any positive divisor. As for $\kappa_k$ and $\kappa^*_k$ in Section~\ref{sec:main-combination}, we then divide by 1. It lowers the final envelope of the best-combination runs from 328.7 to 107.2 on map 1, from 399.4 to 162.5 on map 2 and from 604.7 to 190.6 on two floors, where each floor uses its own divisor, which is valid because the best combination is optimal on each floor; for the runs with Algorithm~\ref{alg:protocol} it lowers the envelope from 254.2 to 54.6 and from 226.8 to 62.2. Dividing by the effective number of colors $\kappa_k$, which needs the best combination, lowers the best-combination envelopes further to 54.8, 116.3 and 133.0. Dividing by the realized number $\kappa^*_k$, which also credits conflicting proposals accepted together, lowers them to 18.1, 87.6 and 69.9, against observed mean gaps of 5.4, 12.8 and 18.2; for Algorithm~\ref{alg:protocol}, its realized number gives 27.5, 33.5 and 50.0, against 5.9, 7.2 and 13.1. Through round 3, the best-combination envelope stays within a factor of 1.5, 3.7 and 1.7 of the observed mean gap with $\kappa^*_k$ and 2.7, 5.7 and 3.1 with $\kappa_k$, against 6.4, 17.8 and 8.9 with $\kappa=3$.

\begin{figure*}[t]
\centering
\includegraphics[width=.96\textwidth]{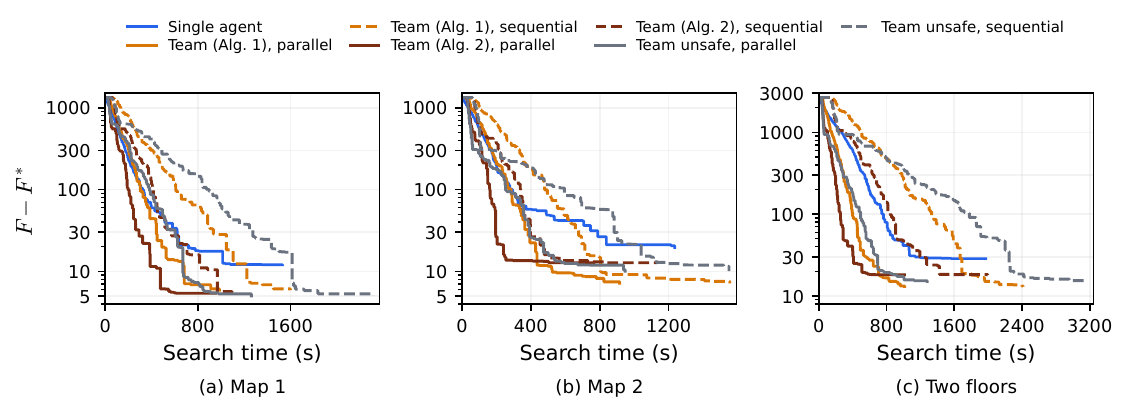}
\Description{Three panels of mean gap to the optimum against search time on logarithmic axes, for map 1, map 2 and two floors. For each of the three team variants, Algorithm 1, best combination and unsafe, the parallel search time (solid) and the sequential replay of the same answers (dashed) are drawn in one color; the single agent is drawn for reference. Each dashed curve reaches the values of its solid curve later, by a factor between about 1.6 and 2.7, and on two floors the sequential replays of the teams end near or beyond the single agent's search time.}
\caption{Parallel search time (solid) against the sequential replay of the same answers (dashed) for Team (Alg.~\ref{alg:protocol}), Team (Alg.~\ref{alg:best-combination}) and Team unsafe; mean gap $F-F^*$ on a logarithmic axis, twenty runs per curve; the single agent for reference.}
\label{fig:routing-sequential}
\end{figure*}

\paragraph{What still needs the block error bound.} The best combination combines the agents' proposals, and an agent returns its current block unless it finds an improvement on its corrected objective. At a single-block trap with exact corrections no block improvement exists, so every agent returns its current block and every merge of the returned proposals stays at the trap. No merge rule over the returned proposals can therefore replace Assumption~\ref{ass:eb}: with exact corrections, a reached trap keeps a positive gap forever, whereas Theorem~\ref{thm:rate} shows that the assumption suffices together with Assumption~\ref{ass:local}. Four and two final states of the best-combination runs on the two maps are such traps (Appendix~\ref{app:routing-merge}), and the unsafe team, which also accepts worse plans, ends in one such trap on each map.

\paragraph{Directions.} Three changes could weaken the assumption. First, a pairwise error bound $\sum_bD_b(x)+\sum_{b<b'}D_{bb'}(x)\geq\mu_2\Delta(x)$, where $D_{bb'}(x)$ is the best improvement from changing blocks $b$ and $b'$ together, holds at every trap that a joint change of two blocks can leave. Exploiting it requires proposals that are not individually improving: each agent could return a short list of candidates, such as its best proposal and alternatives that reroute its robots involved in reported collisions, and the coordinator would check combinations across the lists. A rate would then need a joint counterpart of Assumption~\ref{ass:local}: with some probability, the lists contain both parts of a good two-block change. Second, blocks can be coarsened after a stall: when a round accepts nothing, two agents' blocks are merged for the next round, so one agent controls six robots. A two-block change of the fine partition is a single-block change of the coarse one, and Theorem~\ref{thm:rate} applies to the coarse partition with its own $\mu$. Third, conflicts could be bounded in expectation: Corollary~\ref{cor:realized-rate} needs $\kappa_k\leq\bar\kappa$ in every epoch, whereas a condition such as $\mathbb E[\sum_bg_b/(1+\deg_k(b))\mid\mathcal H]\geq\bar\kappa^{-1}\mathbb E[\sum_bg_b\mid\mathcal H]$ would allow occasional dense conflicts; it couples the gains with the conflicts and would have to be measured like the other constants.

\subsection{Parallelism and decomposition}
\label{app:routing-decomposition}

We distinguish the effect of simplifying an individual search from the effect of executing searches concurrently. The discussion relates these mechanisms to the recorded comparison and the sequential timing replay.

\paragraph{Reducing the difficulty of a model request.}
Each team search writes three routes; with prompt v1 the whole-plan single agent writes nine in every answer and reaches a safe plan in one run of ten. A single agent that still sees and scores the whole plan but replaces only one route per answer is safe in nine runs of ten with prompt v1 and in all twenty with prompt v2 on the first map (Appendix~\ref{app:routing-development}), so much of the difficulty lies in the size of each answer. Within the team, the correction removes collisions between agents: without it, the team is safe in one run of ten with prompt v1, and in none of ten, two of ten and none of ten runs with prompt v2 on map 1, map 2 and two floors.

\paragraph{The same model in both policies.}
Full-plan search can revise several ownership groups jointly and has access to all assignments and full-plan feedback. On one map, our block protocol with Algorithm~\ref{alg:protocol} instead commits at most one proposed block per round, even though three searches run concurrently. External conflicts omitted from the available objective must first be exposed by verification; revising a route against the refreshed peer trajectory can require another response in a later round. Block commitment, conflict discovery and repair are therefore serial steps that parallel search does not remove; from the all-waiting start, the team needs at least three rounds to accept routes for all three agents. A model that reliably writes a correct full plan in one answer avoids this overhead, which is why a single agent that improves one route at a time can fall faster at first.

\paragraph{A sequential block control.}
The sequential replay replays the recorded team histories on a sequential worker (Figure~\ref{fig:routing-sequential}). It completes the three private searches in turn with the same $H$, objective and feedback. The base, retained pairs and peer trajectories remain frozen until all proposals are available, then the verifier applies the team's merge rule. Committing before the next search would change the policy. Replay preserves every answer, selection and round loss, changing only timestamps; it makes no new model requests.

For a fixed sequence of checked responses, let $t_{kb}=\sum_h a_{kbh}$ be block $b$'s response time in round $k$. The two search times would be
\begin{equation}\label{eq:routing-sequential-control}
\begin{split}
T_{\mathrm{sequential}}&=\sum_k\sum_{b}t_{kb},\\
T_{\mathrm{parallel}}&=\sum_k\max_{b}t_{kb}.
\end{split}
\end{equation}
Thus $T_{\mathrm{parallel}}\leq T_{\mathrm{sequential}}$, and for fixed durations $T_{\mathrm{sequential}}$ is at most three times $T_{\mathrm{parallel}}$ (six times on two floors). With prompt v2 the replay gives $853.9\pm374.2$ against $481.0\pm227.3$ seconds on the first map (ratio of means 1.78), $778.2\pm304.6$ against $482.3\pm201.6$ on the second (1.61), and $1632.1\pm468.5$ against $656.4\pm206.1$ on two floors (2.49), where six agents search at once; with prompt v1 it gave $1207.14\pm165.61$ against $589.39\pm64.85$ on the first map (2.05). For the other merge rules the ratios are 1.77, 1.68 and 2.48 with best combination and 1.96, 1.80 and 2.71 for the unsafe team on map 1, map 2 and two floors (Figure~\ref{fig:routing-sequential}). Played in turn, the team with Algorithm~\ref{alg:protocol} would be slower than the one-route single agent on every task, so parallel execution is what makes the team fast; its better plans do not depend on it. Both follow~\eqref{eq:routing-clock}. Actual sequential execution could change serving latencies and samples, so the replay is not a measured variant. Per-round durations, clocks, loss sequences and curves are retained with the analyses.

\section{SQLite: acquiring and reusing feedback}\label{app:sqlite-feedback}

This appendix gives the complete database experiment: the data and query choices, feedback acquisition policies, timing procedure and recorded results. The study isolates the cost of obtaining and reusing execution information.

\subsection{Purpose of the SQLite study}
The robot-routing study (Section~\ref{sec:main-routing}, Appendix~\ref{app:routing-revised}) is our primary experiment with live LLMs. This second study isolates a different question: can execution of a useful task supply reusable information more cheaply than calibrating alternative decisions in advance? We test this question in a real SQLite engine, with constructed data and deterministic planning procedures. The procedures are not language models, and the experiment does not measure parallel-agent inference. The synthetic study in Appendix~\ref{app:synthetic-correction} supplies the controlled connection to the convergence assumptions; this database study supplies a measured application of feedback acquisition.

Prior work already uses execution feedback to improve query planning. LEO corrects cost estimates from observed executions; Bao learns to guide an existing optimizer \citep{stillger2001leo,marcus2021bao}. We do not propose a new optimizer; we use this setting to measure the cost of acquiring and processing selected, reusable observations.

\subsection{Data, queries and available decisions}
Each report combines three aggregate queries, one per source; the main text calls a report simply a query. For source $s\in\{0,1,2\}$, tables $A_s$ and $C_s$ hold endpoint objects with group identifiers, and table $B_s$ holds edges with endpoint IDs and an integer amount. A query selects a group $g_s$ in $A_s$ and a group $h_s$ in $C_s$, then returns the count and sum of amounts of edges connecting the selected objects. The report is one SQL \texttt{UNION ALL} of the three aggregate branches. Correctness is independent of the selected join direction.

Each endpoint table has 4,096 groups and two objects per group, hence 8,192 rows. Half of the $A_s$ groups are randomly designated high-degree: each of their objects has 400 outgoing edges, while every other object has one. Thus $B_s$ contains
\[
2\cdot2{,}048\cdot400+2\cdot2{,}048=1{,}642{,}496
\]
edges. Independently, a randomly chosen half of the $C_s$ groups receives 97\% of the generated edges. Membership is not encoded in group IDs. Edge amounts are integers sampled from 1 through 9,999. Indexes cover endpoint groups and both edge endpoint IDs; ordinary \texttt{ANALYZE} is performed. The three-source database contains 4,927,488 edge rows in total and remains static throughout a workload.

For each branch, the controller chooses one of two physical plans: start from the selected $A_s$ group and traverse toward $C_s$ (left), or start from $C_s$ and traverse toward $A_s$ (right). \texttt{CROSS JOIN} and the relevant edge index enforce these alternatives; the native control lets SQLite choose. All alternatives compute the same aggregate, including when no edges match.

Let $v_s(\ell,g)$ and $v_s(r,h)$ be the numbers of edges incident to the selected left and right groups. Their sum over the three selected directions defines a separable work surrogate,
\begin{equation}\label{eq:sqlite-surrogate}
J_q(a)=\sum_{s=0}^{2}v_s(a_s,q_s),\qquad a\in\{\ell,r\}^3.
\end{equation}
Fan-out predicts work, but is not identical to elapsed time or total VM steps. The experimental outcomes are the independently measured execution and controller costs, rather than treating~\eqref{eq:sqlite-surrogate} as exact runtime. There is no collision-like coupling between branches: a centralized factorized controller can make the same choices without enumerating all eight joint plans.

\subsection{Acquiring a persistent correction}
The controller stores estimates for each source, side and group. Unknown entries start at 401, the source's average fan-out. It chooses the direction with the lower estimate, breaking ties in the order preferred by the native planner. After execution, the engine's \texttt{NVISIT} counter reports how many edge-index entries the chosen branch visited \citep{sqliteScan}. This replaces the estimate for the executed side's group. The estimate is retained for later queries with that group, even if the opposite endpoint predicate changes. Unexecuted alternatives are not revealed automatically.

The \emph{selective-feedback} policy retrieves the detailed profile only if the completed report used more than 1,000 SQLite virtual-machine steps. A trigger updates all three executed branches. The engine collects counters even on reports below this threshold; the gate saves retrieval and processing, not collection. Both instrumentation and retrieval costs remain in the learner's measured time. The threshold was selected during development and frozen before confirmation on the three fresh database seeds.

The observation changes the decision model rather than the answer to the query: it is reused cost information, the learned counterpart of a correction. Because each report brings new predicates, the stream is not a descent on one fixed solution; the study therefore measures the acquisition trade-off of Section~\ref{sec:main-cost} rather than a convergence rate.

\begin{table}[t]
\centering\small
\begin{tabular}{@{}lr@{}}\toprule
Quantity & Fixed value\\\midrule
Sources per report & 3\\
Groups per endpoint table & 4,096\\
Objects per group & 2\\
High/low outgoing degree per object & 400 / 1\\
Edges per source & 1,642,496\\
Probability of a high-degree right group & 0.97\\
Unknown fan-out estimate & 401\\
Profile trigger (strictly greater than) & 1,000 VM steps\\
Fresh database seeds & 3\\
Timing repetitions per seed and workload & 3\\\bottomrule
\end{tabular}
\caption{Fixed database and feedback settings. The 0.97 probability governs generation of edge endpoints, not a success probability of the controller.}
\label{tab:sqlite-settings}
\end{table}

\subsection{Workloads and controls}
The favorable stream has 48 reports. In each source, five-sixths of branch queries pair a low-degree $A_s$ group with a high-degree $C_s$ group, favoring the initial left choice. The remaining sixth repeatedly uses two high-degree $A_s$ groups with varying low-degree $C_s$ groups. Outliers are staggered across sources: half of reports contain one such branch and half contain none. A 192-report stream repeats this same cycle four times, testing how advance calibration amortizes. A balanced 48-report stream mixes the two patterns equally within each source, with less reuse of recurring costly groups. The workloads were deliberately constructed and developed adaptively.

The main controls include native planning, a fixed left plan, samples of 128 or 2,048 edges per source, complete joined histograms, exact cached active-group counts, and bounded cached counting. The strongest counting control first counts the predicted better side, then limits counting of its alternative once that alternative cannot improve the decision. It batches requests across sources and retains exact counts and reusable lower bounds. Calibration queries and their execution costs are charged. This control does not repeatedly identify every possible decision.

Cached plan trials execute both local directions when a predicate pair first appears, retain the lower-VM-step direction and reuse it. The first trial supplies the requested answer and is charged as productive work, while the other is calibration. This avoids charging both alternatives as pure overhead. Further controls retrieve feedback on every report, use centralized selective feedback with the same observations, or freeze feedback after four reports. The centralized controller in the main timing table enumerates eight joint surrogate choices; a separate factorized implementation removes that unnecessary enumeration and independently checks decision equivalence.

\subsection{Execution, timing and correctness}
The engine is SQLite 3.46.1, compiled from pinned public source with statement scan-status support. Fresh database seeds are 51031, 68437 and 89759. Source-specific generation uses seed $s_{\rm db}+18{,}181s$, and the query generator uses $s_{\rm db}+818{,}181$. There are three timing repetitions per database and workload, with randomized policy order. The instrumentation-controlled record contains $3\times3\times3\times12=324$ complete workload timings. These are three independent datasets, not nine independent datasets per table entry.

The timer covers the complete Python/SQLite controller: calibration, execution, instrumentation, profile processing, correctness assertions and trace bookkeeping. Data generation, index construction, ordinary \texttt{ANALYZE}, connection opening and initial native-plan inspection lie outside this timer. Initial database preparation took 10.76--12.21 seconds per seed and produced approximately 193.7\,MB per database; it is separately recorded. Measurements use warm local execution, without artificial network or evaluator delays. Baselines that do not consume profiles have scan-status collection disabled in the timing comparison. VM steps are engine instruction counts, not CPU instructions or a measured hardware time conversion.

During confirmation, every reported answer is checked against a Python reference accumulated independently while generating records. The instrumentation-control run replays those already verified answers. A separate offline audit validates all 483 extracted group-degree measurements in six 48-report streams. A factorized centralized control with the same observations matches all local choices and VM totals in 27 replayed streams. Regenerating the appendix independently recomputes the summary statistics from the 324 saved timings; it does not rerun the benchmark.

\begin{table*}[t]
\centering\small
\begin{tabular}{@{}lrrr@{}}\toprule
Policy & Favorable, 48 reports & Favorable, 192 reports & Balanced, 48 reports\\\midrule
Native planner & 6.64 & 23.44 & 13.27 \\
Always left & 6.54 & 25.05 & 15.77 \\
128-edge sample & 13.61 & 32.40 & 22.76 \\
2,048-edge sample & 36.47 & 51.29 & 40.16 \\
Full joined histograms & 433.34 & 440.16 & 439.11 \\
Exact counts, cached & 10.78 & 21.19 & 10.88 \\
Bounded counts, cached & 7.53 & 18.71 & 10.45 \\
Plan trials, cached & 152.60 & 158.85 & 105.29 \\
Feedback every report & 7.15 & 21.18 & 16.64 \\
Selective feedback & 4.92 & 15.74 & 17.13 \\
Centralized selective feedback & 4.86 & 16.31 & 17.08 \\
Feedback frozen after four & 7.08 & 24.39 & 15.30 \\
\bottomrule

\end{tabular}
\caption{Controller time of all policies, in milliseconds. Each entry is the pooled median of nine timings: three repetitions on each of three independent databases. All answers pass the recorded correctness checks. Calibration, feedback processing and policy-specific instrumentation are included.}
\label{tab:sqlite-times}
\end{table*}

\begin{figure*}[t]
\centering
\includegraphics[width=\textwidth]{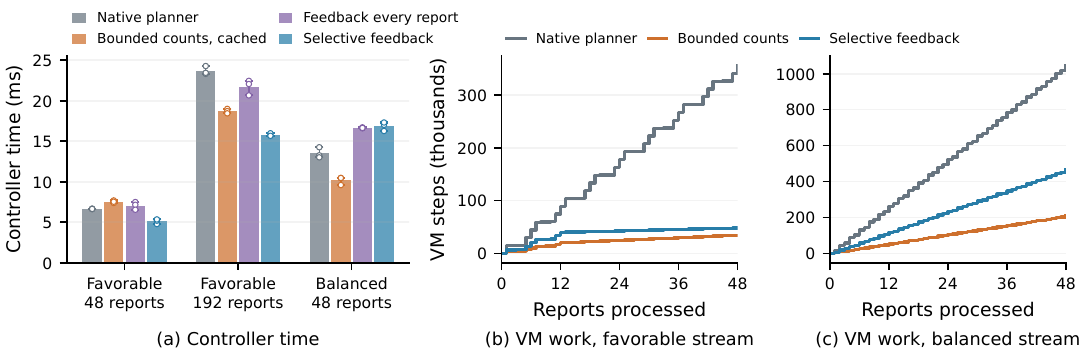}
\Description{Three panels. Left: grouped bars of controller time for the principal policies on three workloads, favorable with 48 and 192 reports and balanced with 48, with white points for the three databases and whiskers for their range. Middle and right: accumulated SQLite virtual-machine work against report index for the favorable and balanced 48-report streams, one line per policy, with light bands for the range over the three databases.}
\caption{(a) Controller time for the principal controls: bars show the mean of the three per-database timing medians; white points are those medians and whiskers their minimum and maximum. Table~\ref{tab:sqlite-times} reports pooled medians, and all policies compute the same correct answers. (b, c) Accumulated SQLite VM work for the favorable and balanced 48-report streams, including calibration, which is charged before its associated report. Lines average the three database trajectories and light bands show their minimum and maximum; repeated timing runs have identical VM counts within a database. VM work measures engine work, not elapsed time: the archive records complete controller time only at the end of a workload.}
\label{fig:sqlite-times}
\label{fig:sqlite-work}
\end{figure*}

\subsection{Results and a recorded correction}
Selective feedback takes 4.92\,ms on the favorable 48-report stream, compared with 7.53\,ms for bounded cached counting: 0.65 of the time, an absolute saving of 2.61\,ms. Per database, the relative time ranges from 0.62 to 0.71. Only six reports per database trigger detailed profiles. At 192 reports, the relative time rises to 0.84 as advance calibration amortizes. Feedback after every report costs 7.15\,ms on the shorter stream, showing why retrieval and processing should be charged.

The balanced stream reverses the result: selective feedback costs 17.13\,ms versus 10.45\,ms for bounded counting, or 1.64 times as much. It retrieves profiles on 48, 47 and 48 reports for the three databases, respectively. Learning from feedback thus pays off when costly groups are few and recur, and advance calibration is preferable when they are frequent; this is the trade-off predicted in Section~\ref{sec:main-cost}.

Time and VM work need not rank policies identically. On the favorable 48-report stream, bounded counting has lower median VM work (34,433 steps versus 48,106 for selective feedback), but performs 96 additional calibration statements per run. Selective feedback needs no calibration statements. Its lower controller time is consistent with saving these invocations; lower VM work alone does not explain the ranking. Figure~\ref{fig:sqlite-times} shows both measurements: controller time in panel (a) and VM work in panels (b, c).

One concrete trace illustrates how reuse changes later choices. On database seed 51031, zero-indexed report 0 uses source-0 groups $(0,2470)$ and chooses left. Its edge-index scan visits 800 entries, replacing the initial left-group estimate 401 with 800. At report 12, group 0 recurs with the different right group 2199; the retained estimate makes the controller choose right. Total work of the three-branch report is 337 VM steps, compared with 6,521 for report 0. These are different predicate pairs, so the difference is a trace illustration, not a controlled single-query speedup. Group 1 shows the same pattern at reports 6 and 18, with 6,521 and 345 VM steps. The later reports stay below the profile threshold and do not request new detailed feedback.

The centralized selective learner makes the same decisions. Its separately audited factorized version has pooled-median controller times 5.35, 16.28 and 16.99\,ms for the three workloads. These replays check implementation and information equivalence. A supplementary SQLite STAT4 control, with richer statistics granted for free and profiling disabled, records 6.14, 22.46 and 12.90\,ms of SQL-only time; it excludes the outer controller and is therefore not directly comparable with Table~\ref{tab:sqlite-times}.

\subsection{Relation to the assumptions}\label{app:sqlite-assumptions}
For a fixed report, the two-direction action spaces are finite, the known additive surrogate has independent source factors, and its blockwise minimization is exact. Thus the planner selects a surrogate minimizer, rather than exhibiting stochastic LLM search quality. If exact fan-outs were supplied for that fixed report, the additive fan-out objective would have block error-bound constant one. These structural facts concern the work surrogate in~\eqref{eq:sqlite-surrogate}, not the controller's actual runtime over the workload.

Assumption~\ref{ass:local} holds for every $\rho_b<1$, since each two-choice surrogate is minimized exactly. Successive reports change predicates instead of improving one solution under a fixed $F$, so Theorem~\ref{thm:rate} does not apply to the stream, and the learned table is updated by direct replacement of observed entries, the coordinate case of the sketch-and-project update of Appendix~\ref{sec:learn}, from per-branch counters rather than from scalar objective differences. The study instead tests the cost model of Section~\ref{sec:main-learning}: whether the information a correction needs is cheaper to acquire from useful work or by measuring alternatives in advance. Theorem~\ref{thm:rate} is checked with measured constants in the routing study and with certified constants in the synthetic study.

\subsection{Scope and retained data}
The workload is constructed to make the trade-off visible. Its joins are highly selective: 431 of 432 audited favorable branch aggregates and 429 of 432 balanced aggregates contain no matching edges, although finding no match still incurs execution work. There are three independent databases, totals are in milliseconds, the branches are separable, and host hardware details were not recorded.

The result is a measured acquisition trade-off: useful executions can reveal a small recurring set of missing costs cheaply enough to outperform a strong cached calibration control, whereas frequent or poorly transferable observations reverse the comparison. The appendix uses the final instrumentation-controlled results.

\section{Synthetic binary optimization}\label{app:synthetic-correction}

This appendix specifies the synthetic task and its four experimental series. It provides the assumption certificates, the cost of private work, the error floor under approximate corrections, and results for learning and reusing corrections.

\subsection{Purpose and experimental scope}
This third study directly instantiates the finite block-optimization protocol with an analytically tractable randomized local solver. It asks how an objective correction changes progress, how many local steps are worthwhile between synchronizations, and whether learning or reusing the correction repays its query cost. Unlike the routing experiments (Appendices~\ref{app:routing-task}--\ref{app:routing-revised}), it contains no LLM inference. Unlike Appendix~\ref{app:sqlite-feedback}, its convergence constants can be established on the entire task domain. All series use one fixed 48-variable task.

The four series are: exact correction with varying private horizon; approximate correction with a persistent bias; correction learned through charged scalar evaluations; and reuse of already-paid proposal evaluations with an observable stopping certificate. Their stopping rules and cost conventions are stated separately. Each series has its own script, and the tables and figures are rebuilt from the saved results.

\subsection{Objective, blocks and all fixed settings}
The objective is~\eqref{eq:main-synthetic-objective}: six blocks $x_b\in\{0,1\}^8$ on the path $1$--$2$--$3$--$4$--$5$--$6$, where neighboring blocks pay for disagreements of corresponding bits and each bit pays for disagreeing with its target $t_{bj}$. The coupling weight is $w=0.2$. Targets are sampled first as independent binary values, then unary weights are sampled uniformly from $[1,1.5]$ using NumPy's default random generator with seed 20260925. The actual targets, all 48 weights and the source are retained; subsequent experiments read this same instance rather than sampling another task. Initialization is $x^0=1-t$. The unique minimizer is $t$, with $F^*=3.4$, and the initial objective gap is $\Delta_0=58.6047775135$.

The private objective $L_b(u)=\sum_ja_{bj}|u_j-\tilde t_{bj}|$ omits coupling and uses the private target $\tilde t_b$. Blocks 1, 3 and 5 use $\tilde t_b=1-t_b$, while the other blocks have $\tilde t_b=t_b$. This deliberately inserts a local/global mismatch. Activity probabilities $\pi_b$ range from 0.5 to 1.0. The complete target and activity assignment is given in Table~\ref{tab:synthetic-task}.

\begin{table}[t]
\centering\small
\begin{tabular}{@{}crrr@{}}\toprule
Block & True target $t_b$ & Private target $\tilde t_b$ & $\pi_b$\\\midrule
1 & \texttt{10100000} & \texttt{01011111} & 0.5\\
2 & \texttt{01000001} & \texttt{01000001} & 0.6\\
3 & \texttt{00100010} & \texttt{11011101} & 0.7\\
4 & \texttt{10000010} & \texttt{10000010} & 0.8\\
5 & \texttt{11011001} & \texttt{00100110} & 0.9\\
6 & \texttt{11111101} & \texttt{11111101} & 1.0\\\bottomrule
\end{tabular}
\caption{The single fixed 48-bit task. Bit strings list coordinates 1 through 8 in order. Random solver trajectories change search randomness, not this task.}
\label{tab:synthetic-task}
\end{table}

\subsection{Exact correction and search}
With neighbors frozen at $x$, the coefficient of candidate bit $u_j$ in the global conditional objective is
\begin{equation}\label{eq:synthetic-coefficient}
A_{bj}(x)=a_{bj}(1-2t_{bj})
+w\!\sum_{r:\,|r-b|=1}(1-2x_{rj}).
\end{equation}
The corrected score is affine. Defining
\[
h_{bj}(x)=A_{bj}(x)-a_{bj}(1-2\tilde t_{bj}),
\]
the anchored correction is $R_b(u;x)=\sum_jh_{bj}(x)(u_j-x_{bj})$. It exactly equals the global conditional change minus the local change. A constant offset could not reverse the wrong private preference; the candidate-dependent vector does. Eight coefficients per block replace a table of 256 candidate values. Neighbor contributions are recomputed at each synchronization and remain fixed during local search.

Each epoch starts from the same shared base for all six searches. At a private step, block $b$ chooses a coordinate uniformly and, with probability $\pi_b$, sets it to minimize its affine corrected score; otherwise it leaves the bit unchanged. After $H$ steps, all six proposed singleton replacements are checked with the exact global objective. Their nonnegative verified gains are summed over the two colors $\{1,3,5\}$ and $\{2,4,6\}$. The larger-gain color is committed, using only strictly improving blocks. Ties select the first color. Same-color edits share no objective factor, so their verified gains add exactly. This differs from routing's complete three-block dependency graph, where each accepted color contains only one block.

The exact-correction series uses the horizons $H=1$, 2, 4, 8, 16, 32, 64 and 128, with 2,000 solver trajectories per horizon. Solver randomness uses seed $20260925+H$. Each corrected run has a ceiling of 512 epochs and stops on exact attainment of the offline-known optimum; every run succeeds before that ceiling. The uncorrected $H=8$ ablation uses only the private unary scores and a ceiling of 128 epochs. It solves none of 2,000 runs and leaves mean gap 30.979, so the correction is necessary on this task.

\subsection{Certifying the assumptions}\label{app:synthetic-assumptions}

We establish correction accuracy, local contraction and block geometry for this fixed task. These certificates determine the constants used in the comparison between the theoretical bound and observed convergence.

\paragraph{Correction accuracy.}
The exact vector satisfies Assumption~\ref{ass:correction} with $\delta_b^*=\delta_b^U=0$; indeed its error is zero for every binary candidate and neighbor context. The saved implementation checks 200 sampled exact-correction identities, with maximum discrepancy $1.33\times10^{-14}$, and 10,000 anchored approximate-correction inequalities. The algebraic identity supplies the uniform guarantee; sampled checks audit its implementation.

If a correction computed at neighbor context $z$ is reused after neighbors change, reanchoring at the current owned block gives the valid bound
\[
\sup_u|\widehat R_b(u;x)-R_b(u;x)|
\leq2w\sum_{r:\,|r-b|=1}\|z_r-x_r\|_1.
\]
The experiment refreshes context every epoch, so this staleness error is zero. Obtaining changed neighbor context is part of synchronization, not an assumed free oracle.

\paragraph{Conditional local contraction.}
Each unresolved coordinate's nonnegative contribution to affine suboptimality disappears with probability $\pi_b/8$ per private step. Linearity of conditional expectation gives Assumption~\ref{ass:local} with $\rho_b=\pi_b/8$, uniformly over bases and coefficient vectors. Thus
\begin{equation}\label{eq:synthetic-q}
\rho_{\min}=0.0625,\qquad
q(H)=1-(1-0.0625)^H,\qquad\kappa=2.
\end{equation}
This is an analytical property of the coordinate solver, not a value fitted to observed trajectories.

\paragraph{Block geometry and reference optimum.}
Let $c_b=w\deg(b)$, which is 0.2 for endpoints and 0.4 for interior blocks. Every $a_{bj}>c_b$. Repairing a wrong bit to $t_{bj}$ therefore saves more unary cost than it can add in incident coupling cost. This proves uniqueness of $t$ and gives a uniform valid block error bound
\begin{equation}\label{eq:synthetic-dominance}
\mu_{\rm dom}=\min_{b,j}\frac{a_{bj}-c_b}{a_{bj}+c_b}
=0.4307855764.
\end{equation}
Indeed, for wrong-bit set $I(x)$,
\[
\sum_bD_b(x)\geq\sum_{(b,j)\in I(x)}(a_{bj}-c_b),
\quad
\Delta(x)\leq\sum_{(b,j)\in I(x)}(a_{bj}+c_b).
\]
Their ratio establishes Assumption~\ref{ass:eb} on the full domain.

The tighter finite-state certificate is $\mu=0.7193550158$. It does not require enumeration of $2^{48}$ joint configurations. The objective separates across the eight bit positions: each feature has just $2^6=64$ block patterns. For each feature we enumerate its exact minimum and all singleton improvements, then minimize the improvement-to-gap ratio over its nonoptimal states. Full-state gaps and best-block improvements sum over features, so the full-state ratio is a weighted average of their ratios; the least feature ratio is valid globally and is attained by making only that feature nonoptimal. The total certificate uses $8\times64=512$ states. A separate appendix builder independently recomputes $F^*$ and this constant from the retained coefficients.

\paragraph{Theory-to-data comparison.}
Using this certified $\mu$, exact corrections and~\eqref{eq:synthetic-q}, Theorem~\ref{thm:rate} gives
\begin{equation}\label{eq:synthetic-rate}
\mathbb E F(x^K)\leq F^*+
\left(1-\frac{\mu q(H)}{2}\right)^K\Delta_0.
\end{equation}
Figure~\ref{fig:main-synthetic-known}(b) compares this sequence with the sample means for $H=8$ and $32$ at integer epochs; no coefficient is fitted. The mean gap stays below the bound at every epoch. Completed trajectories remain at $F^*$.

\subsection{Private work and synchronization}
One parallel private slot costs $c=1$; shared preparation, transfer, verification and integration cost $C$ per epoch. In the known-correction series, verification is included in this common overhead. A run finishing at epoch $K_{\rm end}$ has modeled cost $(H+C)K_{\rm end}$. We show $C\in\{0,5,20,100\}$, including the idealized zero-overhead control. Costs are in modeled work units; six agents use six local slots per private step in total.

\begin{table*}[t]
\centering\small
\begin{tabular}{@{}rrrrrr@{}}\toprule
$H$ & Completion epochs, mean $\pm$ SE & Cost, $C=0$ & Cost, $C=5$ & Cost, $C=20$ & Cost, $C=100$\\\midrule
1 & $61.73\pm0.37$ & 61.7 & 370.4 & 1296.4 & 6235.1 \\
2 & $33.60\pm0.18$ & 67.2 & 235.2 & 739.2 & 3427.2 \\
4 & $18.19\pm0.09$ & 72.8 & 163.7 & 436.6 & 1891.8 \\
8 & $10.34\pm0.05$ & 82.7 & 134.4 & 289.4 & 1116.4 \\
16 & $6.01\pm0.03$ & 96.1 & 126.2 & 216.3 & 696.9 \\
32 & $3.49\pm0.02$ & 111.6 & 129.1 & 181.4 & 460.5 \\
64 & $2.10\pm0.01$ & 134.6 & 145.1 & 176.7 & 345.0 \\
128 & $2.00\pm0.00$ & 256.0 & 266.0 & 296.0 & 456.0 \\
\bottomrule

\end{tabular}
\caption{Exact correction: all 2,000 trajectories per horizon reach the optimum on the same task. Standard errors reflect solver randomness. Cost is modeled parallel work plus synchronization overhead.}
\label{tab:synthetic-known}
\end{table*}

Figure~\ref{fig:main-synthetic-known}(c) plots these costs. At $C=20$, mean cost decreases from 1296.4 at $H=1$ to 176.7 at $H=64$, then rises to 296.0 at $H=128$. The observed best horizons among the eight tested are 1, 16, 64 and 64 for $C=0,5,20,100$. The relaxed cost criterion of Section~\ref{sec:horizon} gives integer minimizers $H_*=1$, 11, 20 and 35; the ceiled expected-gap criterion gives $H=22$ at $C=20$ when $\varepsilon=0.01\Delta_0$ and $H=1,\ldots,256$ are searched. Both move right with $C$, as the observed minima do. Figure~\ref{fig:main-synthetic-known}(c) places each $H_*$ on its cost curve by log-log interpolation between the tested horizons; the interpolated cost at $H_*$ is 1.00, 1.03, 1.16 and 1.29 times the best tested cost for $C=0,5,20,100$. The criterion is smaller for large $C$ because it minimizes a worst-case bound on the expected gap, whereas the experiment measures first exact completion: with $H\geq64$ nearly every block is solved exactly and runs finish in 2.0--2.1 epochs on average, one per color, faster than the geometric rate of the bound.

\subsection{Biased corrections and the floor}\label{app:synthetic-floor}
Theorem~\ref{thm:rate} allows a correction error that does not vanish to leave a floor that more private work cannot remove; this series shows that such a floor appears. The exact-correction series cannot show this, because its error is zero. This series keeps the task, solver, verification and merge above, but agents search the exact conditional coefficients plus a fixed bias,
\[
\widehat A_{bj}(x)=A_{bj}(x)+\nu\,s_{bj},\qquad s_{bj}\in\{-1,1\},
\]
with independent uniform signs drawn once per run and unknown to the agents. The corrected objective remains affine in $u-x_b$ and hence anchored; its error at a candidate $u$ is $\sum_j\nu s_{bj}(u_j-x_{bj})$. We test $\nu\in\{0,0.25,\ldots,2\}$ and $H\in\{8,32\}$ with 2,000 runs per configuration and 160 epochs, using seed $20261009$ offset by $1000H+100\nu$.

For every search the series records the correction errors $e^*_b=|\widehat f_b(u_b^*)-f_b(u_b^*)|$ at a true block minimizer $u_b^*$ and $e^U_b=|\widehat f_b(U_b)-f_b(U_b)|$ at the returned proposal $U_b$, the realized fraction $Q_b$ of the corrected improvement, and the exact block improvement $D_b$; $e^*_b$ and $e^U_b$ are the realized counterparts of $\delta_b^*$ and $\delta_b^U$. The argument of Lemma~\ref{lem:local}, applied to these realized quantities, gives $g_b\geq\max\{0,Q_b(D_b-e^*_b)-e^U_b\}$; the merge commits at least the larger color sum of these lower bounds. Propagating this inequality separately for each run from the initial gap yields an envelope that the observed gap never exceeds. The script asserts the local inequality, the merge inequality and the envelope for every run and epoch, and confirms that the envelope equals the observed gap when $\nu=0$.

\begin{table}[!h]
\centering\footnotesize
\setlength{\tabcolsep}{3pt}
\begin{tabular}{@{}rrrrrrr@{}}\toprule
& \multicolumn{3}{c}{$H=8$} & \multicolumn{3}{c}{$H=32$}\\\cmidrule(lr){2-4}\cmidrule(l){5-7}
$\nu$ & Final gap & Solved & Bound & Final gap & Solved & Bound\\\midrule
0.00 & $0.00\pm0.00$ & 2000 & 0.0 & $0.00\pm0.00$ & 2000 & 0.0\\
0.25 & $0.00\pm0.00$ & 2000 & 10.6 & $0.00\pm0.00$ & 2000 & 6.8\\
0.50 & $0.00\pm0.00$ & 2000 & 20.2 & $0.00\pm0.00$ & 2000 & 13.4\\
0.75 & $0.30\pm0.01$ & 1340 & 26.8 & $0.27\pm0.01$ & 1187 & 19.2\\
1.00 & $3.08\pm0.04$ & 48 & 30.4 & $2.79\pm0.03$ & 40 & 24.7\\
1.25 & $13.12\pm0.07$ & 0 & 36.9 & $13.61\pm0.08$ & 0 & 33.8\\
1.50 & $23.42\pm0.10$ & 0 & 45.8 & $23.55\pm0.10$ & 0 & 44.5\\
1.75 & $26.66\pm0.10$ & 0 & 51.5 & $26.88\pm0.10$ & 0 & 50.5\\
2.00 & $30.01\pm0.09$ & 0 & 55.6 & $29.79\pm0.10$ & 0 & 55.3\\
\bottomrule\end{tabular}
\caption{Approximate corrections: final mean gap after 160 epochs ($\pm$ one standard error), runs reaching the optimum out of 2,000, and the mean per-run recurrence envelope.}
\label{tab:synthetic-floor}
\end{table}

The smallest magnitude of an exact coefficient is $\min_{b,j}(a_{bj}-c_b)=0.605\geq0.6$. A bias below it never reverses the sign of a coefficient, so decisions and trajectories coincide with exact correction and every run reaches the optimum. Above it, a bit whose coefficient magnitude is below the bias and whose bias opposes the required change is never repaired, and the gap settles at a floor that grows with $\nu$. Once the bias exceeds every coefficient magnitude (at most $a_{bj}+2w\leq1.9$), about half of the coefficients are reversed and the floor approaches the uncorrected level 31. The floors for $H=8$ and $H=32$ differ by at most 0.49 at every $\nu$, against values up to 30: longer private search reaches the floor sooner but does not lower it. The envelope is conservative for small biases because it charges both two-point errors in every search, even when they do not change a decision. A uniform bound would be $\delta_b=8\nu$, attained at the initial state in which every bit is wrong, and its floor $B_\delta/\alpha$ is already 64 at $H=8$ and 37 at $H=32$ for $\nu=0.25$ and exceeds the initial gap for every $\nu\geq0.5$; only the two-point errors make the bound informative here.

\begin{table*}[t]
\centering\small
\begin{tabular}{@{}lrrrr@{}}\toprule
Policy & $H$ & Completion epochs, mean $\pm$ SE & Mean verifier queries & Cost, mean $\pm$ SE\\\midrule
Learn ($p=1$) & 8 & $12.78\pm0.07$ & 87.1 & $444.8\pm1.96$ \\
Learn ($p=0.5$) & 8 & $21.32\pm0.13$ & 85.5 & $682.5\pm3.58$ \\
Learn ($p=0.25$) & 8 & $42.35\pm0.29$ & 83.1 & $1268.9\pm8.09$ \\
Refresh every round & 8 & $10.24\pm0.07$ & 524.8 & $811.6\pm5.06$ \\
Identify once & 8 & $10.24\pm0.07$ & 81.1 & $367.9\pm1.89$ \\
Learn ($p=1$) & 32 & $8.86\pm0.02$ & 80.9 & $541.7\pm1.31$ \\
Learn ($p=0.5$) & 32 & $19.57\pm0.12$ & 77.8 & $1095.4\pm6.39$ \\
Learn ($p=0.25$) & 32 & $40.45\pm0.29$ & 76.4 & $2179.6\pm14.90$ \\
Refresh every round & 32 & $3.44\pm0.03$ & 177.9 & $356.8\pm2.59$ \\
Identify once & 32 & $3.44\pm0.03$ & 60.8 & $239.7\pm1.37$ \\
Learn ($p=1$) & 64 & $8.08\pm0.01$ & 76.4 & $755.3\pm0.81$ \\
Learn ($p=0.5$) & 64 & $19.55\pm0.12$ & 75.7 & $1718.0\pm9.78$ \\
Learn ($p=0.25$) & 64 & $40.44\pm0.28$ & 75.1 & $3472.4\pm23.45$ \\
Refresh every round & 64 & $2.10\pm0.01$ & 111.1 & $287.7\pm1.28$ \\
Identify once & 64 & $2.10\pm0.01$ & 58.1 & $234.8\pm0.82$ \\
\bottomrule

\end{tabular}
\caption{Every calibrated configuration in the learned-correction series: 1,000 trajectories each on the fixed task. Queries include initial evaluation, deliberate calibration and changed-proposal verification. Costs use $C=20$, unit parallel private-slot cost and one additional unit per counted scalar evaluation.}
\label{tab:synthetic-learning}
\end{table*}

\subsection{Learning the correction}\label{app:synthetic-learning}
This series tests the learning mechanism of Section~\ref{sec:main-learning} and Appendix~\ref{sec:learn}: agents start without the correction and learn its fixed coefficients from charged scalar evaluations. It keeps the same task, coupling and coordinate solver but hides the static unary correction coefficients. Writing
\[
\vartheta_{bj}=a_{bj}(1-2t_{bj})-a_{bj}(1-2\tilde t_{bj}),
\]
the algorithm starts at $\widehat\vartheta=0$ with known bounds $M_{bj}=2a_{bj}$. The local scores and coupling are available; target bits and true coefficients are used only by the verifier and offline diagnostics. A one-bit probe supplies exactly the scalar residual in~\eqref{eq:bitprobe}, with displacement $\pm1$, and identifies that coordinate. Coefficients are retained forever; changed neighbors affect only the recomputed coupling context.

Before each epoch's $H$ private steps, every block with unobserved coordinates probes with probability $p$, choosing uniformly among those coordinates. The tested probabilities are $1,0.5,0.25$ and horizons are $8,32,64$. The refresh baseline queries all 48 coefficients at every epoch; identify-once queries them only at initialization. Both use the same scalar interface and pay acquisition costs. An uncorrected baseline retains known coupling but never calibrates the unary correction; it differs from the earlier private-score-only ablation.

Each configuration has 1,000 trajectories with seed $20260926+H+100p$. Controls at $p=1$ use that same seed convention. Every calibrated configuration reaches the optimum. The uncorrected $H=32$ case solves none of 1,000 trajectories within 256 epochs. These trajectories stop at the offline-known optimum; the next series replaces this rule with an observable certificate.

The remaining coefficient radius $\Gamma_k$ is observable and upper-bounds candidate correction error. With deliberate probing, Proposition~\ref{prop:radius} applies with $\gamma=p/8$ and at most $d_{\rm tot}=48$ extra calibration queries over a whole run. The all-coordinate excitation assumption is unnecessary here, because an identified coefficient is not probed again; the binary specialization supplies the uncertainty analysis. Separate projection and coefficient-inequality checks audit the relevant identities.

\begin{table*}[t]
\centering\small
\setlength{\tabcolsep}{4pt}
\begin{tabular}{@{}lrrrrrr@{}}\toprule
Policy & $H$ & Epochs, mean $\pm$ SE & Queries & Probes & Reused & Cost, mean $\pm$ SE\\\midrule
Dedicated & 8 & $12.78\pm0.07$ & 87.1 & 48.0 & 0.000 & $444.8\pm1.96$ \\
Verification reuse & 8 & $12.78\pm0.07$ & 83.9 & 44.8 & 3.186 & $441.6\pm1.96$ \\
Identify once & 8 & $10.24\pm0.07$ & 81.1 & 48.0 & 0.000 & $367.9\pm1.89$ \\
Dedicated & 32 & $8.86\pm0.02$ & 80.9 & 48.0 & 0.000 & $541.7\pm1.31$ \\
Verification reuse & 32 & $8.86\pm0.02$ & 80.2 & 47.3 & 0.703 & $541.0\pm1.30$ \\
Identify once & 32 & $3.44\pm0.03$ & 60.8 & 48.0 & 0.000 & $239.7\pm1.37$ \\
Dedicated & 64 & $8.08\pm0.01$ & 76.4 & 48.0 & 0.000 & $755.3\pm0.81$ \\
Verification reuse & 64 & $8.08\pm0.01$ & 76.3 & 48.0 & 0.046 & $755.2\pm0.81$ \\
Identify once & 64 & $2.10\pm0.01$ & 58.1 & 48.0 & 0.000 & $234.8\pm0.82$ \\
\bottomrule

\end{tabular}
\caption{All verification-reuse configurations, with 1,000 trajectories each. Queries count every charged scalar evaluation, probes count deliberate calibration, and reused counts coefficients identified from already-paid proposal evaluations. Epoch and cost errors are standard errors of their means. Identify-once remains the strongest control.}
\label{tab:synthetic-reuse}
\end{table*}

\begin{figure*}[t]
\centering
\includegraphics[width=\textwidth]{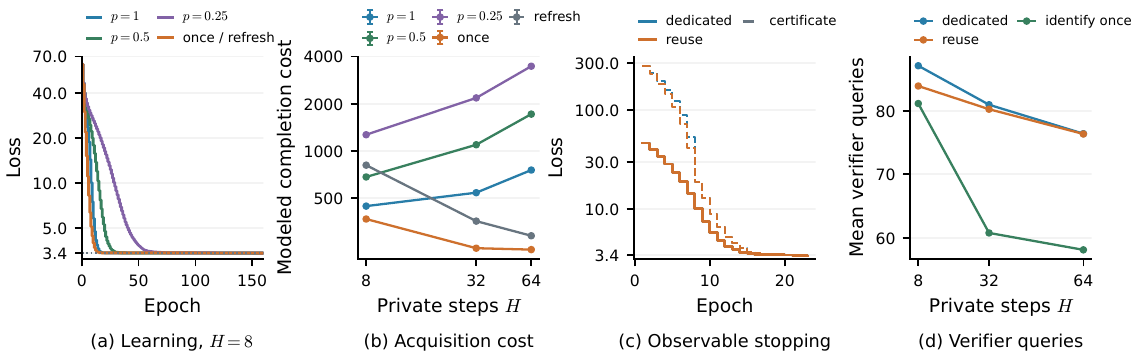}
\Description{Four panels. First: mean loss against epoch at eight private steps for the acquisition policies; refresh and identify-once coincide. Second: mean completion cost of every acquisition policy at eight, 32 and 64 private steps, with one-standard-error bars. Third: true mean loss and the optimum plus the mean online stopping certificate against epoch at eight private steps, for dedicated calibration and verification reuse. Fourth: mean verifier queries against the number of private steps for dedicated calibration, verification reuse and identify-once.}
\caption{Learning the correction and reusing verification. (a) Mean loss at $H=8$ for the acquisition policies; refresh and identify-once have the same trajectory. (b) Mean completion costs of all acquisition policies at the three horizons, with one-standard-error bars. (c) True mean loss (solid) and $F^*$ plus the mean online stopping certificate (dashed) at $H=8$, for dedicated calibration and verification reuse; $F^*$ is added only to display the offline comparison. (d) All counted verifier queries, averaged over 1,000 trajectories, including calibration and the initial evaluation.}
\label{fig:synthetic-learning}
\label{fig:synthetic-reuse}
\end{figure*}

Here a completed run costs $(H+20)K_{\rm end}+N_{\rm q}$, where $N_{\rm q}$ counts the initial evaluation, calibration calls and verification of changed proposals. The base score is cached, unchanged proposals need no new call, and the selected color's value follows from exact additivity. These query charges are explicit, unlike the exact-correction series where verification is included in $C$, so costs are compared within a series. Offline diagnostic evaluations are excluded.

At $H=8,p=1$, gradual learning costs 444.8, compared with 811.6 for repeated refresh and 367.9 for identify-once (Figure~\ref{fig:synthetic-learning}(a, b)). At this horizon, retaining learned coefficients thus nearly halves the cost of reidentifying them every epoch; at $H=32$ and 64, where refresh runs finish in 3.4 and 2.1 epochs while gradual learning needs at least eight epochs to probe each block, repeated refresh is cheaper than gradual learning (356.8 against 541.7 and 287.7 against 755.3). When the hidden part is static and all 48 coefficients can be queried at once, identifying them once is cheapest; at larger $H$ the gap widens because private steps taken before calibration is complete are wasted (gradual learning 541.7 and 755.3, identify-once 239.7 and 234.8). Reducing $p$ reduces total queries only slightly but increases epochs and cost. Gradual learning pays off instead when only a few of many unknown values are ever needed and they recur, as in the favorable SQLite workloads (Appendix~\ref{app:sqlite-feedback}).

\subsection{Reuse and observable stopping}
The fourth series compares dedicated calibration, dedicated calibration plus verification reuse, and identify-once at $p=1$ and $H=8,32,64$, with 1,000 trajectories per configuration. After proposals are fixed, a singleton evaluation can identify one still-unknown coefficient when its displacement contains exactly one unknown coordinate after known contributions are subtracted. Both accepted and rejected proposals are eligible. Measurements with several unknown coordinates are skipped. Updating the estimator does not regenerate this epoch's proposals; identified coefficients are used from the next epoch on. Identified coordinates are removed from future deliberate calibration, and calibration plus reused identifications never exceeds 48 per trajectory.

Stopping now uses an observable gap certificate. At the new committed base let $\widehat A_{bj}$ be the private coefficient plus learned unary correction and refreshed coupling coefficient. Exact affine minimization gives the total surrogate improvement
\[
\widehat D(x)=\sum_{b,j}\big[\widehat A_{bj}x_{bj}-\min\{0,\widehat A_{bj}\}\big].
\]
If $\Gamma(x)$ sums the remaining coordinate radii,~\eqref{eq:stop-inexact} supplies
\begin{equation}\label{eq:synthetic-stop}
\Delta(x)\leq \overline\Delta(x)\eqdef
\frac{\widehat D(x)+\Gamma(x)}{\mu_{\rm dom}}.
\end{equation}
The known weights and coupling determine $\mu_{\rm dom}$ in~\eqref{eq:synthetic-dominance}; the stopping decision does not use the hidden target or optimum. It costs $O(48)$ arithmetic and no new verifier query, and stops when $\overline\Delta(x)<10^{-9}$. This arithmetic and residual processing are included in $C=20$. All 9,000 trajectories stop with zero certified gap, and saved mean certificates dominate saved mean true gaps (Figure~\ref{fig:synthetic-learning}(c)). The simulator's run-time assertions check domination separately for each trajectory.

At $H=8$, reuse lowers mean cost from 444.8 to 441.6 and identifies 3.186 coefficients without extra queries. At $H=32$ and 64, the corresponding reused counts fall to 0.703 and 0.046 and cost savings become negligible. Mean completion epochs are unchanged, and identify-once remains cheaper at every horizon (Figure~\ref{fig:synthetic-learning}(d)). Reuse is limited because an uncalibrated block with the wrong private target proposes no change in the coordinates needing correction, so its evaluation reveals nothing about them; deliberate probing remains necessary.

\paragraph{Summary.}
On a task where every constant is certified, the observed mean gap stays below the bound of Theorem~\ref{thm:rate}, a persistent correction error leaves a floor that longer private search does not lower, the best private horizon grows with the synchronization cost as $J(H)$ predicts, and learned coefficients are retained so that calibration stops after at most 48 queries. Costs are in modeled units on one fixed instance.

\end{document}